\documentclass{aa}  

\usepackage{graphicx}
\usepackage{mathtools}
\usepackage{txfonts}
\usepackage{graphicx}
\usepackage{txfonts}
\usepackage{color}
\usepackage{textcomp}

\begin{document}

   \title{What can be learnt from UHECR anisotropies observations}

\subtitle{Paper II: intermediate-scale anisotropies}

   \author{D. Allard
          \inst{1},
          J. Aublin, %\fnmsep\thanks{Grand vainqueur de la belette de Winchester}, 
          B. Baret
         \and E. Parizot
          }

   \institute{Université Paris Cité, CNRS, Laboratoire Astroparticule et Cosmologie, F-75013 Paris, France\\
             \email{allard@apc.in2p3.fr}
             }

   \date{Received ... ; accepted ... }

    % \abstract{}{}{}{}{} 
% 5 {} token are mandatory

  \abstract
   {%In the recent years, evidences for an anisotropic distribution of ultra-high-energy cosmic rays (UHECRs) have been claimed. Besides a dipole modulation in right ascension reported by the Auger collaboration and discussed in the first paper of this series, various hints of anisotropies at intermediate angular scales have been reported by the Pierre Auger and the Telescope Array (TA) collaborations.
   Various signals of anisotropy of the ultra-high-energy cosmic rays (UHECRs) have recently been reported, whether at large angular scales, with a dipole modulation in right ascension observed in the data of the Pierre Auger observatory (Auger), as discussed in the first paper accompanying the present one, or at intermediate angular scales, with flux excesses identified in specific directions by Auger and the Telescope Array (TA) collaborations.   
   %   Ultra-high-energy cosmic rays have long been a center of great interest in astroparticle physics, due to the possibility they offer to constrain the physical processes and astrophysical parameters in the most energetic sources of the universe. In the recent years various  evidences of the presence of anisotropies in the UHECR sky have been claimed by several experiments, in particular a right ascension dipole modultion passing the $5\sigma$ significance threshold has been reported by the Pierre Auger observatory. These observations bring back to the front of the scene the eventuality of constraining the nature of UHECR sources despite the evidences that the  composition is becoming heavier at highest energies.
    }
   {
   %We investigate the implications of the current data regarding these intermediate scale anisotropies, and examine to what extent they can be used to shed some light on the origin of UHECRs, and constrain the astrophysical and/or physical parameters of the source scenarios. We investigate the possibility and importance of observing the energy evolution of these anisotropies of the UHECR composition and discuss the potential benefit of observing the UHECR sky with larger exposure future observatories.
   We investigate the implications of the current data regarding these intermediate scale anisotropies, and examine to what extent they can be used to shed light on the origin of UHECRs, and constrain the astrophysical and/or physical parameters of the viable source scenarios. We also investigate what could be learnt from the study of the evolution of the various UHECR anisotropy signals, and discuss the expected benefit of an increased exposure of the UHECR sky using future observatories.
%   We investigate the potential constraints which can be brought on the nature of UHECR sources by the various recently reported observations related to  UHECR large scale  anisotropies, in particular the above mentioned observation of a dipole modulation as well as of the absence of significant signal for higher order multipoles. We  investigate the possibility of observing an associated anisotropy of the UHECR composition and discuss, the relevance of a good determination of the composition and of the separation of the different nuclear components of the UHECR dataset in the context of anisotropy studies. We also discuss the interest and relevance of observing the UHECR sky with larger exposure future observatories.
   }
   {
  We simulate realistic UHECR sky maps for a wide range of astrophysical scenarios satisfying the current observational constraints, with the assumption that the UHECR source distribution follows that of the galaxies in the Universe, also implementing possible biases toward specific classes of sources. In each case, several scenarios are explored with different UHECR source compositions and spectra, a range of source densities and different models of the Galactic magnetic field. We also implement the Auger sky coverage, and explore various levels of statistics. For each scenario, we produce 300 independent datasets on which we apply similar analyses as those recently used by the Auger collaboration, searching for flux excesses through either blind or targeted searches and quantifying correlations with predefined source catalogs through a likelihood analysis.
%   We simulated realistic UHECR sky maps for a wide range of possible astrophysical scenarios motivated by the current observational constraints, taking the energy losses and photo-dissociation of the UHE protons and nuclei into account, as well as their deflexions by intervening magnetic fields. Datasets were built assuming various statistics and sky coverages, in particular that of the Pierre Auger observatory at the time of the discovery of the dipole modulation as a reference. The simulated datasets were analysed with methods similar to those used by the Pierre Auger collaboration for the search of large scale anisotropies. A statistical study of the resulting anisotropies was performed for each astrophysical scenario, varying the UHECR source composition and spectrum and the source density and exploring a set of three hundred independent realizations for each choice of a parameter set. We brought a particular attention to scenarios assuming UHECR sources spatial distribution follows that of the galaxies in the Universe.
   }
     {We find that: i) with reasonable choices of the parameters, the investigated astrophysical scenarios can easily account for the significance of the anisotropies reported by Auger%and TA
    , even with large source densities; ii) { the direction in which the maximum flux excess is found in the Auger data differs from the region where it is found in most of our simulated datasets, although an angular distance as large as that between the Auger direction and the direction expected from the simulated models at infinite statistics, of the order of $\sim$$20^\circ$, occurs in $\sim$25\% of the cases;} %however, the direction in which the maximum flux excess is found in the Auger data differs from the region where it is found in most of our simulated datasets; 
    iii) for datasets simulated with the same underlying astrophysical scenario, and thus the same actual UHECR sources, the significance with which the isotropy hypothesis is rejected through the Auger likelihood analysis can be largest either when ``all galaxies'' or when only ``starburst'' galaxies are used to model the signal, depending on which model is used to model the Galactic magnetic field and the resulting deflections ; iv) the study of the energy evolution of the anisotropy patterns can be very instructive and provide new astrophysical insight about the origin of the UHECRs; v) the direction in which the most significant flux excess is found in the Auger dataset above 8~EeV appears to essentially disappear in the dataset above 32~EeV, and, conversely, the maximum excess at high energy has a much reduced significance in the lower energy dataset; vi) both of these appear to be very uncommon in the simulated datasets, which could point to a failure of some generic assumption in the investigated astrophysical scenarios, such as the dominance of one type of sources with essentially the same composition and spectrum in the observed UHECR flux above the ankle; vii) given the currently observed level of anisotropy signals, a meaningful measurement of their energy evolution, say from 10~EeV to the highest energies, will require a significant increase in statistics and a new generation of UHECR observatories.}

  {}


   \maketitle
%
%-------------------------------------------------------------------

\section{Introduction}

   The arrival directions of ultra-high-energy cosmic rays (UHECRs) are a key observable to understand the origin of these particles and to identify their sources. Different signals now indicate with a high level of confidence that the UHECR sky is genuinely anisotropic.
   The most statistically significant, to date, has been reported by the Auger Collaboration \citep{AugerObs} and consists in a dipole modulation in right ascension of the arrival directions of the cosmic rays with energy greater than 8 EeV \citep{AugerDip2017}. In a previous paper (\citet{Paper I}, hereafter Paper~I), we examined the extent to which this large-scale anisotropy signal, together with the reported weakness of higher multipole modulations, could be used to constrain the astrophysical models of the origin of UHECRs. We compared the observations with comprehensive simulations of the UHECR sky exploring a wide range of astrophysical scenarios and taking into account the energy losses, nuclear interactions and deflections of the particles in the extragalactic and galactic media. The common assumption of these scenarios was that the distribution of the UHECR sources in the universe follows essentially the distribution of galaxies (a randomly selected subset of them), although possibly with different weights depending on their luminosity or whether they belong to large galaxy clusters. 

One of our conclusions was that, for suitable choices of the parameters within the range allowed by the astronomical observations, it is relatively easy to reproduce the amplitude of the first-order (dipole) angular modulation observed in the Auger data, as well as its evolution with energy. The situation is, however, highly degenerated since this general agreement can be obtained with different sets of assumptions on the astrophysical and physical parameters, essentially due to the possibility to adjust the amplitude and coherence length of the magnetic fields, which are currently poorly constrained. { Thus, the amplitude of the first-order large-scale anisotropy does not provide, in the present stage, strong constraints about the UHECR source scenarios and their various physical parameters.}
%cannot be used, in the present stage, to draw useful information about the UHECR source scenarios and derive model-independent constraints on the various model parameters individually. 

Another result was that, at least at face value, the direction of the dipole modulation reconstructed from the Auger data appears at odds with the model expectations, for essentially all the scenarios investigated. This calls for a reconsideration of their main assumptions, either regarding the source distribution itself or the assumed magnetic field configuration, especially in our Galaxy. It also calls for some caution about the conclusions of phenomenological studies investigating only one aspect of the observational data, and further suggests that reliable constraints about the nature of the UHECR sources will also require complementary input from other domains of astrophysics.

Although with lower statistical significance, notably below the $5\sigma$ discovery threshold after penalisation, other departures from isotropy have been reported both by the Pierre Auger Observatory (hereafter Auger) and the Telescope Array (TA, \citet{TAObs}) at smaller angular scales, but still larger than $10^\circ$. These potential signals were identified through various types of anisotropy tests, including searches for clusters of events (in excess of the isotropically expected numbers) over a range of angular windows and/or energy thresholds, correlations with identified astrophysical objects considered as potential UHECR sources, or cross-correlation with astrophysical catalogues of specific predefined source populations.

In this paper, we concentrate on three analyses recently conducted by the Pierre Auger collaboration, which we apply to a wide range of simulated UHECR datasets built as described in Paper~I:
\begin{enumerate}[(i)]
    \item a so-called blind search (BS), that is a search for significant excesses in the UHECR flux in some particular directions, without any prejudice about a specific direction in the sky, and also without predefined energy threshold or angular scale.
    \item a search for an excess of UHECR events in correlation with the direction of the radio galaxy Centaurus A (Cen~A), as has been reported over the years by Auger, initially hinted in \citep{AugerAGN2007} and then successively updated in \citep{AugerAGN2010, AugerAni2015, AugerICRC2019, BiteauICRC2021}.
    \item a likelihood analysis of the correlation between the UHECRs arrival directions and some catalogs of candidate sources, as first discussed in \citep{AugerSFG2018} and subsequently updated in \citep{AugerICRC2019, BiteauICRC2021}.
\end{enumerate}

We shall also combine the first two analyses to discuss not only the significance level and angular scale of excesses found in our simulations in the direction of Cen~A, but also the potential implications of the Auger finding that the maximum significance obtained in a blind search of their data appears to correspond to a direction very close to that of Cen~A \citep{AugerICRC2019, BiteauICRC2021}. %Finally, although less significant, the possible implications of the anisotropy signals reported by the TA collaboration \citep{TASpot2014, TAICRC2021, TANewSpot2021} will also be briefly discussed.

In Sect.~\ref{sec:datasets}, we review the models and procedure that we use to produce consistent simulated datasets, notably the various assumptions regarding the spatial distribution of the UHECR sources. In Sect.~\ref{sec:anisoMethods}, we provide some detail about the above-mentioned anisotropy analyses and their application to our simulated UHECR sky maps. The main results are presented and discussed in Sects.~4 through~10, where we confront the different astrophysical scenarios explored in this series of paper with the actual observational data. TA anisotropy blind search analyses are also briefly discussed in the appendix.

\section{UHECR source models and dataset simulation}
%\section{UHECR simulated datasets characteristics and generation methods}
\label{sec:datasets}

\subsection{Model parameters}

A consistent simulation of UHECR datasets at Earth requires definite assumptions about: (i) the physical properties of the UHECR distribution at each source, which includes the nuclear composition, the energy spectrum (shape and maximum energy, possibly dependent on the nuclear species), (ii) the time evolution of the sources, (iii) their spatial distribution, (iv) the photon background seen by the UHECRs along their trajectory, and (v) the cosmic magnetic fields through which they propagate. Given the current lack of knowledge not only about which individual source actually injects UHECRs in the intergalactic medium, but even about which type of sources may contribute, it appears reasonable to reduce the number of free parameters by adopting a number of simplifying assumptions, with the hope that the general features of the resulting datasets be representative of what may be expected in practice, if the basic assumptions underlying the simulated astrophysical models hold, at least on average. While each source is likely to be different, one usually assumes a unique source composition and spectrum, playing the role of an effective average source allowing one to reproduce the main features of the propagated composition and energy spectrum. Likewise, although the concept of a source density may not be the most relevant to describe the actual distribution of sources that happen to be contributing at the present time in our particular location in the universe, one can explore different source distribution scenarios by randomly selecting sources among certain types of astrophysical objects, with a given predefined density.

In Paper~I (Sect. 2--5), we presented in detail the ingredients of the astrophysical models used in our simulations, gave some justification for the various assumptions and ranges of parameters, and described the numerical tools used to generate datasets taking into account the various processes affecting the propagation of the UHECRs from their sources to the Earth (see also \citet{BRDO2014}). We refer the reader to this paper for details, and simply summarize the main ingredients in the rest of this Section.

%The ingredients, as well as the numerical tools used to model the UHECR propagation and arrival direction distribution on Earth are presented in great details in Paper I, in Sects.~2 through~5 (see also \citet{BRDO2014}). We simply remind here some details about our hypotheses on the UHECR source distribution.

\subsection{Source distribution}
\label{sourdist}
%To summarize, assuming that the overall…

Assuming that the overall UHECR source distribution is similar to that of the galaxies, we draw individual realisations of UHECR sources from the 2MASS Redshift Survey catalog (2MRS, \citet{Huchra2012}), and investigate the specific role of cosmic variance by using two complementary approaches: (i) a so-called ``volume-limited approach”, which allows us to work with fixed distributions of sources obtained from a cut on the galaxies $K_\mathrm{s}$-band luminosity, and (ii) a so-called ``mother catalog approach”, where we randomly select sources from the largest volume-limited catalog (i.e. the one with the lowest luminosity cut, which is then the mother catalog), producing many (in most cases 300) realizations with different sources sub-sampled from the mother catalog to reach a given source density. In this selection process, the probability to keep a given source of the mother catalog is thus the ratio between the chosen source density and the density of the mother catalog itself (namely $\sim 7.6\,10^{-3}\,\rm Mpc^{-3}$).

NB: by definition, the volume-limited catalogs are complete only up to a distance $D_{\max}$, which depends on the chosen luminosity cut, $L_{\rm cut}$. To complete the catalogs beyond $D_{\max}$, which is the largest distance at which a source with luminosity larger than $L_{\rm cut}$ would have been detected for sure, we sample with the same source density the 3D distribution of matter in the Universe provided by the large scale structure simulations of \citet{LSSS2018}, which are constrained by the Cosmicflows2 peculiar velocities catalog \citep{CF22014}.

%Various luminosity cuts, which are listed in Tab.~1 of Paper I, correspond to volume-limited catalogs with various source densities. We remind here that each volume-limited subset of the 2MRS catalog is ``complete" up to a distance $D_{\max}$ corresponding to the chosen luminosity cut. The volume-limited catalogs are completed above $D_{\max}$ (to compensate for the radial selection effect) by sampling the 3D distribution of matter in the Universe provided by the large scale structure simulations of \citet{LSSS2018} which are constrained by the Cosmicflows2 peculiar velocities catalog \citep{CF22014}. (ii) For our second approach, the largest density volume-limited catalog is used as a ``mother catalog" in which we randomly pick up sources to produce many (in most cases 300) realizations of a source catalog with a given density. For each realization, the probability to keep a given source of the mother catalog correspond to the ratio between the chosen density of the UHECR source catalog and the density of the mother catalog ($\sim 7.6\times10^{-3} \rm Mpc^{-3}$).\\

The use of these two different approaches allows us to study different aspects of the dispersion in our anisotropy results. In the volume-limited approach, the source distribution is fixed, so each realization provides a new simulation of the exact same underlying scenario, allowing us to explore the evolution of the results with the size of the { UHECR} dataset, as well as the statistical variance of a given dataset. Moreover, it allows us to study the influence of various physical parameters, such as the Galactic magnetic field (GMF) model, its amplitude or its coherence length, while keeping the source distribution unchanged. On the other hand, the mother catalog approach allows us to study the impact of the ``cosmic variance", that is the dispersion in the theoretical expectations resulting from different realizations of the source distribution, within the same general astrophysical scenario.

In addition, it proved interesting to use a modified version the mother catalog approach, in which the sources are indeed randomly selected at each realization, except for the forcing of one particular source of interest, such a Cen~A, M87, M83, Fornax~A or NGC253. In this way, the specific impact of a given source in the simulated anisotropy patterns can be explored. Other source configurations will also be discussed below.

%We note that in the framework of the mother catalog approach, we will also use source configurations for which the presence of some source of particular interest, for instance CenA, M87, M83, FornaxA or NGC253, etc..., is forced in every realizations of a source catalog of a given density, in order to estimate the impact of the presence of a particular source on the anisotropy predictions. More complicated configurations will also be discussed in later sections. 

\subsection{Energy spectrum and composition}

Regarding the source spectrum and composition models, we use the same models A, B, C and D, listed in Tab.~2 of Paper~I. These models represent different variations of mixed-composition ``low-$E_{\max}$ models”, that is mixed-composition models in which the protons do not reach the highest energies and the maximum energy of the different species is proportional to their charge $Z$. We will however mostly show predictions obtained with model A in the following, as our conclusions do not depend strongly on the details of the assumed composition model. 

\subsection{Magnetic field models}

Cosmic magnetic fields, both Galactic (GMF) and extragalactic (EGMF), were also discussed in Sect.~3 of Paper~I. In this paper, we use two different GMF models, which both include a regular and a turbulent component: (i) the parameterizations proposed in \citet{JF2012a, JF2012b} and (ii) the ``ASS+RING” model proposed by \citet{Sun2008, Sun2010}. For both models, we use the parameters as updated after the comparison of their predicted polarized synchrotron and dust emissions with those measured by the Planck satellite mission, as reported in \citet{GMFPlanck2016}. We refer to these models as the ``JF12+Planck” model and the ``Sun+Planck” model, respectively.

\subsection{Size and contours of the datasets}

Finally, the statistics of the datasets must be chosen, as well as the simulated sky coverage. In most cases, we use the same statistics and sky exposure as in the analyses presented by Auger at the International cosmic ray Conference (ICRC) 2019, which corresponds to a total exposure of $\sim$101,400 $\rm km^2\,sr\,yr$ for UHECR showers with a zenith angle lower $80^\circ$ (see \citet{AugerICRC2019}). Accordingly, we fix a statistics of $\sim$42,500 events above an energy threshold of 8 EeV (where the statistical fluctuations of the number of events are small).  This choice allows us, in particular, to make direct comparison with the Auger experimental results, using the numerical tools provided by the Auger collaboration for the likelihood analysis (see below).

\subsection{``Baseline” volume-limited catalog}
\label{sec:baseline}

The most general discussions of the present paper will be carried out in the case of the volume-limited catalog model referred to in Paper~I as our ``baseline model” (see Table 1 there). It corresponds to a luminosity cut that is as stringent as possible, while still not rejecting the local candidate sources that are most often cited, such as Centaurus~A, M81/82 or NGC253 (together with higher luminosity and more distant galaxies such as NGC1068, M87 or Fornax~A). The resulting source density is $\rho_{\mathrm{s}}=1.4\times 10^{-3}\,\rm Mpc^{-3}$. This baseline model also assumes the composition model A and its associated energy spectrum, as well as an EGMF of 1 nG. This model can be used with different choices of the GMF, ``JF12+Planck” or ``Sun+Planck”, with various coherence lengths.

%-----------------------------------------------------------------

\section{Anisotropy analyses}
\label{sec:anisoMethods}

%\subsection{Search for a localized flux excess over the whole skymap and in the direction of CenA}
\subsection{Localised excesses of the UHECR flux}

\subsubsection{Blind search (BS)}
\label{sec:BS}

In the absence of any prejudice about the angular distribution of UHECRs over the sky, it is natural to search blindly for regions where the flux appears higher than what would be expected from an isotropic sky. Furthermore, if no particular energy scale or angular scale can be identified based on {\it{a priori}} theoretical consideration, a scan can be performed over a wide range of energy thresholds, $E_{\rm th}$, and smoothing angles, $\psi$, to identify the scales at which the departure from anisotropy is maximal.

We apply such a blind search (BS) analysis to all our simulated datasets, following closely the scan procedure adopted by Auger and described in \citep{AugerAni2015}. We scan the entire sky map using sharp circular windows (``top hat”) with various angular radii, $\psi$, placing the center of the windows in directions regularly distributed over the celestial sphere using a HEALPix grid \citep{Gors2005} with resolution parameter $N_\text{side}$ = 64. This ``pixelization” is equivalent, from the point of view of the statistical independence of the trials, to the $1^\circ \times 1^\circ$ grid implemented in \citep{AugerAni2015}. The scans run over two different ranges of parameters: (i) the same as used by Auger, to allow direct comparison without additional penalization factors, namely with $E_{\rm th}$ running from 32~EeV to 80~EeV with 1~EeV steps, and $\psi$ running from $1^\circ$ to $30^\circ$ with $1^\circ$ steps; (ii) a wider range, from 8~to 80~EeV with 1~EeV steps and up to $45^\circ$ with $1^\circ$ steps, to have a broader view on the evolution of the anisotropy with $E_{\rm th}$ and $\psi$.

%A blind search for statistical flux excesses is one most natural analysis to perform in case there is no strong prejudice about their location on the skymap. To construct a blind search for excesses on our simulated datasets we closely follow the scan procedure used by Auger and described in \citep{AugerAni2015}. We scan the entire skymap using circular window with various radii, the center of the windows corresponding to directions regularly distributed over the celestial sphere using an HEALPix grid \citep{Gors2005} with resolution parameter $N_\text{side}$ = 64\footnote{We note that this ``pixelization" is totally equivalent to the $1^\circ \times 1^\circ$ grid implemented in \citep{AugerAni2015} in terms of independant trials.}.

%We perform various scans in energy threshold of the dataset $E_{\rm th}$ used for the blind search and angular scale $\Psi$ (that is the radius of the circular window in which the events are counted). The first scan is the same as the one performed by Auger, we scan the energy from 32 EeV to 80 EeV with step of 1 EeV and for the angular window we scan between $1^\circ$ and $30^\circ$ with $1^\circ$ steps. It allows us to trivially compare our predictions with Auger measurement without applying penalization factors.\\
%In addition we perform a wider scan of the $\left(E_{\rm th},\,\Psi\right)$ parameter space, from 10 to 80 EeV with 5 EeV steps and from $3^\circ$ to $45^\circ$ with $3^\circ$ steps to have a better handle the $E_{\rm th}$ and $\psi$ evolution of the anisotropy.\\

For each value of the scan parameters ($E_{\rm th}$,$\psi$,pixel), we calculate the local significance, $N_\sigma$, of the excess in the UHECR numbers above $E_{\rm th}$ and within $\psi$ degrees of the HEALPix pixel central direction, using the \citet{LiMa1983} formula:
\begin{equation}
\label{eu_eqn}
\begin{aligned}
N_\sigma = \sqrt{2} \Biggl( N_{\rm on} &\ln \left[\frac{1+\alpha}{\alpha}\left(\frac{N_{\rm on}}{N_{\rm on}+N_{\rm off}}\right)\right]\\
 + N_{\rm off} &\ln \left[\left(1+\alpha\right)\left(\frac{N_{\rm off}}{N_{\rm on}+N_{\rm off}}\right)\right] \Biggr)^{1/2}
\end{aligned}
\end{equation}
where $N_{\rm on}$ is the number of events in the selected window, $N_{\rm off} = N_{\rm tot}-N_{\rm on}$ (where $N_{\rm tot}$ is the total number of UHECRs above $E_{\rm th}$ in the entire sky), and $\alpha = N_{\rm exp}/(N_{\rm tot}-N_{\rm exp})$, where $N_{\rm exp}$ is the expected number of UHECRs in the selected window, assuming an isotropic distribution of the $N_{\rm tot}$ events and accounting for Auger sky exposure).

For each of our simulated datasets, we perform these scans and register the parameters for which the maximum significance is reached, namely the energy threshold, angular scale and direction.

\subsubsection{Excess around Cen~A}

In the case of the search of a flux excess in the direction of Cen~A, we simply perform the same scans in the $\left(E_{\rm th},\,\psi\right)$ space, but restricted to the actual direction of the radio Galaxy, and register the value of the local significance at each point of the parameter space.

%\subsection{Description of the likelihood analysis}
\subsection{Likelihood analysis and ``test statistics” (TS)}
\label{like}

Our goal is to reproduce the Auger analysis described in \citep{AugerSFG2018} on our simulated datasets. The method used is a maximum likelihood ratio test to distinguish between a signal+background hypothesis, $H_1$, and the null hypothesis, $H_0$, where only background is present. Here and below, ``background” refers to an isotropic distribution, before the application of a given exposure map (depending on the simulated experiment).

The global likelihood $\mathcal{L}(H_\alpha|\mathbf{x})$ of hypothesis $H_\alpha$ ($\alpha = 0$ or 1) associated with the data $\{\mathbf{x}\}$ is the product over all events, $i$, of the individual likelihood $f(H_\alpha|\mathbf{x_i})$:
\begin{equation}
    \mathcal{L}(H_\alpha|\mathbf{x})=\prod_i^{\mathrm{N_\mathrm{events}}} f(H_1 | \mathbf{x_i}) \implies \ln \mathcal{L}(H_\alpha | \mathbf{x})=\sum_i^{\mathrm{N_\mathrm{events}}} \ln  f(H_\alpha | \mathbf{x_i}).
\end{equation}

%The likelihood in the pure background hypothesis $\mathcal{L}(H_0 | \mathbf{x})$ has the same definition.

Following the description of the method in \cite{AugerSFG2018}, we write the likelihood for one event in direction $\mathbf{x_i}$ :
\begin{equation}f(H_1 | \mathbf{x_i})=I_{f_\mathrm{aniso},k} \times \left[ f_\mathrm{aniso} \, S(\mathbf{x_i},k)  + \frac{(1-f_\mathrm{aniso})}{4\pi}\right] \times \mathcal{A}(\mathbf{x_i}) 
 \end{equation}

 %$$ f(H_1 | \mathbf{x_i})=I_{fs,k} \times \left[ f_s \, \sum_{j=1}^{\mathrm{N_{sources}}} \frac{w_j}{W} S(\mathbf{x_i},\mathbf{x_j},k)  + \frac{(1-f_s)}{4\pi}\right] \times \mathcal{A}(\mathbf{x_i}) $$
 
 where $S(\mathbf{x_i},k)$ is the signal term, $f_\mathrm{aniso}$ the signal fraction and $\mathcal{A}(\mathbf{x_i})$ is the Auger exposure function.
 The signal term is computed as the sum of the contributions of the individual CR sources: 
 \begin{equation}
 S(\mathbf{x_i},k)= \frac{1}{W} \sum_{j=1}^{\mathrm{N_{sources}}}w_j \,  s(\mathbf{x_i},\mathbf{x_j},k) \quad \mathrm{with} \quad W=\sum_j^\mathrm{N_{sources}} w_j
 \end{equation}
 where $s(\mathbf{x_i},\mathbf{x_j},k)$ is the expected signal for the jth source that contributes with a weight $w_j$ that takes into account the assumed intrinsic intensity of the source and the CR attenuation due to energy losses.
 For every source, the signal term is written as a Fisher distribution (generalization of the Gaussian distribution on a sphere) :
\begin{equation}
s(\mathbf{x_i},\mathbf{x_j},k)= \frac{k}{4\pi \sinh(k)} e^{k(\mathbf{x_i}\cdot\mathbf{x_j})} 
\end{equation} where $k$ is the concentration parameter that defines the width of the function. For convenience, we report the results in terms of the equivalent variance $\theta^2$ for a symmetrical normal distribution, which is given by the simple relation: $\theta^2= \frac{1}{k}$ ($\theta$ in radians). 
%{\color{red}aie...confusion possible de notation avec les sigma max dans les chapitres d'apres...} 

There are only two free parameters in the fit: the fraction of signal $f_\mathrm{aniso}$ and the smoothing angle $\theta$ defined above. The signal and background probability density functions     must be normalized separately to the same value, and the total likelihood function$ f(H_1 | \mathbf{x_i})$ is normalized to the total number of events in the data set. We thus impose:
\begin{equation}
\int  S(x)\, \mathrm{d}x = 1 \quad \mathrm{and} \quad \int f(H_1 | \mathbf{x_i}) \, \mathrm{d}x = \mathrm{N_ {evts}}
\end{equation}
via the computation of the normalization constant $I_{f_\mathrm{aniso},\theta}$ for each possible value of the parameters $(f_\mathrm{aniso},\theta)$. 

\begin{figure*}[h!]
   \centering
   \includegraphics[width=8.5cm]{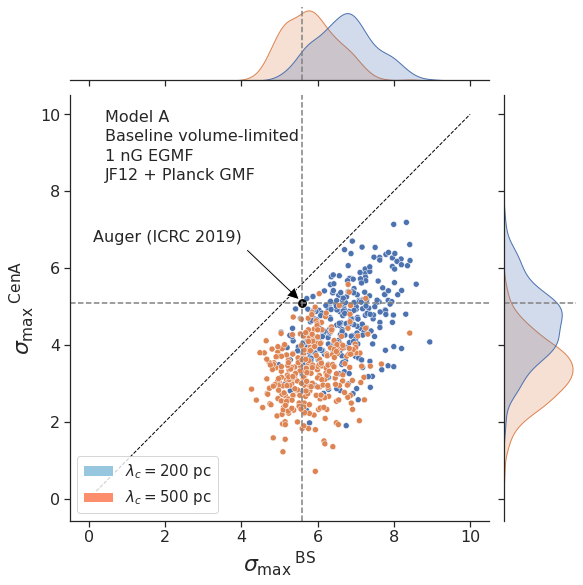}
   \includegraphics[width=8.5cm]{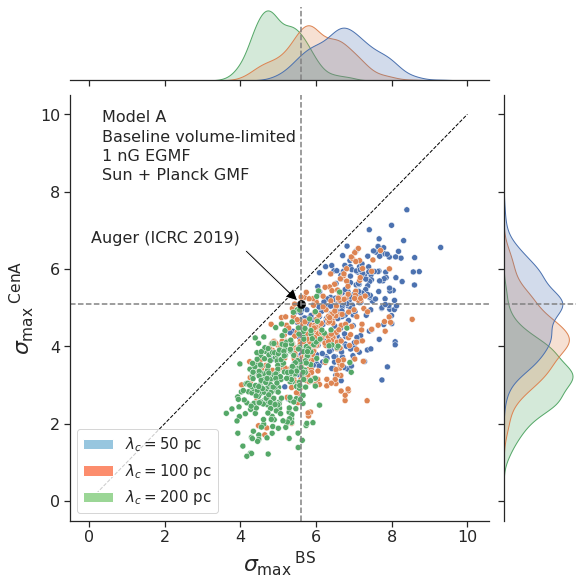}
      \caption{Result of the BS and CenA flux excess analyses for our baseline volume-limited scenario, after scanning over the same parameter space in $E_{\rm th}$ and $\psi$ as Auger (see text). Scatter plot of the CenA flux excess maximum significance versus the BS maximum significance obtained for 300 datasets using the JF+Planck (left) and the Sun+Planck (right) GMF models. Various coherence lenghts $\lambda_{\rm c}$ are considered for the GMF turbulent component (see legends). The values reported for Auger dataset at the ICRC 2019 are shown with a large black circle, the distributions of $(\sigma_{\rm max}^{\rm BS}$ and $\sigma_{\rm max}^{\rm CenA})$ are also shown separately on top of the coordinate axis.
              }
         \label{FigBSVL1a}
\end{figure*}

%The so-called ``test statistic" is then the log of the likelihood ratio, where both numerator and denominator are maximized with respect to their free parameters.
The so-called ``test statistic" is then the logarithm of the likelihood ratio, that is the ratio between the likelihood of the tested model, hypothesis H$_1$, and the likelihood of the pure background hypothesis, H$_0$, where both are maximized with respect to their free parameters. As there is no free parameter in the H$_0$ hypothesis, the maximization concerns only the H$_1$ case, and then the test statistic can simply be expressed as:
\begin{equation}
\mathrm{TS}=\max ( \ln \mathcal{L}(\mathrm{H}_1 | x)) - \ln \mathcal{L}(\mathrm{H}_0 | x ) .
\end{equation}

The test statistic is the variable that is used to reject the H$_0$ hypothesis by looking at the data. The p-value associated with a measurement $\mathrm{TS}_\mathrm{data}$ is then obtained as 
\begin{equation}
p=\int_{\mathrm{TS}_\mathrm{data}}^{\infty} f(H_0 | \mathrm{TS})\, \mathrm{d}(\mathrm{TS})
\end{equation}
where $f(H_0 | \mathrm{TS})$ is the probability density function of the test statistic.

According to Wilk's theorem, for the pure background hypothesis the variable $\lambda=2\times \mathrm{TS}$ should be distributed as a chi-square law of 2 degrees of freedom (which corresponds to the number of free parameters in the likelihood fit). As done in the Auger analysis, we use the chi-square estimation of the p-value in the scan to find the most significant correlation.

As for the search of flux excesses, we reproduce the likelihood analysis of Auger by restricting the datasets to UHECRs with an energy above a threshold scanned from 32~EeV to 80~EeV with steps of 1~EeV. We consider three of the catalogs used in \citep{AugerICRC2019}, namely: (i) a selection of ``starburst galaxy” (SBG) based on \citet{FermiSBG, Becker2009}, weighted by their radio flux, (ii) $\gamma$-ray emitting AGNs selected from the 3FHL catalog \citep{FermiAGN}, and (iii) the 2MRS catalog, from which sources closer than 1 Mpc are removed. For each choice of the energy threshold, $E_{\rm th}$, a weight and attenuation factor are applied to the flux of each source according to the values provided by Auger as supplementary material for \citet{AugerICRC2019}\footnote{https://www.auger.org/science/public-data/data}, for their ``composition model A” (not to be confused with our model A). This ensures a consistent comparison between the likelihood analyses applied to our simulated datasets and to the Auger data. For each realization of each astrophysical model investigated, we search for the highest likelihood at each value of $E_\mathrm{th}$ and register the corresponding ``best-fit” parameters (this procedure, however, is by no means a ``fit” of the data with any predefined model).

%The scan of the dataset energy threshold $E_{\rm th}$ is done from 32 EeV up to 80 EeV by steps of 1 EeV. For the signal term, we consider  three of the catalogs used in \citep{AugerICRC2019}, namely (i) a selection of ``starburst galaxy” mostly based on \citet{FermiSBG, Becker2009} and selected according to their radio flux, (ii) $\gamma$-ray emitting AGN selected from the from the 3FHL catalog \citep{FermiAGN}, and (iii) the 2MRS catalog (removing sources closer than 1 Mpc). The flux weights and the attenuation factors due to energy losses, for each source of the catalogs considered for the signal, are computed for each energy threshold, on the basis the files provided by Auger as supplementary material of \citep{AugerICRC2019}\footnote{https://www.auger.org/science/public-data/data}, for their "composition model A" (not to be confused with our model A). This ensures that the same  ingredients are used in the likelihood analysis applied to our simulated datasets as on Auger data. 
%For each value of $E_\mathrm{cut}$ value, the likelihood is maximised, the lowest p-value and the corresponding best-fit parameters are stored for each realization of each astrophysical model we study.

We note that we did not reprocess all our simulations with the more recent tools provided in \citet{AugerAniso2022}. In the latter publication, which is based on the same dataset as in \citet{BiteauICRC2021}, the astrophysical catalogs used to model the signal component of the likelihood analysis slightly differ from those used in \citet{AugerICRC2019}, which are the ones we consider. However, we checked, using the Auger full UHECR dataset above 32 EeV provided in \citet{AugerAniso2022}, that the results of the likelihood analysis obtained with the three above-mentioned catalogs are %extremely close
almost identical
to those obtained in the most recent Auger analysis with their updated catalogs:  we found essentially identical results for the $(f_{\rm aniso},\theta)$ parameters and TS values, differing in the worst case by $\sim$2 units (that is a factor of $\sim$3 for the local p-value). These differences turn out to be very small compared to the spread of the values obtained due to either the statistical or the cosmic variance, as shown below.\\Likewise, the release of the Auger data allowed us to check that our results for the blind search analysis are compatible with those of Auger. 

         \label{FigBSVL1}
%\end{figure*}

%\section{Results obtained with the baseline catalog}
%\section{Results}
\label{sec:results}

\section{Results of the blind search and flux excess in the direction of Cen~A}
%\subsection{Blind search and flux excess in the direction of Cen~A}
\label{SecBS}

We first examine the results of the blind search and Cen~A excess analyses in the case of our baseline scenario (see Sect.~\ref{sec:baseline}), in comparison with the corresponding Auger results.

\subsection{Significance of the flux excesses}

%A first interesting result can be found in Fig.~\ref{FigBSVL1a}, where each point on the plots corresponds to a different realisation of the model, i.e. a different dataset with the same statistics and exposure as the reference Auger dataset.
We applied the analysis to datasets with the same statistics and exposure as the reference Auger dataset.
For each of them, we determined the highest significance of the blind search, $\sigma_{\max}^{\rm BS}$, and the highest significance of the Cen~A excess analysis, $\sigma_{\max}^{\rm CenA}$, over the above-mentioned range in $E_{\rm th}$ and angular scales, $\psi$ (Sect.~\ref{sec:BS}). Figure~\ref{FigBSVL1a} shows a scatter plot of the 300 realisations of the model in the $(\sigma_{\rm max}^{\rm BS},\,\sigma_{\rm max}^{\rm CenA})$ plane, in the case of the JF12+Planck model on the left panel, and in the case of the Sun+Planck model on the right panel. In each case, different choices of the GMF coherence lengths are shown in different colours, as indicated on the plot. Each dot corresponds to a different dataset. Obviously, all datasets are located below the first diagonal, since the maximum significance of the flux excess in the specific direction of Cen~A cannot be larger than the maximum significance anywhere, irrespective of the direction.

A first important remark is that the results show a very large dispersion both in $\sigma_{\rm max}^{\rm BS}$ and $\sigma_{\rm max}^{\rm CenA}$, which corresponds to orders of magnitude differences in the associated p-value or statistical significance. This is true even though the sources and their relative weight are all exactly the same in each case. 
%{\color{red}[{{En fait ce ne sont rien de plus que de betes fluctuations statistiques donc je n'irais pas plus loin dans la discussion et je virerais la phrase color\'ee]}} This means that, at the present level of statistics, it would be quite hazardous to draw any strong conclusion from the actual values of the maximum significance in the BS or Cen~A excesses in the Auger data.}

\begin{figure*}
   \centering
    \includegraphics[width=8.5cm]{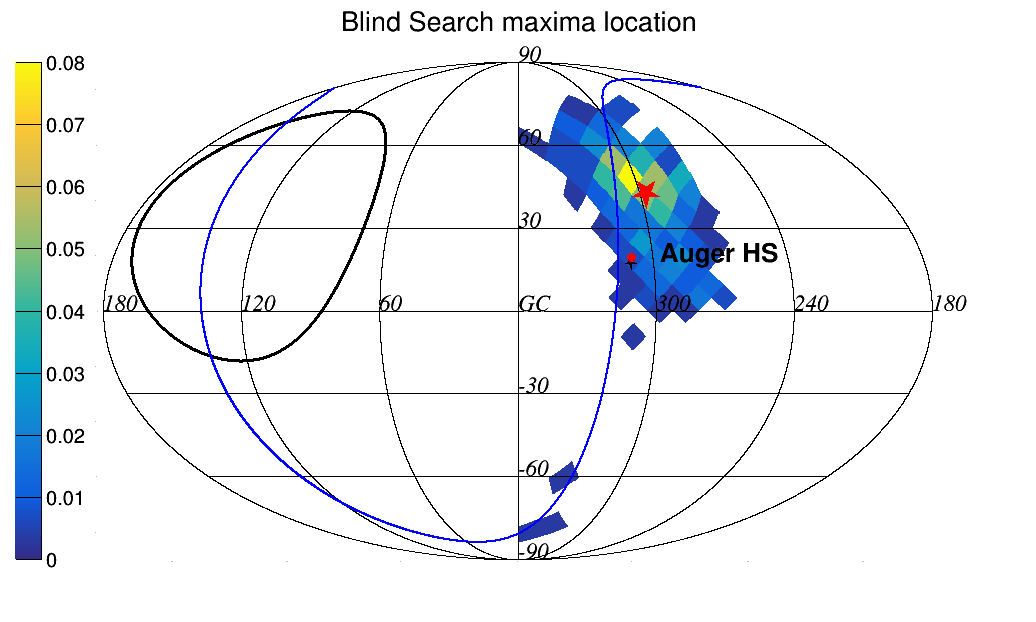}
    \includegraphics[width=8.5cm]{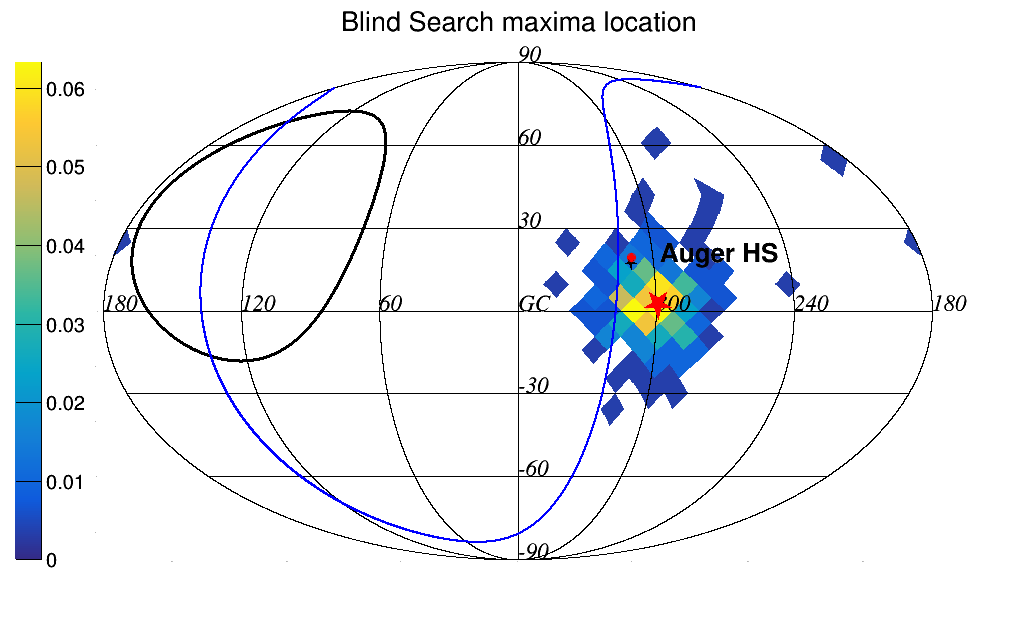}
      \caption{Distribution of the locations of the BS maxima for the 300 datasets. The case of the JF+Planck model with $\lambda_{\rm c}$=200~pc (left) and the Sun+Planck model with $\lambda_{\rm c}$=100~pc (right) are shown. The color scale represents the fraction of realizations for which the BS maximum is found in a given pixel of the sky (the pixel size correspond to Nside=8 on these plots). The position of the CenA radio galaxy as well as the BS maximum location reported by Auger at the ICRC 2019 are shown with a red full circle and black star respectively {(the two markers are practically on top of each other). The large red star shows the direction of the asymptotic BS maximum obtained with a 300 times larger simulated dataset (see text)}.
              }
         \label{FigBSVL1b}
\end{figure*}

It is then interesting to compare the obtained values with those obtained with the Auger dataset, represented on Fig.~\ref{FigBSVL1a} by a thick black circle. As can be seen from the probability distributions shown on the right and the top borders of the plot (in linear scale), the individual values of both $\sigma_{\rm max}^{\rm BS}$ and $\sigma_{\rm max}^{\rm CenA}$ in the Auger data appear to be typical of the those found in our datasets, for both GMF models. Of course, whether the Auger data point is found on the lower end, middle or higher end of the simulated ranges of values depends on the assumed coherence length, $\lambda_{\rm c}$, of the turbulent magnetic field component, but reasonable (in the sense of generically allowed) values of $\lambda_{\rm c}$ can be identified in each case to place the Auger values approximately in the middle of the simulated range (namely $\lambda_{\rm c}=200$~pc and 100~pc for the JF12+Planck and the Sun+Planck models respectively, see Fig.~\ref{FigBSVL1a} for quantitative evaluation). However, what makes the simulation results interesting in this respect is that, on the other hand, the pair of values $(\sigma_{\rm max}^{\rm BS},\sigma_{\rm max}^{\rm CenA})$ obtained with the Auger data is quite unusual in our simulations. As a matter of fact, for the Auger data, $\sigma_{\rm max}^{\rm BS}\simeq \sigma_{\rm max}^{\rm CenA}$, which is related to the fact that the BS maximum in located in the sky at a position very close to that of CenA ($\sim 2^\circ$ away from each other, as reported at the ICRC~2019).

\subsection{Direction of the most significant flux excess}
\label{direction}

\begin{figure}[t]
   \centering
 \includegraphics[width=8.5cm] {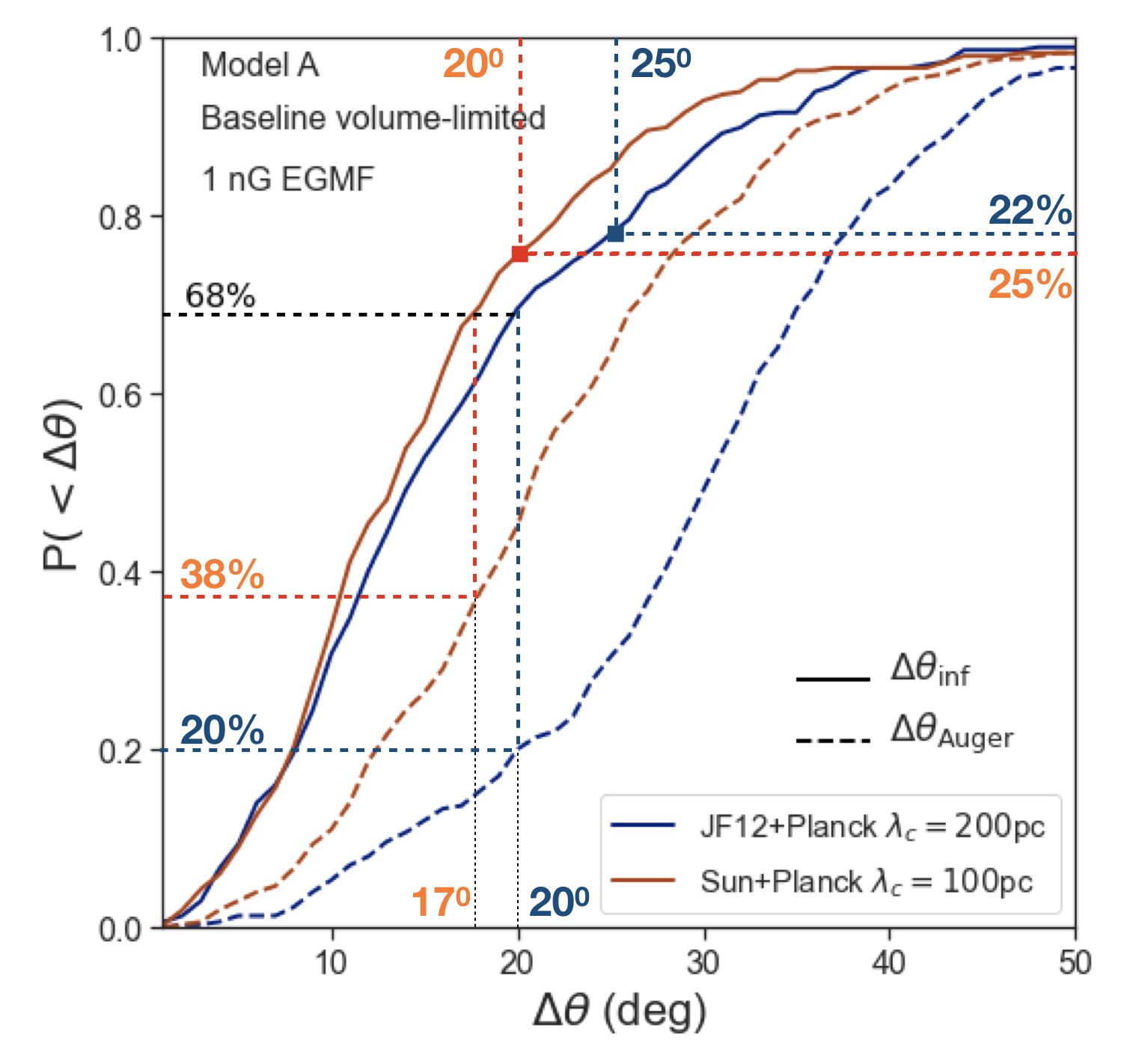}
\caption{Cummulative distribution, built over 300 datasets, of the angular distances $\Delta\theta_{\rm inf}$ (full lines) and $\Delta\theta_{\rm Auger}$ (dashed lines)  defined in Sect.~\ref{direction}. The cumulative functions obtained for the Sun+Planck GMF model with $\lambda_{\rm c}$=100 pc are shown in red, those for the JF12+Planck model with $\lambda_{\rm c}$=200 pc are shown in blue.}
         \label{FigDistoCenBase}
\end{figure}

{ We plotted in Fig.~\ref{FigBSVL1b} the distribution of the locations of the BS maxima for the 300 datasets. The case of the JF+Planck model with $\lambda_{\rm c}=200$~pc is shown on the left, and that of the Sun+Planck model with $\lambda_{\rm c}=100$~pc on the right. %The color scale corresponds to the fraction of the 300 realisations for which the BS maximum is located in the corresponding pixel (using HEALPix with $N_{\rm side} = 8$ on these plots).% 
As can be seen, the distributions obtained with the two GMF models are  different, with a significant shift southwards of the distribution in the case of the Sun+Planck model. This is easily understood as a result of the strong demagnification of the sources in the Virgo cluster region in this case, as discussed below in more detail.  We note however that the two distributions show a large overlap, notably in the region which happens to be where the Auger data indicate the most significant flux excess. The two models can thus not be distinguished on the sole basis of the prediction of the location of the BS maximum at this level of statistics.

The position of the (ICRC 2019) Auger BS maximum is shown on Fig.~\ref{FigBSVL1b} as a black star near Cen~A, represented by a red dot. For both GMF assumptions, that position appears rather uncommon in our simulations, although a position of the BS maximum close to that of Auger can indeed be obtained in some cases with both magnetic field models. %One may of course speculate that another GMF model could be adjusted to place that preferred position very close to Cen~A, but that position would still have to be considered as not necessarily indicative of some exclusive properties of the UHECR sources or source distribution.
%In addition, in view of the large dispersion in the BS maximum positions, at least at this level of statistics, it seems that any specific direction singled out by a given dataset should be considered with caution. 

To better quantify the situation, we examined the distribution of the angular distances between the BS maxima found in our simulated datasets and i) the position of the BS maximum found in the Auger data (angular distance hereafter referred to as $\Delta\theta_{\rm Auger}$), and ii) the position of the BS maximum that would be obtained asymptotically for the same astrophysical model with ``infinite" statistics (hereafter referred to as $\Delta\theta_{\rm inf}$), as indicated with a red star in Fig.~\ref{FigBSVL1b}. To estimate the latter, we apply the BS analysis to a 300 times larger dataset obtained by putting together the 300 different realisations of the model under consideration.  The cumulative distribution functions of $\Delta\theta_{\rm Auger}$ and $\Delta\theta_{\rm inf}$ are shown in Fig.~\ref{FigDistoCenBase}.

As can be seen, concerning $\Delta\theta_{\rm inf}$, the curves are qualitatively and quantitatively similar, which can be understood as a consequence of the fact that both models have similar levels of anisotropy and BS maximum significance as the Auger data (see Fig.~\ref{FigBSVL1a}). One may thus estimate that a similar cumulative distribution function would also be obtained with the actual UHECRs themselves, that is if one had access to a large number of real UHECR data sets with the same statistics as Auger. Specifically, we find that 68\% of the simulated datasets have their BS maximum within $\sim$17$^\circ$ and $\sim$20$^\circ$ of the asymptotic position, respectively for the Sun+Planck model with $\lambda_{\rm c}=100$~pc and for the JF12+Planck model with $\lambda_{\rm c}=200$~pc. These values may thus be considered as representative of the angular distance to be typically expected between the BS maximum direction currently reconstructed with the Auger data, and that which would be obtained at infinite statistics. Given this relatively large ``angular resolution'', the very small angular distance between the direction of the Auger BS maximum and the direction of CenA should be considered with caution: according to our simulations, angular coincidences on scales lower than $\sim$15$^\circ$ cannot be considered meaningful at the current level of statistics.

The $\Delta\theta_{\rm inf}$ cumulative distribution functions also allow us to quantify the compatibility of the simulated models with the Auger data, from the point of view of the direction of the BS maximum. Assuming that the actual UHECR phenomenology is exactly described by (one or the other of) our simulated models, with what probability would a given data set with the Auger statistics have a BS maximum direction reconstructed (at least) as far away from the asymptotic direction as the actual Auger data are found to be? The answer can be read on Fig.~\ref{FigDistoCenBase}. For the  Sun+Planck model, the angular distance between the asymptotic BS maximum direction and the Auger data is $\sim$20$^\circ$, and we find that $\sim$25\% of the simulated data sets are at least as far away as this from the asymptotic direction. In the case of the JF12+Planck model, the angular distance to the Auger BS maximum direction is $\sim$25$^\circ$, which is expected to be the case for $\sim$22\% of the data sets. These numbers suggest that there is no strong contradictions from this point of view between the Auger data and the model expectations.

%The $\Delta\theta_{\rm inf}$ cumulative distribution {\color{red}functions \sout{can}} also allow us to discuss more quantitatively how (a)typical of our simulations the BS maximum direction found by Auger actually is. To that purpose, we compute the number of simulated datasets which have a value of $\Delta\theta_{\rm inf}$ larger than the angular distance between the direction of the Auger BS maximum and the direction of the asymptotic BS maximum found for our simulations. We find  $\sim20^\circ$ and $\sim25^\circ$ for the Sun+Planck and JF12+Planck models respectively. The fraction of simulated datasets for which the angular distance $\Delta\theta_{\rm inf}$ is found to be larger than these values is $\sim25\%$ for the Sun+Planck model, and $\sim22\%$ for the JF12+Planck model. These number suggest that while being in the higher end of the distribution of angular distance to the asymptotic BS maximum for both GMF models, the actual BS maximum direction is not in strong contradiction with the predictions of our simulations.

Conversely, starting with the position of the Auger BS maximum, one may wonder which fraction of the simulated models have their BS maximum direction within an angular distance corresponding to the abovementioned ``angular resolutions''. This can be obtained from the $\Delta\theta_{\rm Auger}$ cumulative distribution functions shown as dashed lines in Fig.~\ref{FigDistoCenBase}. We find that $\sim38\%$ of our datasets yield $\Delta\theta_{\rm Auger}<17^\circ$ for the Sun+Planck model, while $\sim20\%$ of our datasets yield $\Delta\theta_{\rm Auger}<20^\circ$ for the JF12+Planck models. These fractions remain sizeable, which again suggests that the BS maximum direction observed by Auger above 32 EeV is still marginally compatible with the simulated models, particularly for the Sun+Planck GMF model.

%On the other hand, using the $\Delta\theta_{\rm BS}$ cumulatives shown in Fig.~\ref{FigDistoCenBase}, one can also estimate the fraction of simulated datasets for which the value of $\Delta\theta_{\rm BS}$ is lower than the abovementioned resolutions estimated for both GMF models with the $\Delta\theta_{\rm inf}$ cumulatives. We find that $\sim38\%$ of our datasets yield $\Delta\theta_{\rm BS}<17^\circ$ for the Sun+Planck model, while $\sim21\%$ of our datasets yield $\Delta\theta_{\rm BS}<20^\circ$ for the JF12+Planck models. These fractions are significantly lower than the $\sim68\%$ which should be expected at best but remain sizeable. This again suggests that the BS maximum direction observed by Auger above 32 EeV is in agreement, although marginal, with the predictions of our simulations once their dispersion is taken into account for both GMF models. 
 %while the predictions of our simulations are in marginal agreement with the Auger data concerning the direction of the BS maximum direction above 32~EeV, this observable cannot be used to reject these models. 
 Finally, beyond the discussion of the BS maximum direction, it is worth nothing that datasets produced with the baseline model and the JF12+Planck GMF tend to predict high UHECR fluxes in the region of the sky near the Virgo cluster, seemingly in tension with the Auger observations. This property has important implications for the discussions below. 
}

%\begin{figure}[t]
%   \centering
   
%   \includegraphics[width=8.5cm]{Figs/Figure4.png}
%\caption{Cummulative distribution, built over 300 datasets assuming Auger statistics, of the angular distance between the BS maxima postions obtained after a scan on $E_{\rm th}$ and $\psi$ similar to that performed on Auger data,  and the one which would be obtained with an arbitrarily large statistics (see text) for the same scan of the parameter space.}
      %   \label{FigDistoInf}
%\end{figure}

\subsection{BS energy and angular scales}
\label{sec:SBEandPsiScales}

\begin{figure}
   \centering
     \includegraphics[width=8.5cm]{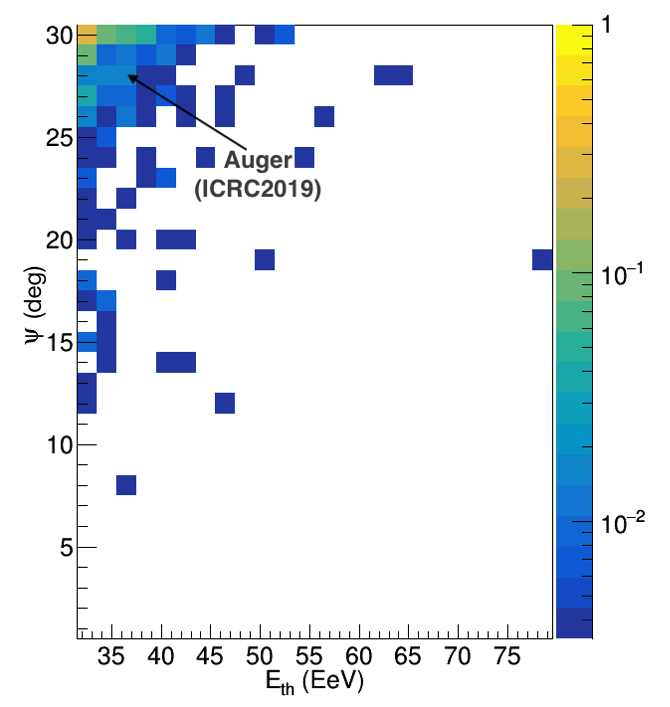}
  \includegraphics[width=8.5cm]{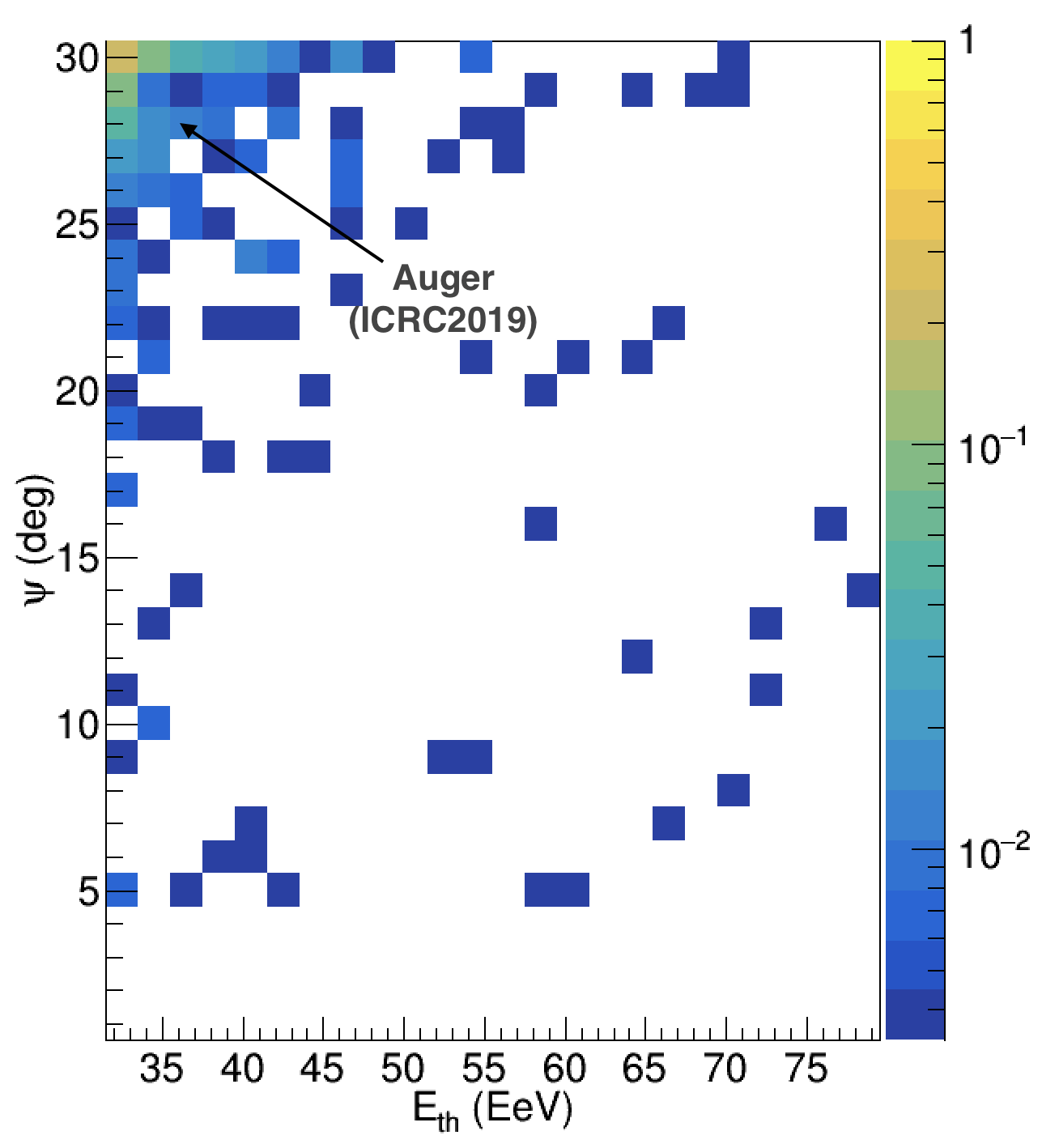}
\caption{Distribution of the threshold energy $E_{\rm th}$ and angular scale  $\psi$ of the BS maxima obtained after the analysis of 300 datasets assuming the same astrophysical models as in Fig.~\ref{FigBSVL1b}, the JF12+Planck model with $\lambda_{\rm c}$=200 pc (top panel) and the Sun+Planck model with $\lambda_{\rm c}$=100 pc (bottom panel). The color scale represents the fraction of realizations for which the BS maximum is obtained in a given ($E_{\rm th}$, $\psi$) bin.
              }
         \label{FigEthPsi}
\end{figure}

We also examined the distribution of the energy thresholds and angular scales at which the BS maxima were found, for the 300 realisations of the simulated models. The result is shown in Fig.~\ref{FigEthPsi} for both GMF models. We note that the distributions obtained for the two models are very similar.
As can be seen, the BS maximum shows a clear tendency to be located on the boundaries of the parameter space, that is at the lowest values of the energy threshold and the largest angular scale. The same is true for the search of a flux excess in the direction of Cen~A. This clearly suggests that higher significances could actually be found if one enlarged the range of scan parameters (see below). However, as Fig.~\ref{FigEthPsi} shows, even for an astrophysical scenario that does not favor the Auger values of the BS maximum parameters, these values or others similarly distant from the most { likely} ones for that scenario can be obtained from time to time. In such cases, finding the BS maximum away from the borders of the scanned parameter space may lead to the wrong impression that one does not need to extend the search further. %In other words, if the Auger dataset is one such realisation of an underlying model whose preferred parameter values would be at a lower energy and/or larger angular scale, it could be misleading to draw general conclusions from the actual experimental values, and not even explore the behaviour of the flux excess significance in an extended range of parameters. %As a matter of fact, there is little doubt that, had Auger found a BS maximum at the border of the searched parameter space, that space would have been extended to search for the ``true” maximum.
Now, given the variance in the BS results at the current level of statistics, already shown in Figs.~\ref{FigBSVL1a} and~\ref{FigBSVL1b}, it seems difficult to exclude the possibility that it is somewhat by chance that the parameters of the BS maximum of the current dataset are located inside the { arbitrary} limits of the predefined parameter range. { For instance, for the models displayed in Fig.~\ref{FigEthPsi}, $\sim 17\%$ and $\sim 27\%$  of the realisations, for the JF12+Planck and the Sun+Planck models respectively, have their BS maximum in the part of the parameters space delimited by the 2D interval $E_{\rm th} \geq 35$~EeV and $\psi \leq 28^\circ$.
As a consequence the fact that the Auger BS maximum is not found at the lowest energies and largest angles of the explored parameter space cannot be used at this point to reject with high confidence level the type of scenarios that we consider.}

%Even though individual datasets can exhibit BS maxima with different values of the parameters, in general the most significant flux excesses in the parameter range explored by Auger are to be expected at the lowest energies and the largest angles, at least for the astrophysical scenarios that we investigated. Unfortunately, the fact that the Auger BS maximum does not satisfy this property cannot be used at this point to reject these types of scenarios with a high confidence level.
%{\color{red}[NB: je me suis permis d'ajouter un arc de cercle rouge sur la figure, mais d'une part c'est mal fait (avec powerpoint!), et d'autre part on peut bien sûr revenir à la figure initiale (que j'ai laissée dans le dossier "Figs"). C'était plutôt pour montrer ce que j'avais en tête dans ma proposition de discussion (paragraphe précédent). What do you think?]}

The above considerations clearly suggest that a significant increase in the UHECR statistics would be desirable. In the meantime, until larger datasets become available, it is advisable to extend the discussion, taking into account the intrinsic dispersion expected in the BS results for a given astrophysical model. In this respect, it is instructive to study in a more systematic way the evolution of the significance of the flux excesses as a function of the BS parameters (including outside the limited range used by Auger). This is what we do in the next two sections.

%Concerning the values of $E_{\rm th}$ and $\psi$ where the BS maximum is found we find that most dataset lie on the boundaries of the parameter space, as can be seen in Fig.~\ref{FigEthPsi}. That is, low value of $E_{\rm th}$ and large values of $\psi$ are preferred (the same is true for the CenA flux excess search) suggesting that higher significance could be found by enlarging the scan parameter space (see below). As in the case of the significance and the location of the BS maximum, the dataset to dataset dispersion is found to be quite large so that the ($E_{\rm th},\,\psi$) couple found for Auger data although quite marginal is still  found for a few datasets.\\

%The dispersion we observe for the various quantities we discuss here (maximum significance, location of the maximum, energy and angular scale of the maximum) which are obviously important to characterize the UHECR sky anisotropy, already strongly suggest that a significant increase of the UHECR statistics would be desirable. We note that this actually depends on the genuine anisotropy level of a given astrophysical scenario therefore, for these first astrophysical scenarios considered, we chose to set our model parameters so that the median value of the maximum significance $\sigma_{\rm max}^{\rm BS}$  of our datasets found after a BS analysis corresponds more or less to that reported by Auger for a similar statistics.  

\begin{figure*}[!ht]
   \centering
   \includegraphics[width=8.5cm]{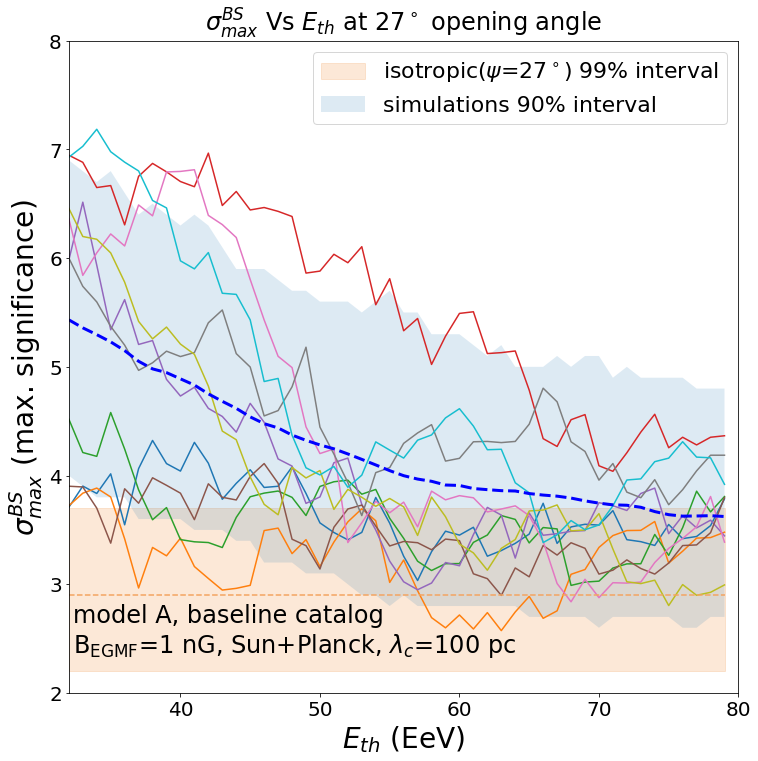}
   \includegraphics[width=8.5cm]{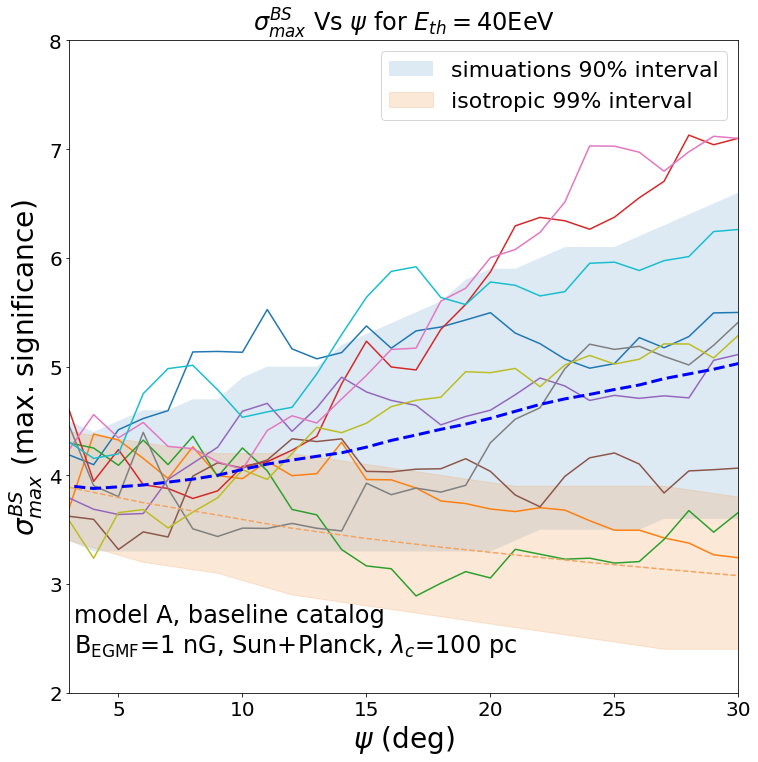}
   \includegraphics[width=8.5cm]{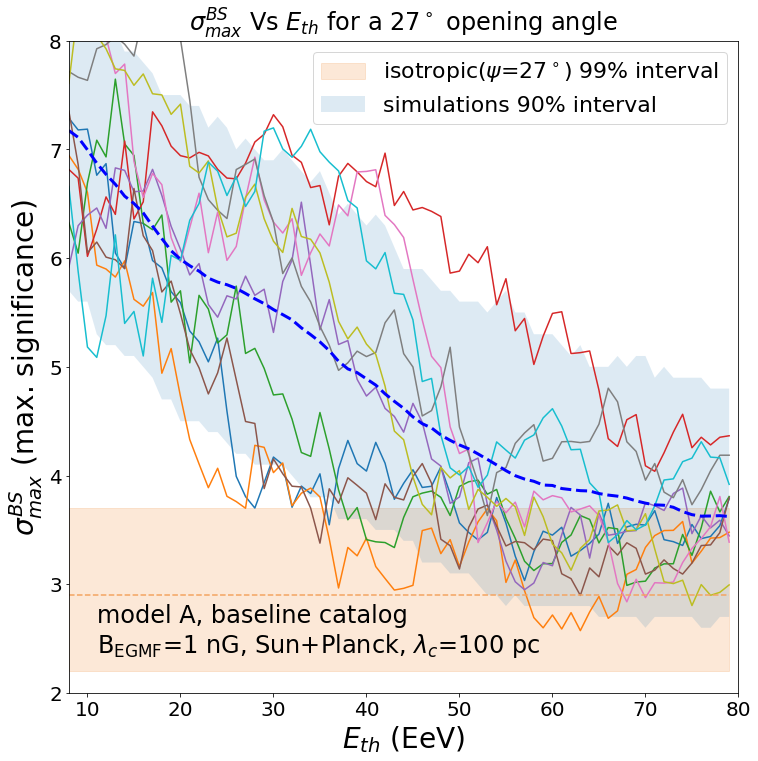}
  \includegraphics[width=8.5cm]{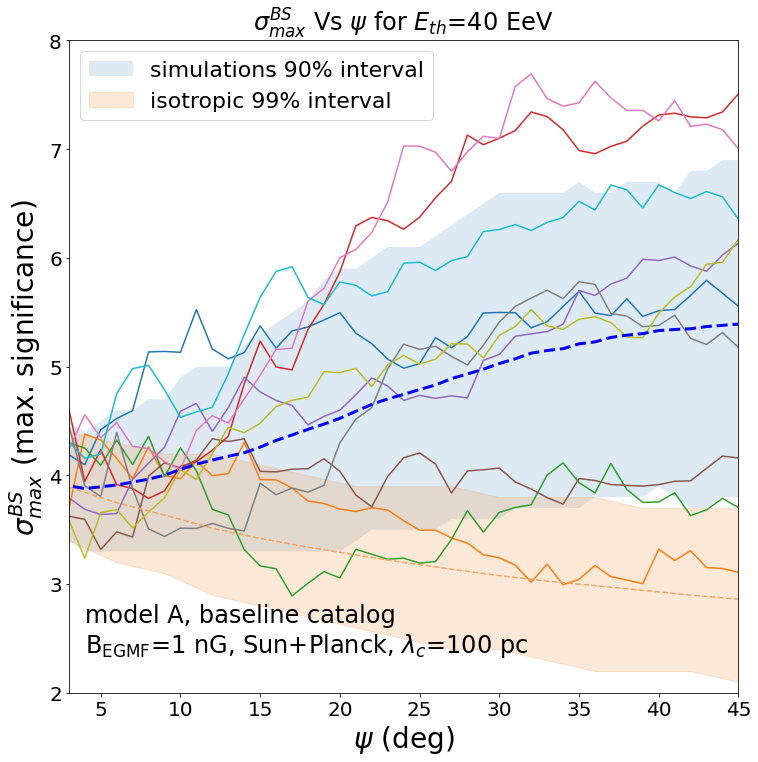}    
      \caption{Evolution of the BS maximum significance, $\sigma_{\rm max}^{\rm BS}$, with $E_{\rm th}$ and $\psi$, for the baseline astrophysical scenario with source composition model~A, 1~nG EGMF and the Sun+Planck GMF model with $\lambda_{\rm c}$=100~pc. Top left: evolution of $\sigma_{\rm max}^{\rm BS}$ with $E_{\rm th}$ for a smoothing angle $\psi = 27^\circ$. The thick blue dashed line shows the evolution averaged over 300 datasets. The thin lines correspond to 10 individual datasets chosen randomly. The light blue shaded area shows the 90\% interval of our simulated datasets. The brown shaded area shows the 99\% interval for isotropic datasets. Top right: evolution of $\sigma_{\rm max}^{\rm BS}$ with $\psi$, for a threshold energy of $E_{\rm th} = 40$~EeV. Bottom panels: same as the top panels, for a wider range of the $E_{\rm th}$ and $\psi$ scans.
      }
         \label{FigSigVs1a}
\end{figure*}

%\subsubsection{Evolution of the flux excess significance with the energy threshold}
\subsection{Evolution with the energy threshold}
\label{SecEvolE}

As discussed above, for a given astrophysical scenario there is a large dispersion in the expected values of the various quantities characterising the BS maximum in a dataset with the Auger statistics, namely the value of the maximum significance, the central direction of the corresponding flux excess, its energy threshold and its angular scale. This dispersion, however, also depends on the level of the true anisotropy associated with the model, that is the anisotropy that would be observed with infinite statistics. { The larger the underlying anisotropy, the lower the dispersion (for a given UHECR statistics)}. For this reason, in the following study we choose the parameters of the model in such a way that the median of the values of the maximum significance in the simulated datasets, $\sigma_{\rm max}^{\rm BS}$, is roughly similar to the value reported by Auger (for comparable statistics).  

We first examine the evolution of the BS maximum significance as a function of the energy threshold, $E_{\rm th}$, for our baseline scenario with composition model A and the Sun+Planck GMF model with $\lambda_{\rm c}$=100~pc. We thus leave the angular scale unchanged, at $\psi = 27^\circ$, and look for the most significant flux excess at each value of $E_{\rm th}$ from 32 to 80~EeV, with steps of 1~EeV. The result is shown on the top left panel of Fig.~\ref{FigSigVs1a}. %Each thin lines with a different colour corresponds to one of 10 individual datasets, randomly picked out of the 300 datasets simulated from the same astrophysical model, while the thick dashed blue line shows their average, energy bin per energy bin. The blue shaded area shows the 90\% interval for all 300 datasets (energy bin by energy bin), and the brown shaded area shows the 99\% interval of control datasets built from a purely isotropic distribution (with the same exposure map and statistics).

As can been seen, the average value of the maximum significance, $\sigma_{\rm max}^{\rm BS}$, steadily decreases as the energy thresholds increases. This is due to the rapid decrease of the UHECR statistics as a function of energy, which is not compensated by significantly larger intrinsic anisotropies, since the energy evolution of the particle rigidities remains weak (see for instance the discussion in \cite{Lemoine2009}) for the assumed composition, consistent with the measurements of Auger.
However, when looking at individual realisations, the situation appears much more erratic. From almost any single simulated dataset with the current statistics, the general trend, although very clear on average, cannot really be guessed. In particular, some local maxima are often found for intermediate energy thresholds, which may then wrongly seem to reveal a preferred energy scale, while the global view on the 300 datasets clearly shows that this energy scale has nothing to do with the underlying astrophysical model, but only with the particular dataset under examination.

Following the discussion of the previous section, we also extended the range of the energy scan, from 8~EeV to 80~EeV, with steps of 1~EeV. The results are shown on the bottom left panel of Fig.~\ref{FigSigVs1a}. They confirm that larger values of $\sigma_{\rm max}^{\rm BS}$ are obtained on average at lower energy thresholds. 

It is interesting to note that the direction in the sky in which these maxima are found is very similar in the case of the wider scan, compared to the more restricted one. However, the dataset-to-dataset dispersion is smaller in the former case. This results from the fact that the anisotropies at lower energy generically have a larger significance, due to the larger statistics (even though they are not necessarily intrinsically stronger).

\subsection{Evolution with the angular scale}

We now repeat the analysis by varying the angular scale, $\psi$, instead, while keeping the energy threshold fixed. In the top-right panel of Fig.~\ref{FigSigVs1a}, we show the evolution of the maximum significance of a flux excess as a function $\psi$, in the case of $E_{\rm th} = 40$~EeV and for the same data sets as above. Again, the average value of $\sigma_{\rm max}^{\rm BS}$ shows a steady evolution with the angular scale, with larger values at larger smoothing angles. { This is related to the predominance of the large scale structures of the source distribution in the observed anisotropy patterns. Tight UHECR multiplets from individual sources are essentially absent because of the large magnetic deflection (again consistent with the Auger composition measurements).  %This is related to the predominance of the large scale structures in the source distribution over the small scale structures of UHECR multiplets that would arise from the contribution of individual sources -- the latter being essentially absent because of the large magnetic deflections (again as implied by the Auger composition measurements). However, individual dataset with the current statistics do not show such a clear trend as the angular scale increases.

As for the energy evolution, we also extended the range of values of $\psi$, scanning up to 45 degrees (with 1-degree steps). The results are shown on the bottom right panel of Fig.~\ref{FigSigVs1a}. They confirm the indicated trend, but do not seem to provide additional information at the level of individual datasets, given the erratic behaviour of the excess significance as a function of $\psi$. We note that the results shown on Fig.~\ref{FigSigVs1a} are perfectly in line with the findings of Sect.~\ref{sec:SBEandPsiScales} and Fig.~\ref{FigEthPsi}.}% even though individual datasets can exhibit BS maxima with different values of the parameters, in general the most significant flux excesses in the parameter range explored by Auger are to be expected at the lowest energies and the largest angles, at least for the astrophysical scenarios that we investigated. Unfortunately, the fact that the Auger BS maximum does not satisfy this property cannot be used at this point to reject these types of scenarios with a high confidence level, given the large dispersion in the parameters and the expected erratic behaviour of the excess significance as a function of both $E_{\rm th}$ and $\psi$.

\begin{figure*}[h!]
   \centering
    \includegraphics[width=8.5cm]{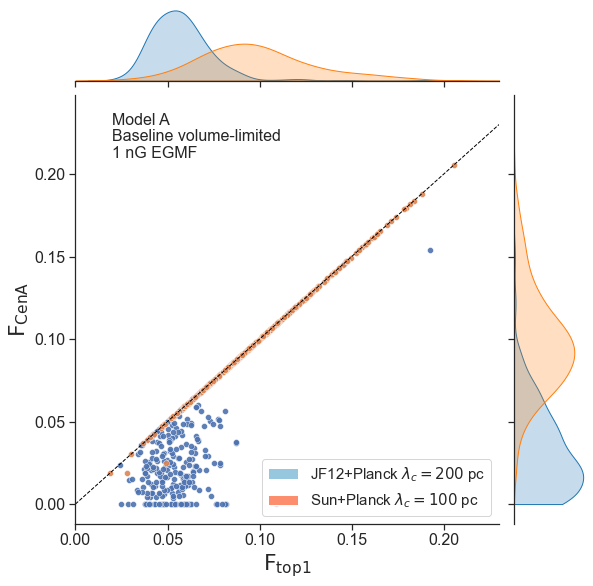}
    \includegraphics[width=8.5cm]{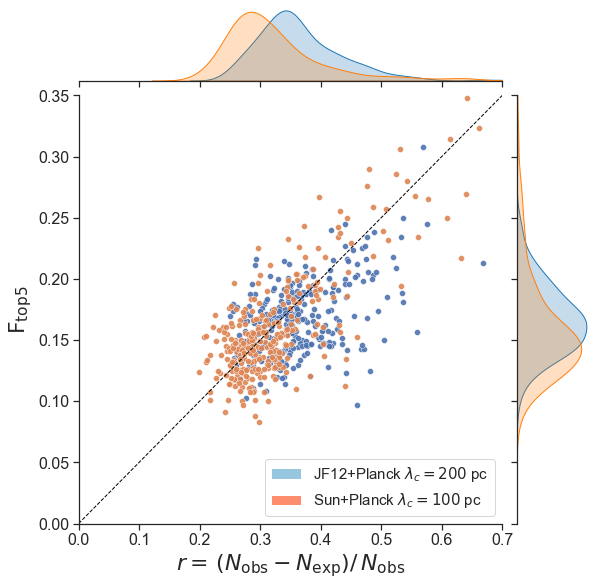}\\
    \includegraphics[width=8.5cm]{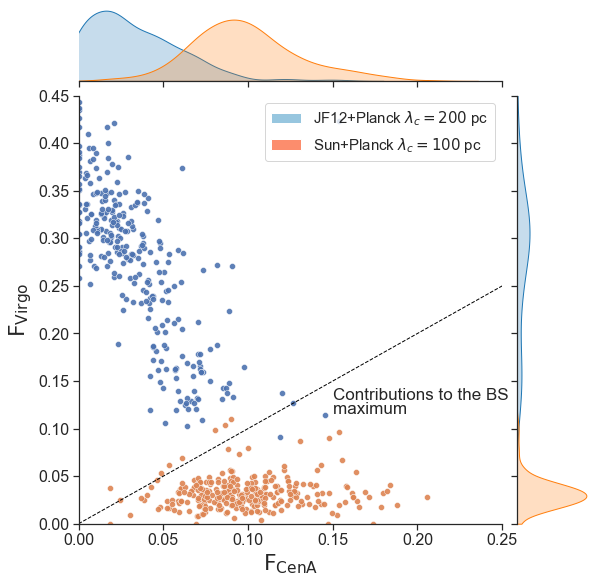}
    \includegraphics[width=8.5cm]{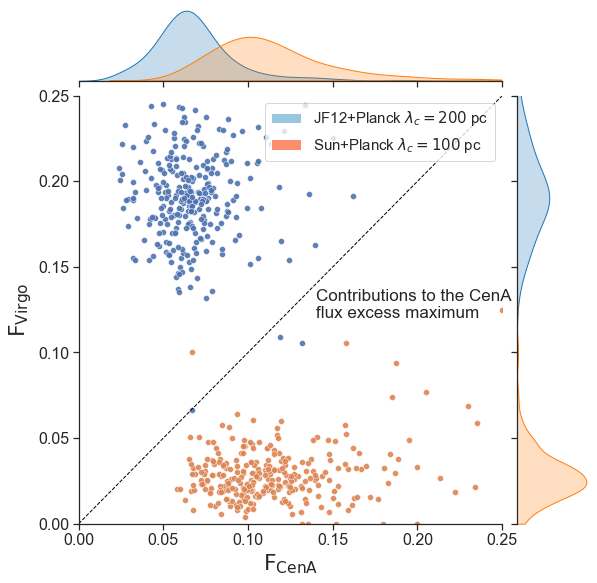}
      \caption{Scatter plot of the contributions of different sources to the BS maximum, for 300 realisation of the baseline scenario, with the JF12+Planck GMF model (blue dots) or the Sun+Planck GMF model (orange dots). The corresponding probability distributions are shown along the top and right borders of the plots. Top left: fraction of events coming from Cen~A vs. the fraction of events coming from the dominant source (i.e. the source contributing the most to the flux in the BS maximum angular window). Top right: fraction of events coming from either of the 5 dominant sources in the BS maximum window vs. the relative flux excess, $r$ (see text) in that window (the dashed line displayed to guide the eye is of equation $y=0.5x$). Bottom left: fraction of events in the BS maximum window coming any source in the Virgo association vs. the fraction of events coming from Cen~A. Bottom right: same as bottom left, but for the excess in the Cen~A direction instead of the BS maximum direction. The plot thus shows the fraction of events coming from any source in the Virgo association that are found in the angular window centered on Cen~A for which a flux excess has the largest significance vs. the fraction of events in that window coming from Cen~A.
              }
         \label{FigContribVL1}
\end{figure*}

\subsection{Distinct contributions to the flux excesses}
 \label{SectContrib}

It is instructive to examine the relative contribution of different sources to both the BS maximum and the Cen~A region, at the energy and angular scales where a flux excess is the most significant. This cannot be done with actual data, of course, but we can easily extract from our simulated datasets which sources contribute the most, and at which level. In the case of our baseline scenario, the hierarchy of sources or sky regions contributing to the observed excesses are found to depend critically on the assumed GMF model, even when a fixed distribution of sources is considered. Here, we focus on the most significant excesses obtained with the BS analysis and in the direction of Cen~A, regardless of the value of $E_{\rm th}$ or $\psi$ at which they occur within the limits of the Auger scan.

The top-left panel of Fig.~\ref{FigContribVL1} is a scatter plot of two quantities computed for each of the 300 datasets simulated for the model under consideration: i) in abscissa, we show the fractional contribution of the dominant source, $F_{\rm Top1}=N_{\rm Top1}/N_{\rm obs}$, where $N_{\rm obs}$ is the total number of events in the angular window defining the BS maximum (same as $N_{\rm on}$ in Eq.~\ref{eu_eqn}), and $N_{\rm Top1}$ is the number of events coming from the source which contributes the most to this maximum; ii) in ordinate, we show the corresponding fractional contribution of the source Cen~A, $F_{\rm CenA} = N_{\rm CenA}/N_{\rm obs}$. The results obtained with our two reference GMF models are shown with different colours (see legend)%, and the probability distributions of the different contributions are shown as continuous curves on the top border and right border of the panel
. The datasets located on the main diagonal correspond to datasets in which Cen~A is indeed the dominant source in the BS maximum window, which is almost always the case when the Sun+Planck GMF model is used, and almost never the case with the JF12+Planck model (NB: M104, located slightly south of the Virgo cluster, is often the dominant source in that case, despite being more distant than Cen~A, because of its much larger K-band luminosity, accordding to 2MRS). In any case, the contribution of the dominant source remains low, which indicates that the observed excess cannot be associated with a specific, individual source. Even when the dominant source is Cen~A and the BS maximum is indeed located in a direction close to its position, its relative contribution has a median value between 5 and 10\%, and never exceeds 20\% of the flux in that direction. %In other words, at least 80\% of the events recorded in the angular window corresponding to the most significant flux excess come from other sources than the dominant one, and this fraction increases to $\sim 95\%$ is the case of the JF12+Planck GMF model.

Similarly, it is interesting to investigate the contributions of dominant sources to the part of the UHECR flux in the BS maximum window that appears in excess of the isotropic expectation. For this, we define the flux excess ratio, $r$, as $r=(N_{\rm obs}-N_{\rm exp})/N_{\rm obs}$. In the top right panel of Fig.~\ref{FigContribVL1}, we show the cumulated fraction of events in the top 5 sources as a function of this ratio, for the same 300+300 simulated datasets. The results, which are similar for both GFM models, show a flux excess between, say, one fifth and one half the isotropically expected flux. However, the contribution of the top 5 sources is always significantly lower than this, between $\sim$10\% and $\sim$25\%, and typically account for only about one half of the apparent excess (as represented by the dashed line on the plot).

Thus, it appears that the analysis of a flux excess cannot easily be used, by itself, to suggest one or even several dominantly contributing sources, but rather reflects a preferred direction in the sky where a large number of sources happen to contribute and build, together, a significant excess. Our simulations therefore suggest that this type of analyses should not be expected to pinpoint a particular source (at least with the current statistics), but could potentially constrain some general features of the source distribution. However, this can only be done if an assumption regarding the GMF can be made reliably. In particular, a key feature of the Auger dataset is the absence of a significant flux excess in the direction of the Virgo cluster, where many source candidates could be expected in principle. But this observation cannot be turned into a constraint on the UHECR source distribution until we have a reliable GMF model.
%{\color{red}[J'ai ajouté ce petit paragraphe pour deux raisons : 1) pour poser une premier élément de discussion/conclusion – car je pense que c'est mieux d'égrainer les remarques de ce genre, plutôt que tout reporter à la fin, à la fois pour aider le lecteur qui pourrait saturer un peu avec la quantité de résultats qui arrivent les uns après les autres et se demander où on veut en venir avec tout ça, et pour éviter de faire des remarques à la fin dont la justification aura été vue de nombreuses pages avant, et que le lecteur aura oublié ; et 2) pour faire la transition avec l'étude des contributions spécifique de Virgo, qui vont tout à fait dans ce sens. What do you think?]}

This is clearly demonstrated by the results shown in the two bottom panels of Fig.~\ref{FigContribVL1}, where we compare the weight of the sources in the Virgo cluster (more precisely the Virgo association, as defined in Kourkchi \& Tully, 2017) with the weight of Cen~A among the UHECR events in the angular window corresponding to the BS maximum (bottom left) or the Cen~A excess (bottom right). The difference between the two GMF models is striking, but easily understood from our earlier remark that the Sun+Planck GMF model strongly demagnifies the sources in the Virgo direction. In this case, indeed, the contribution of these sources is always much smaller than that of Cen~A, not only to the signal around Cen~A, but also to the signal around the BS maximum, wherever it may be. It is typically between 1 and 5\%, while Cen~A contributes between 5 and 15\%.

Conversely, the JF12+Planck GMF model does not strongly affect the flux of the UHECRs entering the Galaxy from directions around that of the Virgo association, so the corresponding sources contribute together a large fraction of the BS maximum signal, adding up to roughly one third of the events, while Cen~A only contributes a few percent at most. In this case, of course, the BS maximum is indeed strongly ``attracted" toward the direction of Virgo, as already shown in Fig.~\ref{FigBSVL1b}, which explains the small contribution of Cen~A. Yet, even when considering the signal in the direction of Cen~A itself (in the angular window corresponding to the most significant flux excess), Cen~A as a source contributes between $\sim$3 and 10\%, while the sources in the direction of Virgo contribute significantly more, namely between 15 and 25\%. This larger contribution, however, is a collective effect of several sources, since Cen~A generally remains the dominant source in this angular window even in the case of the JF12+Planck model.

\subsection{Dependence on UHECR composition}
\label{sec:compo}
\begin{figure*}
   \centering
   \includegraphics[width=8.5cm]{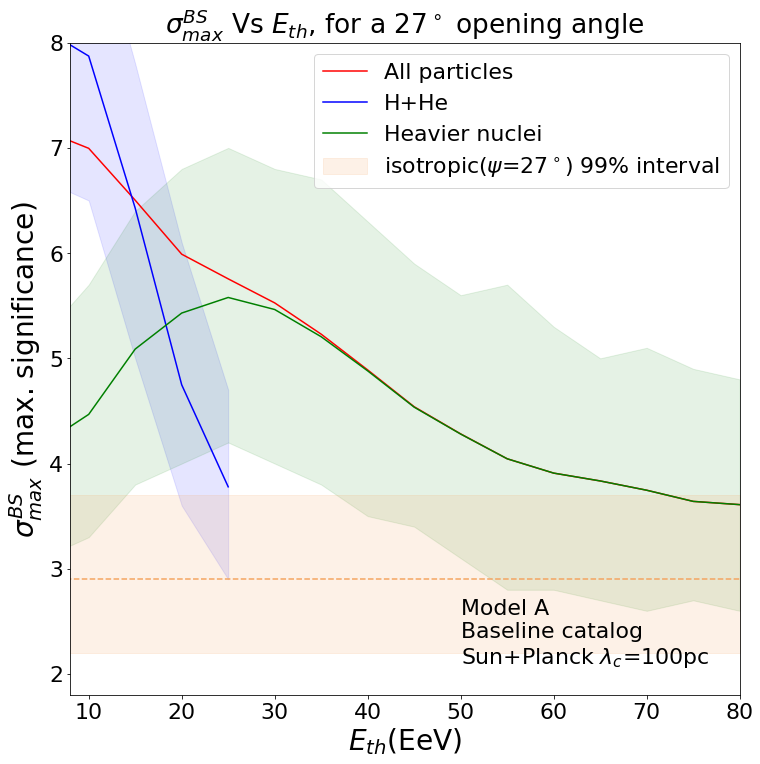}
   \includegraphics[width=8.5cm]{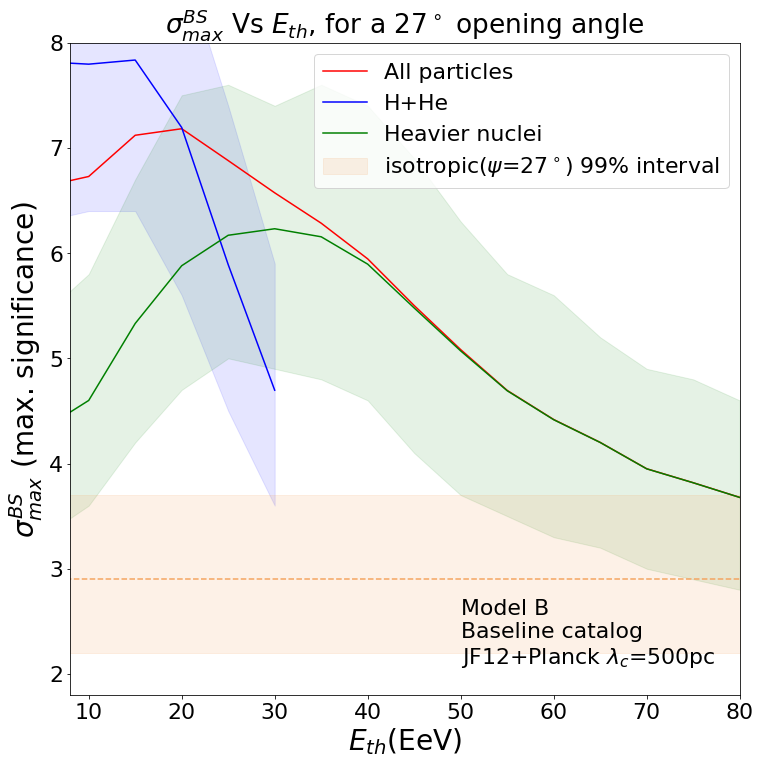}
 \caption{Averaged $E_{\rm th}$  evolution of the BS maximum significance for different source composition models. The $E_{\rm th}$ evolution averaged over 300 datasets (setting $\psi$ to 27$^\circ$) is shown with a thick red line for the all particles dataset while the light (H+He) and heavier components are shown respectively in blue and green (the shaded areas show the 90\% dispersion of the 300 datasets). The brown shaded area shows the 99\% dispersion of isotropic datasets. For both panel, the baseline volume-limited catalog and a 1~nG EGMF are assumed. The left panel shows the case of the source composition model A and the Sun+Planck GMF model with $\lambda_{\rm c}$=100 pc, and the right panel the case of model B and JF12+Planck model with $\lambda_{\rm c}$=500 pc.
              }
         \label{FigSigComp}
\end{figure*}

Since the magnetic deflections depend on the rigidity of the nuclei, and thus at a given energy on their charge, the distribution of the arrival directions of the UHECRs could potentially provide a handle on their composition, independently of the measurement of the depth of the shower maxima in the atmosphere. In principle, specific signatures of the composition and its evolution with energy could thus be found in the associated evolution of the anisotropy patterns. As we did in Paper~I with the evolution of the amplitude of the first two harmonics of the angular modulation (dipole and quadrupole), we show in Fig.~\ref{FigSigComp} the evolution of the significance of the BS maximum as a function of the energy threshold, $E_{\rm th}$, for two different composition models: model~A (on the left), and model~B (on the right), which is characterised by a larger maximum rigidity (see Table~2 of Paper~I) and a weaker component of light nuclei (although this light component has a larger proton-to-helium ratio than in model A). In the case of Model~B, the GMF coherence length is increased to 500~pc, to recover similar values of the average significances above 30~EeV as with model A, and thus allow comparison.

On Fig.~\ref{FigSigComp}, the red lines show the evolution of $\sigma_{\rm max}^{\rm BS}$, averaged over the 300 datasets for each model. Thus, the red line of the left panel is by definition the same as the dashed blue line on the bottom left panel of Fig.~\ref{FigSigVs1a}. As can be seen, models A and B show different behaviours, with a maximum significance reached at higher energy for model B, although still below 30 EeV (i.e. the limit of the Auger scan). %However, as above, the current level of statistics does not allow to discriminate between the models, due to the large dataset-to-dataset fluctuations (see Fig.~\ref{FigSigVs1a}).

To better understand this general behaviour, it is interesting to isolate the ``light component” (H and He) and the ``heavy component” (all other nuclei) within the all-particle datasets, and to look at the energy evolution of $\sigma_{\rm max}^{\rm BS}$ for these two components separately. The results are also shown on Fig.~\ref{FigSigComp}. The blue lines correspond to the light component, and the green ones to the heavy component. The shaded areas of the same colours show the 90\%-interval over which the values for individual datasets fluctuate above and below the average. %Note that the blue lines end at somewhat different, but always low energy, due to the disappearance of the corresponding nuclei among the UHECRs (a generic feature of the low-$E_{\max}$ models).

For a given nuclear species, it is natural to expect an evolution of the corresponding values of $\sigma_{\rm max}^{\rm BS}$ that is first increasing with energy, due to the proportional increase of the rigidities, and then reaching a maximum before decreasing as a result of the sharply decreasing statistics. The increasing part is missing on the plots for the light component, because of the low energy cutoff for H and He (although the ``plateau” around the maximum is visible in the case of model B, which has a slightly higher maximum rigidity). Not surprisingly, the energy evolution of $\sigma_{\rm max}^{\rm BS}$ for the all-particle datasets depends on the UHECR composition through the balance between the light and heavy components (and at higher order on additional details of their composition). This is how complementary handles on the composition could be obtained, in principle, provided the statistics is large enough. { The energy evolution of individual datasets with the current Auger statistics (depicted in Fig.~\ref{FigSigVs1a}) as well as our discussion of larger statistics datasets (see Sect.~\ref{Sec:stat}  below) suggest that a substantial increase of exposure would however be required for that purpose. Finally we note that on average, our simulations suggest that the most significant flux excesses are likely to be found with energy thresholds below 30~EeV for astrophysical models that account for the evolution of the composition implied by the Auger data.}

\begin{figure*}
   \centering
   \includegraphics[width=9cm]{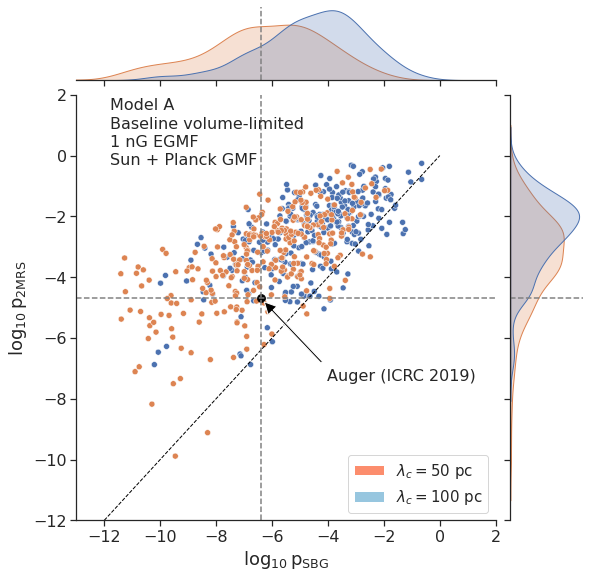}
   \includegraphics[width=9cm]{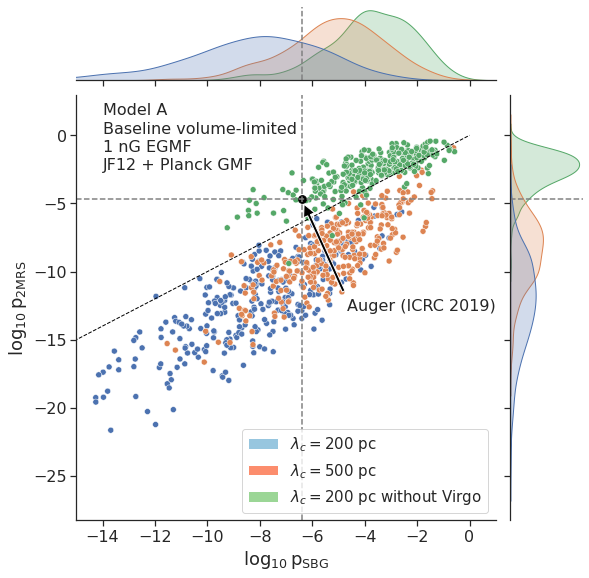}
 \caption{p-values obtained after performing the likelihood analysis on our datasets. The logarithm of the p-value corresponding to the maximum likelihood obtained for the SBG catalog ($P_{\rm SBG}$) catalog is plotted against that obtained for 2MRS ($P_{\rm 2MRS}$) for each dataset. The astrophysical model considered to build our datasets corresponds to model A, the baseline catalog and a 1 nG EGMF. The values reported by Auger in \citet{AugerICRC2019} are shown with large black full circle. The individual distributions of the different quantities plotted are shown on top of the coordinate axis.
 The left panel shows the case of the Sun+Planck GMF model for two different values of the coherence length. The right panel shows the case of the JF12+Planck model. In this case we also show datasets produced after excluding galaxies from the Virgo association from the source catalog (see legend). }
         \label{LL_base_Pval}
\end{figure*}

\section{Results of the likelihood analysis}
\label{Seclike}

We now turn to the results of the likelihood analysis on our simulated datasets for the same (baseline) scenarios as above. For the signal component of the likelihood fit, we focus on two catalogs: the 2MRS catalog (used as a proxy to trace ordinary galaxies) and the starburst galaxy (SBG) catalog (see Sect.~\ref{like}).

\subsection{Significance of the rejection of isotropy}
\label{sec:signifRejectIso}

In Fig.~\ref{LL_base_Pval}, we show the p-values corresponding to the maximum of the likelihood function for the SBG catalog (in abscissa) and the 2MRS (in ordinate), for two different models of the GMF (Sun+Planck on the left, JF12+Planck on the right) and two different values of $\lambda_{\rm c}$ for each of them. Each data point corresponds to one of the 300 datasets simulated from our astrophysical scenarios. The values obtained with the Auger data are shown with a black circle.

An important lesson can be drawn from the comparison of the results obtained with these two GMF models. It is indeed striking that the p-values obtained with the Sun+Planck model are (almost) always larger when correlating the datasets with the 2MRS catalog than when correlating them with the SBG catalog, whereas the exact opposite is true with the JF12+Planck catalog. To make this easier to see, we plotted a diagonal dashed line corresponding to equal p-values for both source models: the data points are above that line on the left plot, and below it on the right plot. %In other words, the significance of the rejection of the null (isotropy) hypothesis can be either systematically higher or systematically lower for the SBG catalog than for the 2MRS catalog, depending on the assumption regarding the GMF models. 
This fact should be welcome as a warning against premature interpretations of the Auger data as suggesting that a given catalog somehow provides a better description of the UHECR sources than another.
%As a matter of fact, one should always keep in mind that a stronger rejection of the isotropy does \it{not} imply a better description of the data by any mean. But in the case under study, we can even witness a complete reversal of the situation depending on the choice of the magnetic field model, both of which are similarly consistent with the available data, to the best of our knowledge. It is worth stressing again that the datasets shown on both panels of Fig.~\ref{LL_base_Pval} are obtained with the exact same underlying astrophysical scenarios, i.e. with the very same sources, having the same intrinsic power and the same UHECR spectrum and composition at injection. Only the assumed GMF model is different.
Although a typical realisation on the left panel of Fig.~\ref{LL_base_Pval} appears very different from a typical realisation on the right panel from the point of view of the likelihood analysis, all the datasets shown on both panels are obtained with the exact same underlying astrophysical scenarios, that is with the very same sources, having the same intrinsic power and the same UHECR spectrum and composition at injection. Only the assumed GMF model is different.

The reason for the observed difference in the likelihood values is mostly related to the weight of the sources in the direction of the Virgo cluster, which happens to be strongly demagnified by the magnetic field configuration of the Sun+Planck GMF model. The SBG catalog is thus more efficient in rejecting the isotropy hypothesis in that case, since the UHECR flux from the direction of Virgo is then much lower than what would be expected from the 2MRS catalog if one simply assumes a gaussian blurring of the UHECR arrival directions, without systematic deflections, as in the Auger likelihood analysis.

The specific role of the Virgo cluster and its possible demagnification by the GMF is further seen on the right panel of Fig.~\ref{LL_base_Pval}, where we have added the result of the likelihood analysis for 300 additional datasets simulated with the JF+Planck GMF model, but from a modified version of our baseline scenario in which we removed the sources belonging to the Virgo cluster (based on their identification in \cite{Kourk2017}), as we also did for some scenarios in Paper~I.
In the non-modified case (red points), while the p-value obtained by Auger with the SBG catalog is quite typical of the values obtained with our simulated datasets, their p-value is almost always much lower than that of Auger when performing the likelihood analysis with the 2MRS catalog. However, when one removes the Virgo galaxies from the possible sources (green points), the resulting datasets are found to reject the isotropy hypothesis with larger significance when being analysed against the SBG catalog than against the 2MRS catalog, just as in the case of the Sun+Planck GMF model. This confirms the role of the Virgo cluster demagnification in the latter case. We also checked that the situation does not change, neither qualitatively nor quantitatively, when one removes the Virgo galaxies in the case of the Sun+Planck GMF model, as expected since the contribution of these galaxies is already strongly attenuated in that case.

We also note that, in these latter cases, a more significant rejection of the isotropy hypothesis is obtained when analysing the Auger data against the SBG catalog despite the absence of CenA from this catalog, even though CenA provides the strongest contribution to the flux excesses in our datasets. In the SBG model, excesses in the direction of the sky close to that of CenA are indeed accounted for by the expected contribution of NGC4945 and M83. Interestingly, these two sources are absent from the baseline source catalog that we use to produced our datasets, since they do not pass the $K_{\rm s}$-band luminosity cut applied to produce this volume limited catalog.

\subsection{Parameters of the maximum likelihood anisotropy signal}

\begin{figure}
   \centering
   \includegraphics[width=8.5cm]{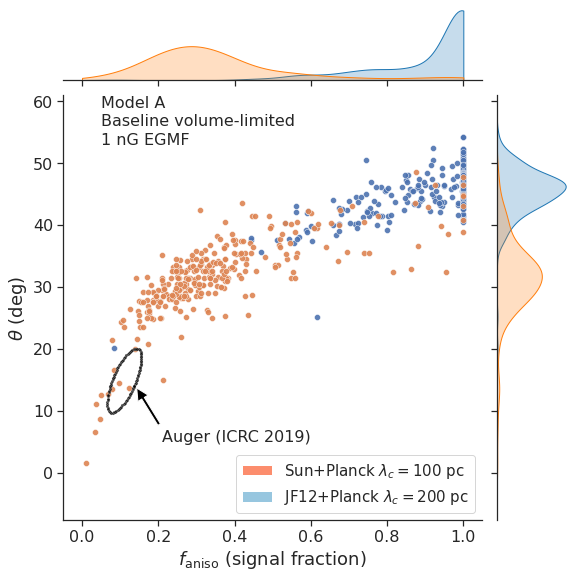}
   
 \caption{Best-fit parameters obtained after performing the likelihood analysis for the SBG catalog on our datasets. The value of $f_{\rm aniso}$ and $\theta$ allowing to maximize the likelihood for each dataset are plotted against each other for the astrophysical model considered in Fig,~\ref{LL_base_Pval} and two GMF models (see legend). The 1$\sigma$ ellipse reported in \citet{AugerICRC2019} from Auger data is shown. The individual distributions of the different quantities plotted are shown on top of the coordinate axis.
              }
         \label{LL_base_Par}
\end{figure}

For each choice of a candidate catalog, the likelihood analysis identifies the values of the parameters for which the largest correlation signal is obtained, namely $f_{\rm aniso}$, which is the fraction of UHECRs that may be associated with sources in the catalog, and $\theta$, which is the angular scale over which the catalog is to be smoothed (see Sect.~\ref{like}). A scatter plot of the values obtained in the case of the SBG catalog is given in Fig.~\ref{LL_base_Par}, for 300 datasets simulated with the  Sun+Planck GMF model (in red), and 300 datasets simulated with the JF12+Planck model (in blue). In the latter case, large values of the signal fraction, $f_{\rm aniso}$, and large blurring angles, $\theta$, are clearly preferred for the astrophysical model considered (with very small dependence on the assumed coherence length of the turbulent component of the GMF, in the range we considered). This is quite different from what was reported by Auger with their dataset, even though the p-values are of the same order as those found by Auger for this source catalog, as shown above. On the other hand, lower values of both $f_{\rm aniso}$ and $\theta$ are found for the Sun+Planck model, with a majority of datasets in the range $f_{\rm aniso}\sim 0.2-0.4$ and $\theta\sim25^\circ-35^\circ$. 
{ The values  of $f_{\rm aniso}$ and $\theta$ are  clearly correlated, with large values of $f_{\rm aniso}$ systematically associated with large smoothing angles. The likelihood function can thus occasionally yield values of $f_{\rm aniso}$ reaching 1 even though the intrinsic anisotropy of the simulated skymap under study is weak.}
Only a few datasets appear to lie within the 1$\sigma$ ellipse reported in \citep{AugerICRC2019}. 
%Further discussion of this apparent mismatch requires to take into account that the actual distribution of the UHECR sources contributing to the observed flux at the present time is subject to cosmic variance. This is addressed in the next section.

Further discussion of this apparent mismatch requires consideration of the cosmic variance, this is of the fact that not all the sources above a given luminosity may contribute at all times, so variations depending on the specific subset of sources actually contributing to the current UHECR flux should also be explored. This is addressed in the next section.

Finally, we note that most datasets reject the isotropy hypothesis with the largest significance when the threshold energy, $E_{\rm th}$, is set at the lowest values, that is close to the 32~EeV scan boundary. This is reminiscent of what we found for the BS and the Cen~A flux excess analyses (cf. Figs.~\ref{FigEthPsi} and~\ref{FigSigVs1a}). However, there are indeed some datasets for which the largest significance is obtained for values of $E_{\rm th}$ close to 38~EeV or larger, in which case one might be tempted not to extend the scan range, although it could be relevant. Likewise, the $E_{\rm th}$ evolution of the p-value is found to have a large dataset-to-dataset variability at the current level of statistics, again reminiscent of what we found for the BS search (cf. Fig.~\ref{FigSigVs1a}).
%{\color{red}[Ne devrait-on pas montrer un plot de la distibution  es $E_{\rm th}$? Sinon, c'est un peu abstrait…] Reponse de Denis ---> Ca ne me parait pas capital de rajouter une figure de plus apres avoir dit que c'etqit qualitativement similaire a ce qu'on a trouve lors du BS ... Donc dites moi si vous y tenez vraiment ...}

%We finally note, concerning the threshold energy of the dataset $E_{\rm th}$ at which the isotropy hypothesis is best rejected, that like for the BS and the CenA flux excess analyses, the lowest values of $E_{\rm th}$, close the 32 EeV scan boundary, are usually preferred for most datasets. However values of $E_{\rm th}$ close to 38 EeV or larger for the maximum of the likelihood are not rare. Moreover, the $E_{\rm th}$ evolution of the p-value is found to have a large dataset to dataset variability at the current level of statistics as we experience for the BS search (top letf panel of Fig.~\ref{FigSigVs1a}).  

\section{Cosmic variance}

To study the typical cosmic variance associated with our results, we adopt the mother catalog approach presented in Sect.~\ref{sourdist} and already used in Paper~I. In this section, we restrict our discussions to source densities larger than $10^{-4}$ $\rm Mpc^{-3}$, which lead to better agreement with the observed level of anisotropy, at least for the  considered range of extragalactic and Galactic magnetic field intensities. Lower source densities will be briefly discussed later. %We start by setting the source density to $10^{-3}$ $\rm Mpc^{-3}$. 

%We now turn to the discussion of the influence of the cosmic variance on the dispersion of our predictions. To that purpose, we use the mother catalog approach presented in Sect.~\ref{sourdist} and in Paper~I. In this section, we restrict our discussions to source densities larger than $10^{-4}$ $\rm Mpc^{-3}$, which lead to better agreement with the observed level of anisotropy, at least for the  considered range of extragalactic and Galactic magnetic field intensities. Lower source densities will be briefly discussed in later sections. We start by setting the source density to $10^{-3}$ $\rm Mpc^{-3}$. 

\subsection{Flux excesses from the blind search and around Cen~A}

%Unsurprisingly, we find that the spread of the predictions for the 300 realizations of the source distribution is larger than what we obtained with our baseline catalog, either in terms of the dispersion of the maximum significance or that of the celestial location of the maximum.

As expected, the dispersion of the results for 300 datasets obtained with different realizations of the UHECR source distribution is larger than for datasets obtained with the baseline catalog, both regarding the significance of the maximum flux excess and its location over the sky.

\begin{figure}
   \centering
   \includegraphics[width=8.5cm]{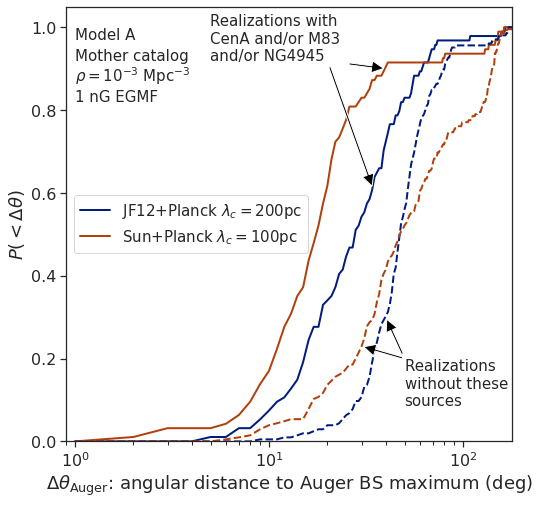}
\caption{Cumulative distributions of the angular distance $\Delta\theta_{\rm Auger}$ (see Sect.~\ref{direction}), obtained for the mother catalog approach with model A, a source density of $10^{-3}$ $\rm Mpc^{-3}$ and a 1 nG EGMF. The red and blue curves correspond to  different GMF models, as indicated. Separate distributions are shown for source realizations including either Cen~A, NGC4945 or M83 (solid lines) and source realizations including none of those (dashed lines). 
              }
         \label{FigDistoCenMC}
\end{figure}

It is interesting to consider separately the realizations that include either Cen~A, NGC4945 or M83 (or to impose one of these galaxies to be present among the sources). Indeed, this subset of realizations has a larger probability to produce a BS maximum localized close to { that found by Auger}. This is true for both GMF models, with a shift toward the direction of the Virgo cluster in the case of the JF12+Planck model (as anticipated from the results discussed in the previous sections).
On Fig.~\ref{FigDistoCenMC}, we show the cumulative distributions of the angular distance %between the position of the BS maximum and the direction of Cen~A, 
$\Delta\theta_{\rm Auger}$ defined in Sect.~\ref{direction}, in the case of model A and a 1~nG EGMF, using the mother catalog approach with a source density of $10^{-3}$ $\rm Mpc^{-3}$. %The red and blue curves correspond to the JF12+Planck and Sun+Planck GMF models, respectively. 
Separate distributions are shown for source realizations including either Cen~A, NGC4945 or M83 (solid lines) and source realizations including none of those (dashed lines).

In the former case, the median angular distance $\Delta\theta_{\rm Auger}$ is $\sim17^\circ$ for the Sun+Planck model, and $\sim27^\circ$ for the JF12+Planck model. The corresponding distribution of the BS maximum position over the sky is similar to that obtained above with the baseline catalog.
%, that is, only a few realizations out of 300 have a BS maximum as close as to Cen~A as in the Auger data. 
On the other hand, for the realizations without any of the quoted sources, this median angular distance is around $50^\circ$ for both GMF models, with a lower significance of the BS maximum flux excess, on average. Thus, a position of the BS maximum in the close vicinity of the location of Cen~A may be seen as somewhat favoring scenarios in which one of the quoted nearby sources is among the UHECR accelerators. However, it is important to keep in mind that this may be GMF model dependent, and that even when including these nearby sources, { the expected dispersion in the BS maximum directions does not allow to draw strong conclusions from the direct comparison with the direction obtained from the Auger data at the current level of statistics, as already commented in Sect.~\ref{direction}.}  
%agreement between the BS position in our simulations and in the Auger data remains marginal { peut etre on devrait surtout mettre qu'il ne permet pas de faire de fortes conclusions}.

%When isolating realizations which include either CenA, NGC4945 or M83 (or by imposing one of these sources to be present in the source distribution), one finds that this subset of realizations has a larger probability to produce a BS maximum localized close to the direction of CenA for both GMF models. As discussed in the previous sections, the BS maximum is shifted toward the direction of the Virgo cluster in the case of the JF12+Planck model.  

%As can be seen in Fig.~\ref{FigDistoCenMC}, the median angular distance of the BS maximum to CenA is $\sim20^\circ$ for the Sun+Planck model and $\sim30^\circ$ for the JF12+Planck model, for the realizations which include either CenA, NGC4945 or M83. We note that their distribution on the sky is similar to what we found earlier for the baseline catalog, that is, only a few realizations result in a BS maximum as close as to CenA as what was found in Auger data. \\
%On the other hand, for the rest of the realizations, the median the angular distance of the BS maximum is around $50^\circ$ for both GMF models and the significance are in average lower a for a given assumed GMF model. The position of the BS maximum in the close vicinity of the location of CenA can thus be seen as a weak indication of the presence of one of the abovementioned nearby sources among UHECR accelerators. It is  however important to keep in mind that the latter statement may be somewhat GMF model dependent and that even when including these nearby source, the agreement between are simulations and Auger data remains quite marginal as far as the position of the BS maximum is concerned.

Essentially the same results are obtained when lowering the source density to a $10^{-3.5}$ $\rm Mpc^{-3}$, except that the assumed coherence length of the GMF and/or the intensity of the EGMF have to be increased not to overshoot the observed significance of both the BS and the Cen~A flux excess maxima, in particular for realizations in which at least one of the three above-mentioned nearby sources is present in the source distribution.

Finally, as expected, the realization-to-realization dispersion of the results also increases as the source density decreases.  

\subsection{Likelihood analysis}

\begin{figure}
   \centering
   \includegraphics[width=8.5cm]{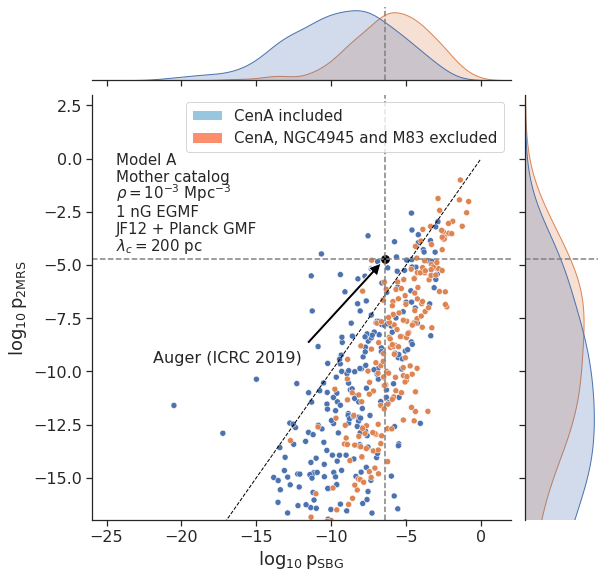}
   \includegraphics[width=8.5cm]{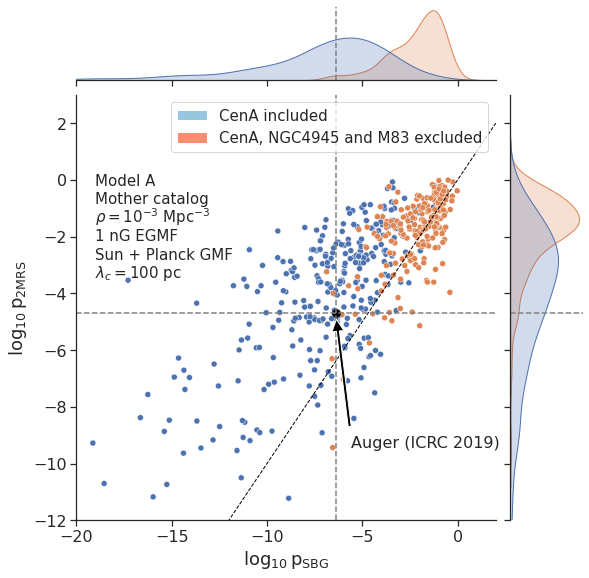}
\caption{Same as Fig.~\ref{LL_base_Pval} in the framework of the mother catalog approach. The top panel shows the results of the likelihood analysis obtained for the JF12+Planck GMF model while the bottom panel shows those obtained for the Sun+Planck GMF model. On each plots, two categories of realizations are shown : source distributions in which the presence of CenA is imposed and source distributions which include neither CenA nor NGC4945 or M83.
              }
         \label{FigLLMC_30}
\end{figure}

Allowing additional dataset-to-dateset fluctuations by taking into account the cosmic variance (i.e. drawing new sets of sources for each realization) does not change the general finding that the reference catalog leading to the most significant rejection of isotropy in the likelihood analysis depends on the assumed GMF model. This is shown on Fig.~\ref{FigLLMC_30}, where the same trend as in Fig.~\ref{LL_base_Pval} can be seen: the datasets obtained with the JF12+Planck model are essentially located below the "equal rejection line", while the opposite is true for the Sun+Planck model. However, the cosmic variance creates a few more outliers.

In Fig.~\ref{FigLLMC_30}, we also distinguish between realizations of the source catalog in which the presence of Cen~A is forced, and realizations in which neither CenA nor NGC4945 or M83 is present. In the case of the JF12+Planck GMF model, the presence of Cen~A leads to somewhat smaller p-values in general (the same was observed by forcing one of the other two sources instead of Cen~A). This could be expected, since the presence of a very nearby source tends to produce datasets with larger anisotropies on average (unless that source is strongly demagnified by the GMF). The presence of Cen~A also appears to result more often in realizations for which $P_{\rm SBG}$<$P_{\rm 2MRS}$, but such cases remain exceptional.

In the case of the Sun+Planck model, the impact of the presence of Cen~A is much stronger, as can be clearly seen on Fig.~\ref{FigLLMC_30}. This is because in that case, the contribution of Cen~A to the overall anisotropy is larger than in the case of the JF12+Planck model, because the Virgo cluster is anyway strongly demagnified (see Sect.~\ref{SectContrib} and Fig.~\ref{FigContribVL1}). The vast majority of the realizations showing significant anisotropy (mostly those having Cen~A as a source, but the same would be true for NGC4945 or M83) reject isotropy with a larger significance when using the SBG catalog, as discussed in Sect.~\ref{Seclike}.

Regarding the distribution of the parameters of the likelihood maximum, $f_{\rm aniso}$ and $\theta$, we find that including cosmic variance does not modify the main results, as they were presented in Fig.~\ref{LL_base_Par}. Even in the most optimal cases, that is using the Sun+Planck model and imposing the presence of Cen~A among the sources, only a handful of realizations (out of 300) fall within the 1$\sigma$ ellipse extracted from the Auger data. Most realizations lead to parameters in the range $f_{\rm aniso}\sim 0.2-0.4$ and $\theta \sim 20^\circ-30^\circ$.

Finally, we found that forcing the galaxy NGC253 to be among the sources does not significantly modify any aspect of the above discussion. As noted in Sect.~\ref{Seclike}, when using the JF12+Planck GMF model, a more significant rejection of isotropy is obtained with the SBG catalog for most realizations when the Virgo cluster sources are barred from UHECR sources.

\section{Source distributions biased toward starburst or star-forming galaxies}
\label{SecForceSBG}

\subsection{Motivation}

In this Section, we further explore the role of the most nearby sources in shaping the UHECR angular distribution, and examine how a bias toward specific classes of sources could be used to reach a better agreement between the simulated datasets and the Auger data at the highest energies. For this case study, we impose among the UHECR sources a sample of nearby star-forming galaxies (SFGs), mostly originating from the HCN survey of \citet{Gao2004} and also used by the Fermi collaboration in \citet{FermiSBG} as a list of targeted sources to study the GeV $\gamma$-ray emission from SFGs. The main motivation for this analysis is that this sample, to which we added the Circinus galaxy located close to the Galactic plane, was used in the original Auger likelihood analysis to model the signal component in the ``starburst galaxies (SBG) hypothesis''.\footnote{The most important sources of this sample were also present (and had a dominant contribution to the signal component) in the subsequent iterations of the likelihood analysis performed by the Auger collaboration, based either on the \citet{Becker2009} or the \cite{Lunar2019} samples.} We note that essentially all of these sources are present in the 2MRS catalog, and thus were randomly selected in some realizations of the UHECR source distribution in the ``mother catalog approach'' discussed above. Some of them (notably, NGC253 and NGC1068) are also present in our baseline catalog (provided that their K-band luminosity is large enough). In this Section, however, they are all jointly present in all realizations.

%\subsubsection{Production of source catalogs}
%\subsubsection{Source selection procedure}
%\subsubsection{UHECR source catalog production}
\subsection{SFG-biased catalog generation}
\label{Sec:SFG-bias}

In addition to the above-mentioned SFG sample, we complete the effective source catalogs for this study by randomly picking sources from our volume-limited catalog with the largest density, as we do in the mother catalog approach. We assume that these sources inject UHECRs with the same energy spectrum and composition as in our model~A, with an intrinsic luminosity proportional to their total infrared (IR) luminosity, $L_{\rm IR}$, which is expected to be a good proxy for the star formation rate of the galaxy. For the galaxies in the SFG sample, we use the IR luminosity given in \citet{Gao2004}, corrected whenever needed to account for the more recent distance estimates that we are using (see Paper~I). For the other sources, drawn from the mother catalog, we randomly assign to them an IR luminosity by sampling the IR galaxy luminosity function obtained from the observations of the Spitzer satellite in \citet{Rodi2010}.

In practice, we fix the density of the source catalogs built in this way by setting a threshold in the IR luminosity, above which we keep all the sources. This applies also to the galaxies in the SFG sample, which is assumed to include all the potential UHECR sources within 10 Mpc, once Circinus is added. The number of sources that need to be added from the mother catalog to complete the sources beyond 10 Mpc depends on the chosen source density (or equivalently the chosen $L_{\rm IR}$  threshold). For instance, a cut on $L_{\rm IR}$ at $2\times 10^{10}\,L_\sun$ (where $L_\sun$ is the IR luminosity of the Sun) corresponds to a source density of $\simeq 10^{-3}\,\rm Mpc^{-3}$ and allows to keep in particular M82, NGC253, M83, NGC4945, Circinus and NGC1068 from the initial SFG sample.

We also explore another possible type of biases in the UHECR source distribution, by implementing a second way to complete the source catalog beyond 10~Mpc, namely by excluding {\it{a priori}} all the galaxies located inside large galaxy clusters with a mass larger than >$10^{14}\,M_{\sun}$, as identified in \citep{Kourk2017}). This is motivated, although in a very simplified way, by the possible lack or deficit of SFGs or SBGs in rich galaxy clusters (see, e.g., \citet{Guglielmo2015,Boselli2016}).

%In practice, we fix the assumed source density of the source catalogs built with this procedure by fixing a threshold IR luminosity to our simulated UHECR sources. This threshold luminosity corresponds to a source density according the above-mentioned IR luminosity function. Only the sources with a IR luminosity larger than the threshold are kept among the SFG sample (which is assumed to include all the UHECR source within 10 Mpc once Circinus is added). The number of sources added from the mother catalog above 10 Mpc to complete the source distribution depends on the source density to reach, according to the assumed $L_{\rm IR}$  threshold.  To give an example, a $L_{\rm IR}$ cut at $2\times 10^{10}\,L_\sun$ (where $L_\sun$ is the IR luminosity of the sun) corresponds to a source density of $\sim10^{-3}\,\rm Mpc^{-3}$ and allows to keep in particular M82, NGC253, M83, NGC4945, Circinus and NGC1068 among the initial SFG sample.\\
%Finally, we note that this completion process above 10 Mpc is done in two different ways, i) by peaking  any source in the mother catalog, ii) by picking only among sources located outside of large (>$10^{14}\,M_{\sun}$) galaxy clusters ( according to their identification in \citep{Kourk2017}). The latter method is used to account in a an oversimplified way for the possible lack or deficit of SFG or SBG in rich galaxy clusters (see e.g, \citet{Guglielmo2015}).

\begin{figure}
   \centering
   \includegraphics[width=8.5cm]{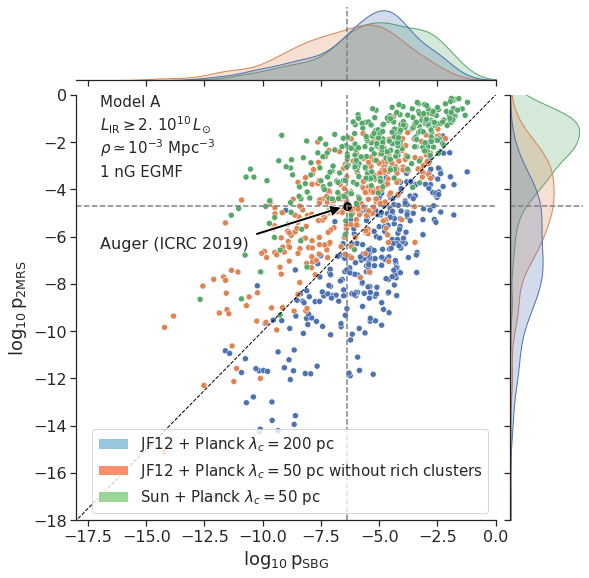}
   \includegraphics[width=8.5cm]{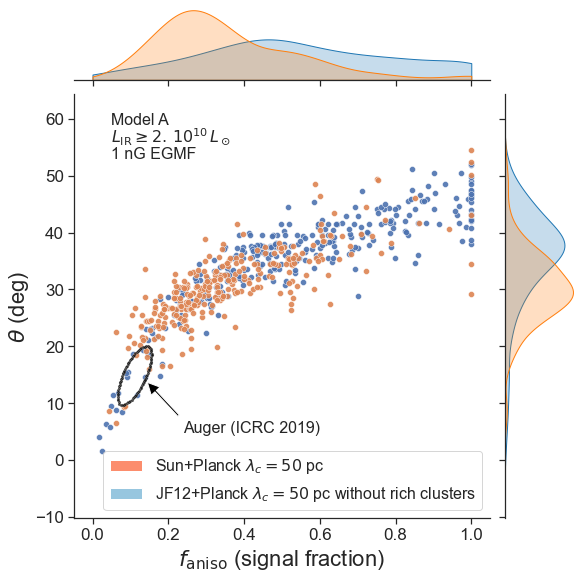}
\caption{Results of the likelihood analysis obtained for our special hypothesis on the source distribution of Sect.~\ref{SecForceSBG} (see text). The top panel is built as Fig.~\ref{LL_base_Pval} and shows the cases of the JF12+Planck GMF ($\lambda_c$=200 pc), the Sun+Planck GMF ($\lambda_c$=50 pc) and a special case assuming the exclusion of galaxies members of nearby rich galaxy clusters from the source distribution and using JF12+Planck GMF ($\lambda_c$=50 pc). On the bottom panel, similarly to Fig.~\ref{LL_base_Par}, the $f_{\rm aniso}$ and $\theta$ parameters maximizing the likelihood analysis (for the SBG catalog), for the last two cases shown on the top panel, are plotted against each other. 
              }
         \label{FigLLMC_SBG}
\end{figure}

\subsection{Results}
The p-values obtained for various models are displayed on the top panel of Fig.~\ref{FigLLMC_SBG}.
When the Sun+Planck GMF model is used, the SFG-biased source model is found to produce similar results as reported above, without any significant improvement in the agreement between the simulated datasets and the Auger data. This is true both for the search of a flux excess (either through a blind search or in the direction of Cen~A) and for the likelihood analysis. This confirms that NGC4945 and/or M83 can produce an anisotropy at intermediate angular scales that is similar to that produced by Cen~A, and vice versa.

In the case of the JF12+Planck model, a more significant rejection of isotropy is obtained when the SBGs are used as the underlying signal in the likelihood analysis, in the case when the second catalog completion method is used, that is when galaxies in large galaxy clusters are excluded. This is again reminiscent of what was reported in Sect.~\ref{sec:signifRejectIso}. However, when the first completion method is used, although the proportion of the datasets that reject isotropy better when the likelihood analysis is performed with the SBG catalog is found to be larger than in the case of non-biased datasets, this proportion remains below 25\% for all the values of the $L_{\mathrm{IR}}$ threshold that we considered.

Overall, for both GMF models and for the various cuts on the IR luminosity of the galaxies, no significant improvement is found regarding the similarity between a typical simulated datasets and the Auger data compared to what was obtained earlier with our baseline catalog. %Indeed, there is still a systematic offset between the position of the BS maximum flux excess and the position of Cen~A, and in general larger values of $f_{\rm aniso}$ and $\theta$ are found for the likelihood analysis. 
Only a few realizations are found within the 1$\sigma$ ellipse obtained from Auger data (see the bottom panel of Fig.~\ref{FigLLMC_SBG}). 

\subsection{General comment about the likelihood analysis results}

The above results allow us to underline another important caveat regarding the likelihood analysis, in addition to the strong dependence of its results on the choice of a GMF model. Even though we forced the most nearby UHECRs sources, which are indeed those which determine or influence the most the resulting anisotropies, to be exactly the same as those used to model the signal component in the likelihood analysis (i.e. the SBG model used by Auger), the parameters reconstructed from the resulting datasets do not necessarily show a strong concordance with the underlying model. This is because the likelihood analysis, as constructed in \citet{AugerSFG2018}, assumes a simple gaussian blurring of the UHECR arrival directions around the SBG sources, which is not a realistic pattern to be expected from magnetic deflexions in the GMF. Even in the absence of significant magnification or demagnification, the GMF leads to systematic shifts of the UHECR image of given source over the sky, which depend both on its actual position and on the UHECR rigidity. Given the current lack of knowledge of the actual magnetic deflections, it is important to keep in mind that the values of the ``signal fraction'', $f_{\rm aniso}$, and angular scale, $\theta$, providing the largest likelihood signal in such analyses should not be mistaken as an estimate of the fractional contribution of the dominant sources and of the typical magnetic deflection. Their meaning cannot be extrapolated beyond their technical role in the analysis, namely they are the set of parameters which allow to reject the isotropy hypothesis with the largest significance. As already discussed, caution is also required when it comes to the comparison of the p-values obtained for the various catalogs used to model the signal component.

\begin{figure*}[!ht]
   \centering
   \includegraphics[width=8.5cm]{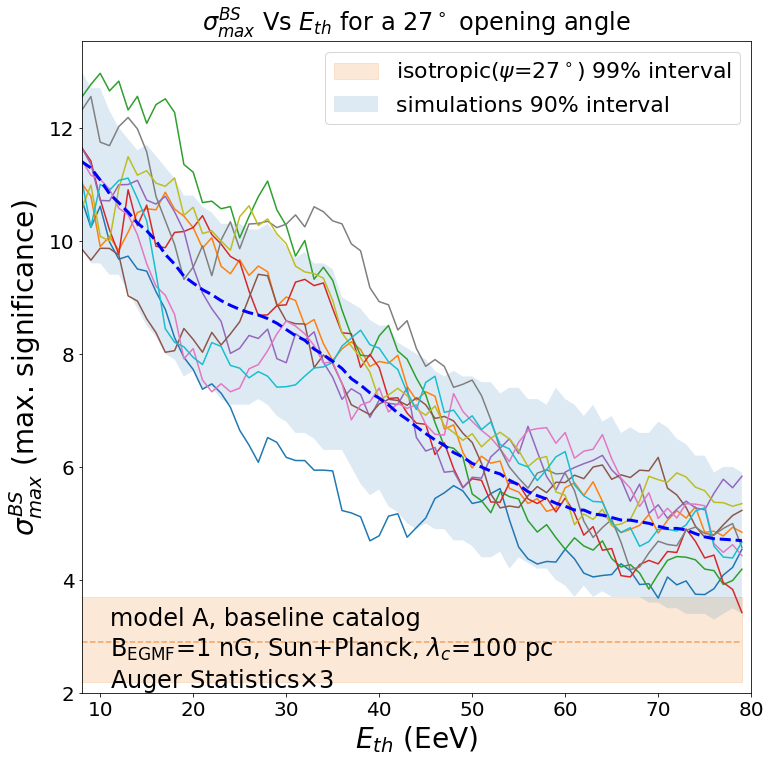}
   \includegraphics[width=8.5cm]{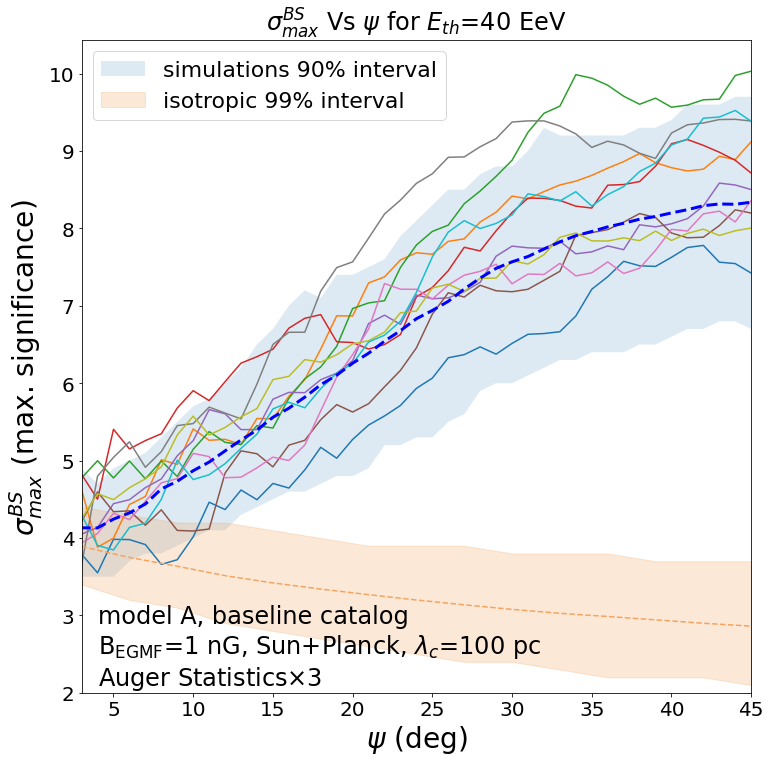}
   \includegraphics[width=8.5cm]{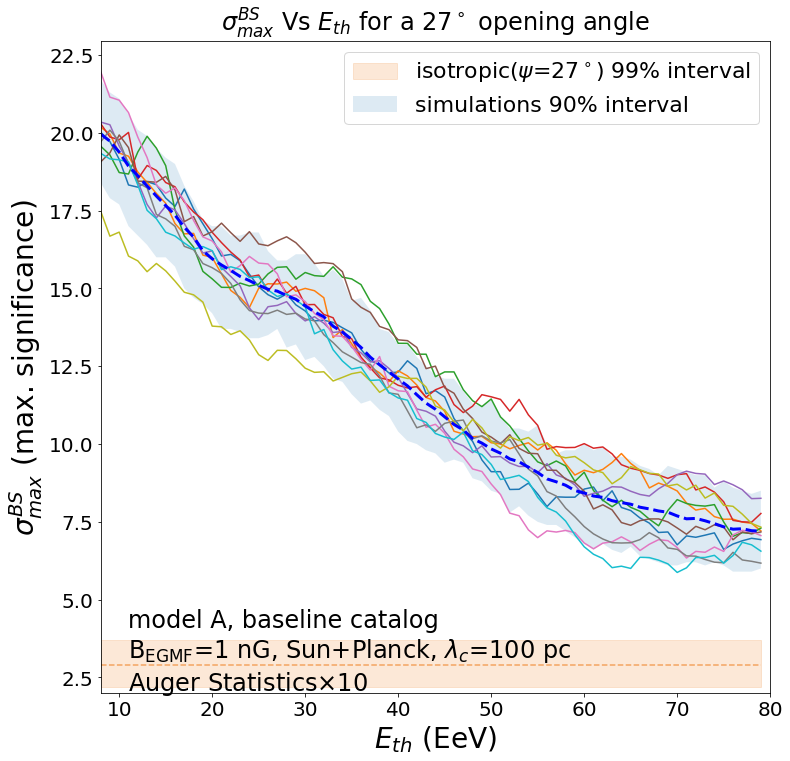}
   \includegraphics[width=8.5cm]{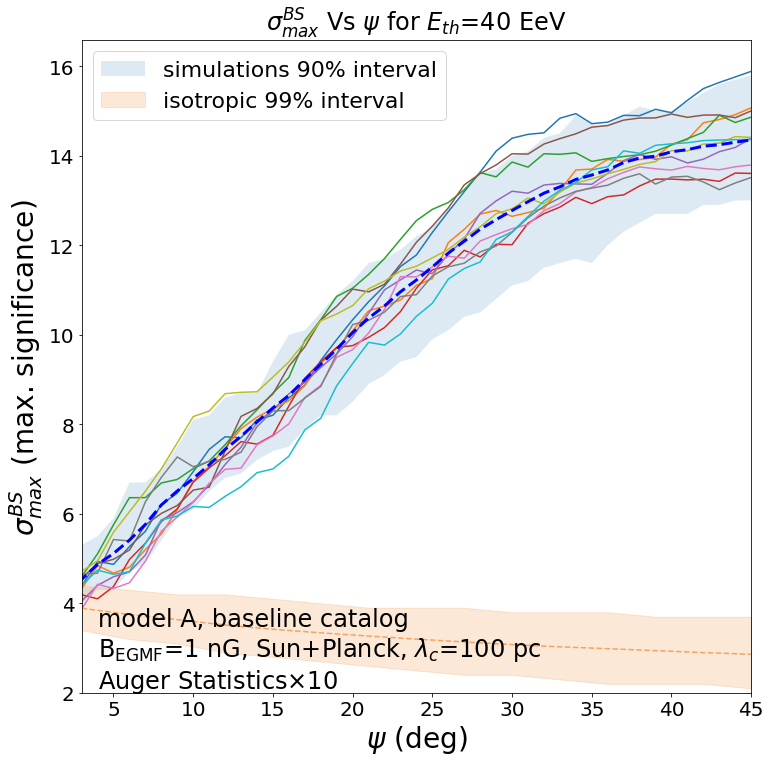}
    
      \caption{Same as the bottom panel of Fig.~\ref{FigSigVs1a}, the top and bottom panels consider respectively statistics 3 and 10 times larger than that of Auger at ICRC 2019.
              }
         \label{FigSigVsStat}
\end{figure*}

%\begin{figure}
%   \centering
%   \includegraphics[width=7.5cm]{Figs/Stat1.png}
%   \includegraphics[width=7.5cm]{Figs/Stat2.png}
%   \includegraphics[width=7.5cm]{Figs/Stat3.png}
% \caption{Evolution of the characterization of the BS maximum with the datasets statistics. The top panel shows the evolution of the value of the significance obtained at the BS maximum for 3 different values of $E_{\rm th}$=10, 40 and 80 EeV, using for each case an angular window of $27^\circ$. The astrophysical model considered is based on model A, the baseline catalog a 1 nG EGMF and the Sun+Planck GMF model with $\lambda_c$=100~pc (for which the Auger BS maximum significance above 32EeV lies close to the median value obtained for 300 datasets at the Auger statistics). The central panel shows the corresponding evolutions of the ratio $R=n_{\rm obs}/n_{\rm exp}$. The bottom panel shows the corresponding evolution with statistics of the 68\% upper limits of the separation angle between the location of the BS maximum at a given statistics and the one that would be obtained with an arbitrarily large statistics. 
%              }
%         \label{FigStat}
%\end{figure}

\begin{figure*}
   \centering
   \includegraphics[width=8.5cm]{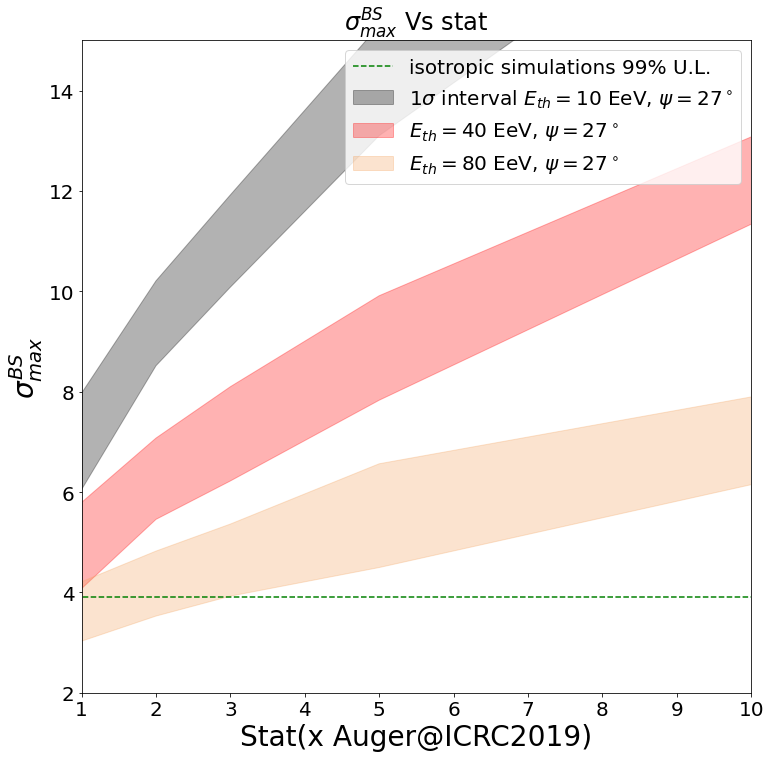}
   \includegraphics[width=8.5cm]{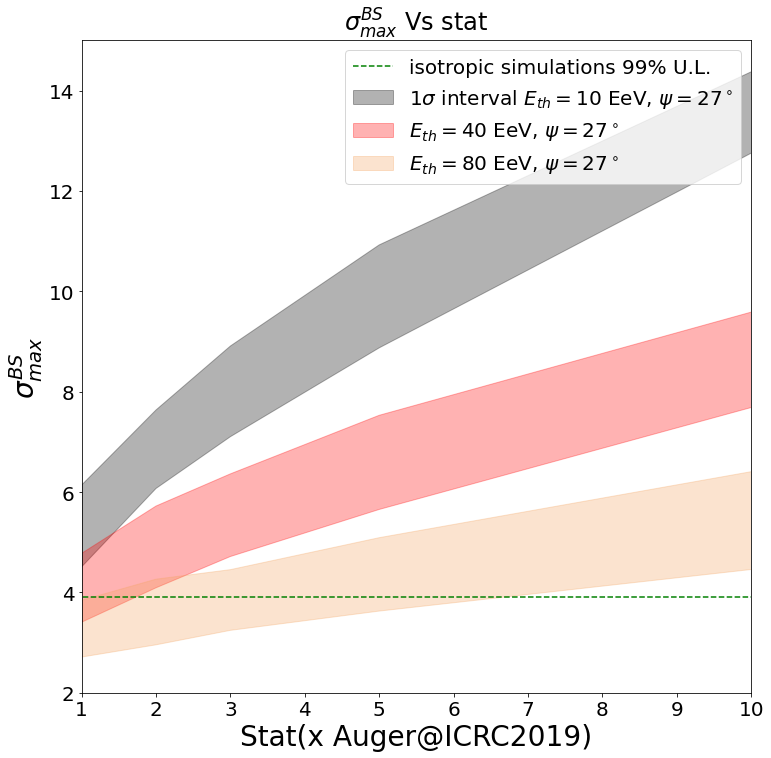}
\caption{Evolution of the significance of the BS maximum flux excess at an angular scale of $27^\circ$ for 3 different values of the energy threshold, $E_{\rm th}$=10, 40 and 80 EeV, as a function of the size of the datasets, in units of the ICRC 2019 Auger statistics. The underlying astrophysical scenario is described in the text. The Sun+Planck GMF model is assumed, either with $\lambda_c$=100~pc (on the left), or $\lambda_c$=200~pc (on the right). The shaded areas show the 1$\sigma$ interval around the average values of $\sigma_{\rm max}^{\rm BS}$, obtained from 300 independent datasets. The green dashed line represents the 99\% upper limit of the BS maximum significance obtained for isotropic datasets for the same  $\psi=27^\circ$ angular scale.}
         \label{FigStatSigmaMax}
\end{figure*}

\begin{figure*}
   \centering
   \includegraphics[width=8.5cm]{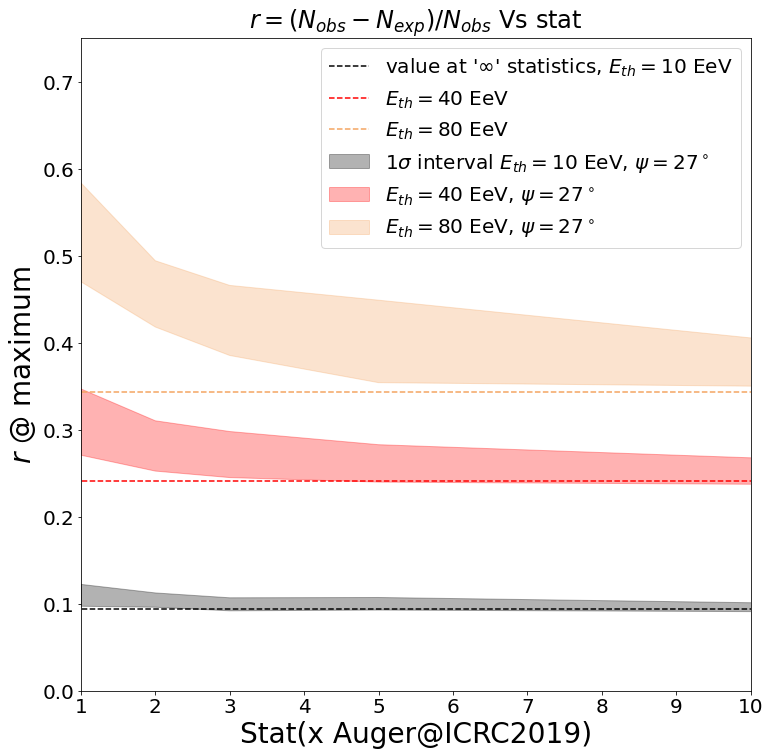}
   \includegraphics[width=8.5cm]{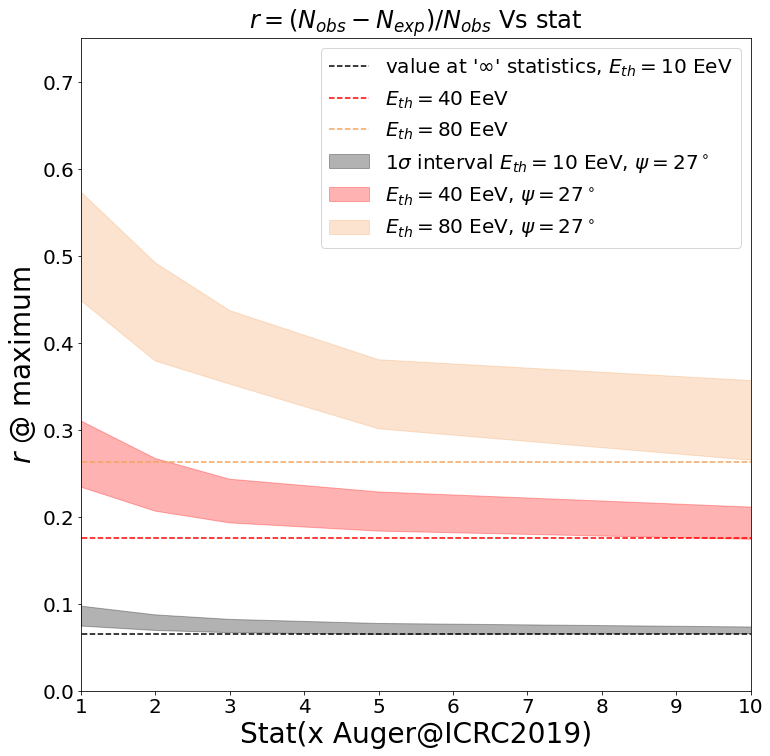}
\caption{Same as Fig.~\ref{FigStatSigmaMax}, for the evolution of the flux excess ratio $r=(N_{\rm obs}-N_{\rm exp})/N_{\rm obs}$. The dashed lines show the asymptotic values corresponding to datasets with infinite statistics }
         \label{FigStatFluxExcess}
\end{figure*}

\begin{figure*}
   \centering
   \includegraphics[width=8.5cm]{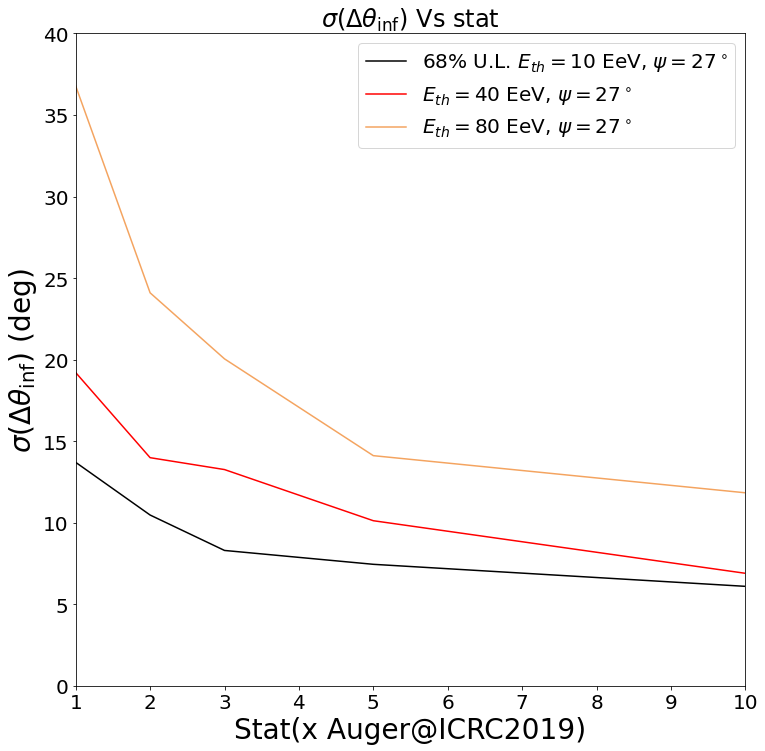}
   \includegraphics[width=8.5cm]{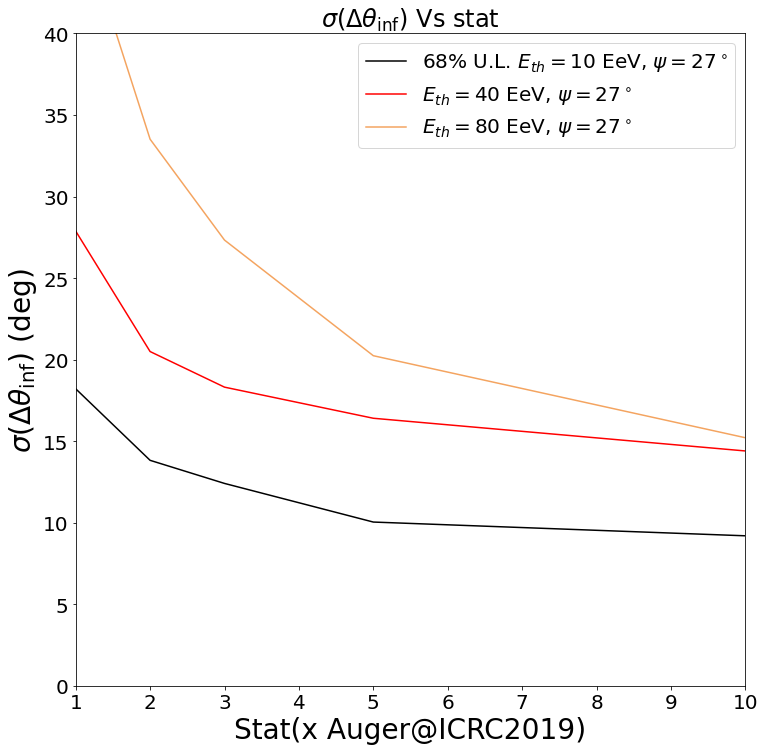}
\caption{Evolution with statistics of $\sigma(\Delta\theta_{\rm inf})$, the 68\% upper limits of $\Delta\theta_{\rm inf}$ (separation angle between the location of the BS maximum at a given statistics and the one that would be obtained with  arbitrarily large statistics). The two same models as on Fig.~\ref{FigStatSigmaMax} are considered.}
         \label{FigStatAngularDistance}
\end{figure*}

%\begin{figure}
%   \centering
%   \includegraphics[width=7.5cm]{Figs/Stat12.png}
%   \includegraphics[width=7.5cm]{Figs/Stat22.png}
%   \includegraphics[width=7.5cm]{Figs/Stat32.png}
% \caption{Same as Fig,~\ref{FigStat}, but we are now considering the Sun+Planck GMF model with $\lambda_c$=200 pc. 
%              }
%         \label{FigStat2}
%\end{figure}

%\section{Expectations for higher statistics datasets}
\section{Anisotropy expectations with larger datasets}
\label{Sec:stat}
%As discussed in previous sections, some important aspects of the anisotropies of the UHECR sky, which go beyond the simple estimate of the significance of the rejection of the isotropy hypothesis, such as the real celestial position of the BS maximum, the level of the anisotropy (see below) or its evolution with the energy and angular scale, remain quite poorly characterized at the current level of statistics, even though the Pierre Auger observatory has cumulated an unprecedented amount of data at the highest energy. 

As discussed in the previous sections, even when considering a wide range of generic astrophysical models and varying their main parameters, one does not seem to be able to find a satisfactory account of the entire set of information currently available about the UHECR anisotropies, which goes beyond the simple estimate of the significance of the rejection of an isotropy hypothesis, and includes the direction of the BS maximum, the amplitude of the anisotropy (see below) and its evolution with energy and angular scale. Although the Pierre Auger observatory has accumulated an unprecedented amount of good quality data at the highest energy, it is currently difficult to assess whether there is one or more key ingredients missing in the models, or whether the data are still subject to important statistical fluctuations that somehow blur the message or hide some important information that would only be accessible with still larger statistics. To explore this possibility and evaluate what could be gained from datasets containing several times more events, we examine in this section how the main anisotropy observables are expected to evolve as a function the accumulated statistics, at least in the framework of our generic models, and more specifically in the case of our baseline scenario, discussed through Sects.~\ref{SecBS} and \ref{Seclike}. To this end, we performed the BS analysis on datasets with statistics ranging from 1 to 10 times the Auger statistics reported at the ICRC~2019, which is our reference value throughout this paper (namely 42500 events above 8~EeV, corresponding to an exposure of 101,400 $\rm km^2\,sr\,yr$). 

%In this section we examine the expected evolution of some quantities of interest for the characterization the UHECR sky anisotropy as a function the accumulated statistics. To this purpose, we come back to our baseline scenario, discussed through Sect.~\ref{SecBS} to \ref{Seclike}, and perform the BS analysis for various datasets statistics ranging from 1 to 10 times that of Auger reported at the ICRC 2019 which we have considered so far. 

%The $E_{\rm th}$ (assuming a $\psi=27^\circ$ angular window) and $\psi$ (assuming $E_{\rm th}$=40 EeV) evolutions of the BS maximum significance are shown in Fig.~\ref{FigSigVsStat} for dataset statistics 3 (top panels) and 10 times (bottom panels) larger than that of Auger. The BS analysis is performed on the wider scan of the space parameter we used in Sect.~\ref{SecEvolE} and therefore Fig.~\ref{FigSigVsStat} is directly comparable with the bottom panel of Fig.~\ref{FigSigVs1a}. Besides the obvious fact that the BS maximum significance is expected to increase with the dataset statistics, one sees that the energy evolution of the individual datasets become less noisy as the statistics increases and more representative of the mean energy evolution obtained by overaging over all the datasets at a given statistics.  Unsurprisingly, this suggest that a measurement of the $E_{\rm th}$ and $\psi$ evolution of the blind search maximum become more meaningful and constraining with large statistic datasets.\\

\subsection{Energy dependence of the BS maximum significance}

In Fig.~\ref{FigSigVs1a}, we had shown the evolution of the significance of the BS maximum flux excess, $\sigma_{\rm max}^{\rm BS}$, as a function of the energy threshold, $E_{\rm th}$, for a given angular scale (namely $\psi=27^\circ$), and as a function of the angular scale for a given energy threshold (namely $E_{\rm th}$=40 EeV), where these fixed values of $\psi$ and $E_{\rm th}$ are close to those corresponding to the BS maximum in the Auger dataset. These plots show a clear evolution of the values to be expected on average, but also that individual datasets with the statistics of Auger usually do not exhibit such an obvious trend, and may have a maximum at some particular energy or angular scale which is essentially fortuitous, and could thus be misleading if taken at face value to derive definite constraints about the underlying UHECR model. In Fig.~\ref{FigSigVsStat}, we show the same evolutions of $\sigma_{\rm max}^{\rm BS}$, but for datasets 3 times (top panels) and 10 times (bottom panels) larger than the Auger dataset. The underlying model and the meaning of the individual curves and shaded areas are the same as in the bottom plots of Fig.~\ref{FigSigVs1a}.

%The BS analysis is performed on the wider scan of the space parameter we used in Sect.~\ref{SecEvolE} and therefore Fig.~\ref{FigSigVsStat} is directly comparable with the bottom panel of Fig.~\ref{FigSigVs1a}. Besides the obvious fact that the BS maximum significance is expected to increase with the dataset statistics, one sees that the energy evolution of the individual datasets become less noisy as the statistics increases and more representative of the mean energy evolution obtained by overaging over all the datasets at a given statistics.  Unsurprisingly, this suggest that a measurement of the $E_{\rm th}$ and $\psi$ evolution of the blind search maximum become more meaningful and constraining with large statistic datasets.

As can be seen, in addition to the obvious increase of the significances due to the larger statistics, the energy and angular scale evolutions of $\sigma_{\rm max}^{\rm BS}$ for the individual datasets appear less ``noisy'' when the size of the datasets increases, and are more and more representative of the average trends, from which interesting information could in principle be drawn. This is of course all the more true with ten times the Auger statistics.%With ten times the Auger statistics, the statistical fluctuations appear low enough and very few datasets depart significantly.{\color{red}C'est mal dit je pense, il y en a toujours autant qui depart significantly mais c'est la departure relative a attendre qui est plus petite, non ?}

\subsection{Evolution of the BS maximum significance with statistics}

Conversely, one can draw the evolution of $\sigma_{\rm max}^{\rm BS}$ with the size of the dataset, for given values of the energy threshold and angular scale. This is done in Fig.~\ref{FigStatSigmaMax}, for an angular scale of $27^\circ$ and  3 different values of the energy threshold, namely $E_{\rm th} = 10$~ EeV, 40~EeV and 80~EeV. The shaded areas show the intervals in which 68\% of the 300 datasets are found in each case, and the size of the datasets, in abcissa, is expressed in units of the ICRC 2019 Auger statistics. The underlying astrophysical scenario assumes the energy spectrum and composition of model A, our baseline UHECR source catalog, a 1~nG EGMF and the Sun+Planck GMF model with two different choices of the coherence length of the turbulent component. On the left plot, $\lambda_c$=100~pc, for which the BS maximum and Cen~A flux excesses of the simulated datasets have typical significances comparable to those reported by Auger ({ $5.6\sigma$ and $5.1\sigma$ respectively}), as seen in Fig.~\ref{FigBSVL1a} (i.e., the Auger BS maximum significance above 32~EeV lies close to the median value obtained for 300 Auger-like datasets). On the right plot, $\lambda_c$=200~pc, for which the Auger significances correspond to the higher end (positive fluctuations) of the simulated distribution.

 Of course, the values of all BS flux excess significances increase with the size of the dataset, but it is also interesting to note how the gap between the shaded areas increases, that is the differences between the BS maximum significances for different values of $E_{\rm th}$ are larger and larger. Thus, at least for the models under consideration, it becomes less and less likely to obtain a dataset for which the largest BS flux excess significance is found at high energy (say around 40~EeV, as in the Auger dataset) rather than at lower energy ($\sim10$~EeV for model~A or $\sim20$~EeV in the case of model~B, as shown in Fig.~\ref{FigSigComp}). This is particularly true for the lower value of the coherence length (left plot) for which the expected level of anisotropy is indeed higher.

\subsection{Flux excess ratio at the BS maximum}

In Fig.~\ref{FigStatFluxExcess}, we show the evolution of the flux excess ratio, $r=(N_{\rm obs}-N_{\rm exp})/N_{\rm obs}$, associated with the BS maximum, for the same three values of $E_{\rm th}$ and the angular scale again fixed at 27$^\circ$. As before, the shaded areas show the interval in which 68\% of the realizations are found. The dashed lines show the asymptotic value of the flux excess, $r$, for the model under consideration, as would be obtained with datasets of ``infinite'' statistics. Comparing the two panels, one can see that the values of the maximum flux excess are on average larger for the smaller coherence length of the turbulent magnetic field, which follows from the corresponding higher level of anisotropy in that case.

Unsurprisingly, the average maximum flux excesses observed in finite datasets are most of the time larger than the ``true" (asymptotic) value. This is because the scanning procedure tends to always pick an upward fluctuation of the UHECR flux, in a region of the sky (probably) already characterized by an intrinsically larger flux than average. This overestimation effect is larger for smaller datasets and thus also larger for higher energy thresholds. For $E_{\rm th}=10$~EeV, the reconstructed value with the Auger statistics should be already within a few percent of what would be found with an arbitrarily large dataset. At $E_{\rm th} = 40$~EeV, and even more so at $E_{\rm th} = 80$~EeV, significantly larger statistics are required to better estimate the $R$ ratio and characterize the actual anisotropy level. At $E_{\rm th} = 80$~EeV, an order of magnitude larger dataset might be necessary to combine a large enough value of the anisotropy significance and a reliable estimate of the flux excess in the BS maximum region.

Another clear effect is the increase of the expected flux excess ratio as a function of $E_{\rm th}$, even though the UHECR rigidity does not strongly evolve in this energy range, because of the composition change. This is in line with the larger expected anisotropies, as already discussed, and a natural consequence of the lower horizon distance of the UHECRs at higher energy, which reduces the number of sources. However, of course, a larger intrinsic anisotropy does not imply a larger significance, since the latter also depends on the size of the dataset. This is why, in the case of our model A, the values of $\sigma_{\rm max}^{\rm BS}$ and $r$ are clearly seen to evolve in opposite ways as a function of energy over the entire energy range above $\sim 10$~EeV (or above $\sim 20$~EeV for model~B, as seen in Sect.~\ref{sec:compo}).

%It is worth highlighting that a more accurate characterization of the anisotropies can be much more valuable than a larger significance of the rejection of isotropy, which by itself may provide only very little clue about the origin of the UHECRs. Therefore, although the desirable gain in statistics requires important experimental efforts, with a new generation of either ground-based \citep{GRAND2020, GCOS2021} or space-based  \citep{Bertaina2019, POEMMA2021} UHECR observatories, increasing the statistics at the very end of the UHECR spectrum is particularly valuable, since the corresponding anisotropies are intrinsically larger at the highest energies, with a larger relative contribution of the brightest sources and a lower background from the more distant universe. The identification of the nature of the UHECR sources should thus greatly benefit from a significant gain in exposure specifically in the highest energy part of the spectrum.

{ Although the desirable gain in statistics requires important experimental efforts, with a new generation of either ground-based \citep{GRAND2020, GCOS2021} or space-based  \citep{Bertaina2019, POEMMA2021} UHECR observatories, it is worth highlighting that increasing the statistics at the very end of the UHECR spectrum is particularly valuable, since the corresponding anisotropies are intrinsically larger at the highest energies, with a larger relative contribution of the brightest sources and a lower background from the more distant universe. The identification of the nature of the UHECR sources should thus greatly benefit from a significant gain in exposure, particularly in the highest energy part of the spectrum.}

\subsection{Direction of the BS maximum}

Finally, Fig.~\ref{FigStatAngularDistance} shows how the precision of the reconstruction of the BS maximum direction evolves with the size of the dataset. This is estimated from the difference between the reconstructed direction of each individual dataset at a given statistics and the direction obtained with an extremely large dataset (gathering 300 Auger-like datasets). On the plots, we show the angular distance, $\Delta\theta_{\rm inf}$, within which 68\% of all datasets are found at the statistics considered, for a GMF coherence length of 100~pc (left) or 200~pc (right).

As expected, $\Delta\theta_{\rm inf}$ evolves with the energy threshold and the statistics in the same way as the maximum significance $\sigma_{\rm max}^{\rm BS}$. As a result, in the case of model A, for a given exposure, the BS maximum location is found on average closer to its true position when $E_{\rm th} = 10$~EeV, than for larger threshold. The same is true for model B, except that the cases for $E_{\rm th} = 10$~EeV and $E_{\rm th} = 40$~EeV appear closer to one another. For $E_{\rm th} = 80$~EeV, an approximately 10 $\times$ larger exposure is needed to reach the same precision on the position of the BS maximum as at 10~EeV. Regarding the influence of the GMF coherence length, one sees that the precision is higher for the smaller value of $\lambda_{\rm c}$, due to the correspondingly higher intrinsic anisotropy.

From the quantitative point of view, the statistics of a dataset may be considered sufficient when the corresponding angular precision of the BS maximum reconstruction is small compared to its angular scale (here, $\psi = 27^\circ$). At the highest energies, this requires a larger exposure. For instance, at $E_{\rm th} = 80$~EeV, exposures of the order of $10^6\rm\,km^2\,sr\,yr$  (that is $\sim 10$ times the current Auger statistics) are likely to be required to pinpoint the BS maximum location with an accuracy of 15 to 20$^\circ$.

\section{Comparison between the intermediate scale anisotropies at 8 EeV and above 32~EeV}

\begin{figure}[t!]
   \centering
   \includegraphics[width=9cm]{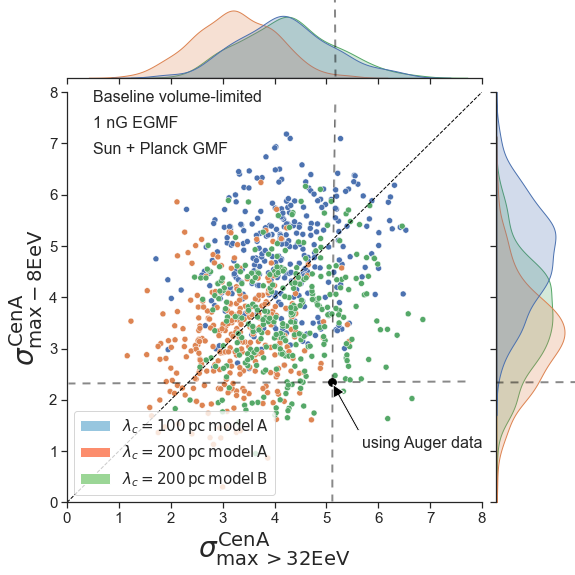}
\caption{Scatter plot of the maximum significances of a flux excess in the direction of Cen~A with a fixed energy threshold at $E_{\rm th} = 8$~EeV (in ordinate) and with a scan over $E_{\rm th} > 32$~EeV. The different colours correspond to different underlying astrophysical scenarios, as indicated on the plot. The two values of $\sigma_{\max}$ obtained with the Auger data are indicated by the black dot.}
    \label{FigCenAHighLow}
\end{figure}

As discussed above as well as in Paper~I, our simulations show that there is potentially a lot information to be gathered from the evolution of the anisotropies with energy, both at large angular scales, with the amplitude and direction of the dipole, and at smaller angular scales, with the direction, significance and excess ratio of the BS maximum flux excess, or the evolution of the flux excess around a specific direction. While we do not have access to the full Auger dataset, and thus cannot develop this study with real data, the data made available by the Pierre Auger Collaboration on various occasions allow us to make a first comparison of the situation at $E_{\rm th} = 8$~EeV (using the dataset released at the time of the publication of the Auger dipole anisotropy in \citep{AugerDip2017}) and $E_{\rm th} > 32$~EeV which we have discussed and compared to our simulations so far.

The dipole dataset includes UHECR events above 8~EeV and corresponds to a cumulated exposure of 76,800 $\rm km^2\,sr\,yr$. Regrettably, the release does not include information about the energy of the listed events, so the full BS analysis cannot be performed on this dataset. We thus scanned only over the angular scales, with $\psi$ ranging from 1$^\circ$ to 30$^\circ$ in steps of $1^\circ$, with a fixed value of $E_{\rm th}=8$~EeV. The most significant flux excess resulting from this analysis was found at the largest angular scale, namely $\psi = 30^\circ$, with a significance $\sigma^{\rm BS}_{\rm max-8EeV} \simeq 4.7$, in the direction $(l = 267^\circ, b = -44^\circ)$ in Galactic coordinates. As for the targeted search of a flux excess in the direction of Cen~A, the largest significance is found at $\psi=28^\circ$, with $\sigma^{\rm CenA}_{\rm max-8EeV}\simeq2.4$.

\subsection{Comparison of the maximum significances in the Cen~A direction}

It is interesting, and perhaps very informative, that the flux excess in the direction of Cen~A is much more significant above 32~EeV (namely $\sigma^{\rm CenA}_{\rm max>32EeV}\simeq5.1$ with the ICRC 2019 statistics) than when the threshold is set at 8~EeV. Of course, the possibility to scan over two parameters ($\psi$ and $E_{\rm th}$) instead of one leads to necessarily larger maximum significances. However, the dataset above 32~EeV is also much smaller than above 8~EeV, and the extra scanning dimension is not enough explain such a large difference, between 2.4$\sigma$ and 5.1$\sigma$. To show this, we compare in Fig.~\ref{FigCenAHighLow} the two values $\sigma^{\rm CenA}_{\rm max-8EeV}$ and $\sigma^{\rm CenA}_{\rm max>32EeV}$ obtained with our simulated datasets, for 3 different astrophysical models: the two models shown in Fig.~\ref{FigStatSigmaMax}, that is Model A with a GMF coherence length of 100~pc (blue dots) or 200~pc (red dots), and Model B with $\lambda_{\rm c} = 200$~pc. We chose the Sun+Planck GMF model is all cases, { since it leads to an  overall better agreement with the Auger anisotropy measurements above 32~EeV when considering the baseline catalog (see above)}, but similar conclusions are obtained with the JF12+Planck model. The Auger data are represented by a black circle, which appears very far from the typical location of the simulated datasets in this 2D plot. Indeed, even though the significance of the maximum flux excess at $E_{\rm th} = 8$~EeV is often lower than the signifance above 32~EeV (datapoints below the diagonal in Fig.~\ref{FigCenAHighLow}), the difference is almost never as large as in the case of the Auger data. (NB: in the simulated datasets, we reproduced the same exposure difference as in the available Auger data.)

%As can be seen, with the chosen astrophysical parameters, the Auger dataset would correspond to both an upward fluctuation of the flux excess above 32~EeV, and an downward fluctuation at 8~EeV. For Model~A and $\lambda_{\rm c} = 200$~pc (red dots), the Auger data point appears rather representative of the simulated datasets regarding $\sigma^{\rm CenA}_{\rm max-8EeV}$ (moderate downward fluctuation), but exceptional regarding $\sigma^{\rm CenA}_{\rm max>32EeV}$, as can be seen from the distributions respectively on the right and upper sides of the plot. Conversely, in the case of Model~A and $\lambda_{\rm c} = 100$~pc (blue dots), the Auger values are a moderate upward fluctuation of $\sigma^{\rm CenA}_{\rm max>32EeV}$, but an exceptionally large downward fluctuation of $\sigma^{\rm CenA}_{\rm max-8EeV}$. A somewhat intermediate compromise would be obtained with values of $\lambda_{\rm c}$ around 150 pc, but the Auger data would still appear very different from essentially all the model realizations.

The case of Model B with $\lambda_{\rm c} = 200$~pc appears slightly more favorable, with the Auger data around Cen~A corresponding to a moderate downward fluctuation of the low energy flux and moderate upward fluctuation at high energy. This can be understood as a consequence of the different energy evolution of average maximum significance between 8~EeV and 32~EeV for model~A and model~B, as already observed in Fig.~\ref{FigSigComp}. However, these data would still appear quite unsual for such models, when both energy ranges are considered together. This shows once more the interest of examining the UHECR data globally, with the entire set of observables including their evolution with energy, as already emphasized in Paper~I.

Incidentally, we note that, in the case of Model~A, the larger value of $\lambda_{\rm c}$ (red dots) is also the one which better reproduces the value of the dipole amplitude and its energy evolution (see Paper~I). In the framework of the simulated scenarios, this would thus tend to favor an interpretation of the apparent mismatch between $\sigma^{\rm CenA}_{\rm max-8EeV}$ and $\sigma^{\rm CenA}_{\rm max>32EeV}$ according to which the later benefits from an upward fluctuation, rather than the opposite.

\begin{figure}
   \centering
   \includegraphics[width=9cm]{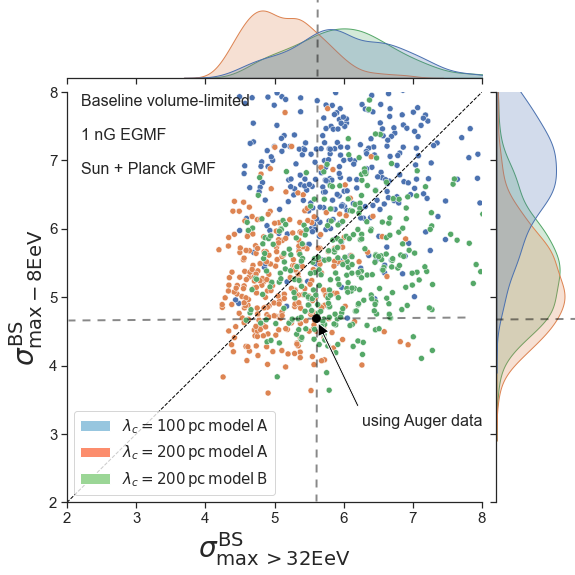}
\caption{Same as Fig.~\ref{FigCenAHighLow}, but for a blind search (BS) of the maximum flux excess, instead of targeted search in the direction of Cen~A (see text).}
    \label{FigBSHighLow}
\end{figure}

\subsection{Comparison of the maximum significances of the BS maximum flux excesses}

%Although the large difference between $\sigma^{\rm CenA}_{\rm max-8EeV}$ and $\sigma^{\rm CenA}_{\rm max>32EeV}$ in the Auger data appears very uncommon in our simulated datasets. The specific direction of CenA  has a special status in the   Auger data, since it is very close to the BS maximum flux excess above 32~EeV, while it is more distant from the corresponding maximum in most of our simulated datasets (see, e.g., in Fig.~\ref{FigBSVL1b}). Therefore it is interesting to extend the study to the comparison between the significances of the BS (blind, i.e. non targeted search) maximum flux excess at $E_{\rm th} = 8$~EeV and above 32~EeV.

We extended the study to the comparision between the significances of the BS (blind, i.e. non targeted search) maximum flux excess at $E_{\rm th} = 8$~EeV and above 32~EeV.
{ The results, shown in Fig.~\ref{FigBSHighLow}, confirm that the Auger dataset is much less atypical from this point of view, compared to the simulated datasets. %Although still somewhat on the side of a downward fluctuation of the BS maximum significance at $E_{\rm th} = 8$~EeV and upward fluctuation above 32~EeV in the case of Model A with $\lambda_{\rm c} = 200$~pc, many realization of the scenario lead to values of $\sigma^{\rm BS}_{\rm max-8EeV}$ and $\sigma^{\rm BS}_{\rm max>32EeV}$ similar to those obtained with the Auger data. 
In the case of Model A with $\lambda_{\rm c} = 200$~pc, many realizations of the scenario lead to values of $\sigma^{\rm BS}_{\rm max-8EeV}$ and $\sigma^{\rm BS}_{\rm max>32EeV}$ similar to those obtained with the Auger data. It is also interesting to note that lower values of the GMF coherence length are again disfavored, at least for the models under study. Indeed, for these values (blue dots in Fig.~\ref{FigBSHighLow}), although the significance of the maximum flux excess above 32~EeV appears on average very similar to the Auger value, the significance at 8~EeV is almost always much larger.}
From a general point of view, a lower value of $\lambda_{\rm c}$ allows the gain in statistics at low energy to be translated into a larger gain in significance, despite the (moderately) lower magnetic rigidity and larger number of sources (more distant horizon).

\begin{figure}
   \centering
   \includegraphics[width=9cm]{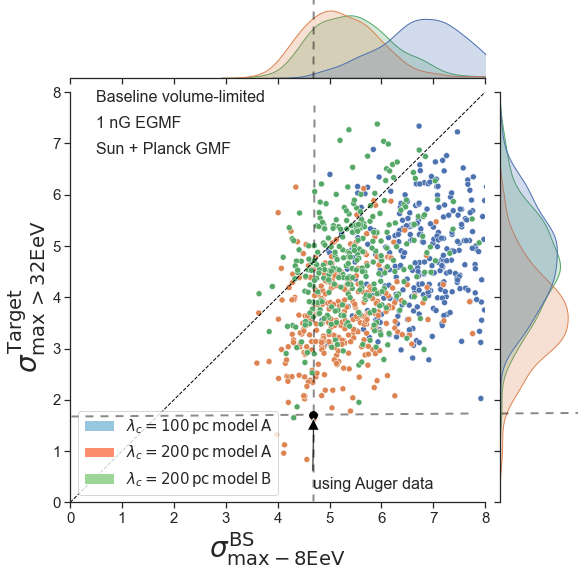}
\caption{Scatter plot of the maximum significances for $E_{\rm th} > 32$~EeV of the flux excess in the direction of the BS maximum at 8~EeV, $\sigma^{\rm target}_{\rm max>32EeV}$ vs. the corresponding value of $\sigma^{\rm BS}_{\rm max-8EeV}$ (see text). The different colours correspond to different underlying astrophysical scenarios, as indicated on the plot. The two values of $\sigma_{\max}$ obtained with the Auger data are indicated by the black dot.}
         \label{FigTargetLowE}
\end{figure}

\subsection{Energy evolution of the flux excess significance at the position of the BS maximum at 8~EeV}
\label{sec:energyEvol}
%\subsection{Collapse of the significance of the localised low-energy flux excess}

In the last two sections, we saw that the magnitude of the Auger flux excesses at 8~EeV and above 32~EeV is well reproduced by our simulations if one does not specify a given direction in sky, but is strongly different from the expectations of our generic models when the search is performed in the specific direction of Cen~A. Since the region very close to Cen~A does not play such a prominent role in our simulations as it appears to be the case in the Auger data above 32~EeV, it is interesting to address a more general question, focusing not on a predefined direction in the sky, but on a direction that is identified from the datasets themselves, independently of any preconception, namely the direction of the BS maximum at low energy.
%{ Je n'aime pas trop cette petite intro pour plusieurs raisons}

We applied this idea to the same simulated datasets as above, with the following procedure. First, we performed the BS maximum study with $E_{\rm th} = 8$~EeV for each individual dataset, determining both its significance and direction. Then, we performed a targeted search for a flux excess above 32~EeV in the very direction where the maximum significance was found, scanning over $E_{\rm th}$ and the angular scale, $\psi$, but not over the positions in the sky. We recorded the largest significance, $\sigma^{\rm target}_{\rm max>32EeV}$. The results are shown in Fig.~\ref{FigTargetLowE}, for the same three models as above, through a scatter plot of the three times 300 realizations of $\sigma^{\rm target}_{\rm max>32EeV}$ vs. $\sigma^{\rm BS}_{\rm max-8EeV}$. The position of the Auger data in this plot is indicated by a black dot. To obtain it, we used the dataset of \cite{AugerAniso2022} up to a total exposure of the ICRC 2019 ($\sim$101,400 $\rm km^2\,sr\,yr$) to match our simulated dataset statistics, and found a maximum significance of the targeted search of $\sigma_{\rm max\,>32\,EeV}^{\rm target} \simeq $1.7.)

The result is particularly striking, as it shows how exceptional the situation of the Auger data is. Unsurprisingly, the significance of the maximum flux excess above 32~EeV is generally smaller than what it was at $E_{\rm th} = 8$~EeV, since i) the dataset is smaller, and ii) the targeted position was optimized by the BS at 8~EeV, and not allowed to vary anymore. It thus usually corresponds to an upward fluctuation of an underlying high flux region, which has no reason to be also present with a similar fluctuation in the high-energy data (which amount to a negligible fraction of the dataset above 8~EeV, given the rapid decrease of the spectrum). However, in the case of the Auger data, the significance of the flux excess above 32~EeV appears to have largely collapsed at the position where it was maximum at $E_{\rm th} = 8$~EeV.  Indeed, between the two Auger datasets, the significance dropped from 4.7$\sigma$ to 1.7$\sigma$, much more than for the vast majority of the simulated datasets.

From Fig.~\ref{FigTargetLowE}, one may conclude that, among the generic models that we studied, the Auger data would be best understood as very negative fluctuations of the flux expected above 32~EeV in the direction of the maximum flux excess at 8~EeV, with a remote preference for model~A and $\lambda_{\rm c}$=200~pc, which is the least incompatible with the observations. However, the strong apparent disagreement between the data and our simulations on this particular exercise allows to speculate that the actual UHECR phenomenology in this energy range may be fundamentally different from the simulated one. Indeed, the expectation that a genuine and significant flux excess in some direction at low energy remains measurably significant at high energy is a natural and quite generic expectation in the framework of low proton-$E_{\rm max}$ scenarios, with the additional assumption that all sources emit UHECRs with roughly similar compositions and energy spectra. In such models, the average UHECR rigidity evolves slowly with energy, so the regions of the sky corresponding to larger UHECR fluxes remain the same in various energy ranges. There is no clear mechanism, in such a framework, that would allow a significantly larger flux in a given region to rapidly disappear in a neighbouring energy range.

Therefore, we are left with the following possibilities: i) the signal at low energy is a strong positive fluctuation of the UHECR flux, in the direction of the maximum that we obtained, namely (l=267, b=-44); ii) the low maximum significance of the corresponding excess above 32~EeV corresponds to a strong negative  fluctuation of the UHECR flux in this direction; iii) a contribution to the UHECR flux that is responsible for the position of the BS maximum excess location at 8~EeV simply vanishes at high energy. %This, obviously, would represent a very important astrophysical information.

The first possibility is interesting in the perspective of our study (both this paper and Paper~I), as it would tend to reconcile the generic type of models that we are analysing with the Auger data, not only from the point of view of the BS flux excess maximum at 8~EeV, but also from the point of view of the dipole direction, which was discussed in Paper~I. Indeed, while the direction in which the Auger data have the most significant flux excess at a threshold energy of 8~EeV is in the vicinity of the Fornax cluster, which to leads to a somewhat larger flux than average in our simulations, this excess is never dominant in essentially all of our simulated datasets, whatever the GMF model and the other astrophysical parameters. Only a strong upward fluctuation could change this situation. But it is also worth noting that the presence of this particular flux excess in the Auger data at above 8~EeV has necessarily a direct influence on the direction where a dipolar modulation is found in this same energy range, namely in the two energy bins from 8~EeV to 16~EeV and from 16~EeV to 32~EeV. The same upward fluctuation that would produce a BS maximum flux excess where Auger finds it would also drag the direction of the dipolar modulation southwards in Galactic coordinate, thereby reducing the disagreement with our simulations on this aspect (see Paper I). However, the fact that a fluctuation of the required amplitude never occurred in our simulations (with 300 datasets for each set of parameters) suggests that this possibility, although interesting for the mentioned reasons, remains quite unlikely.

Regarding the second possibility, namely a downward fluctuation above 32 EeV, only additional data may provide useful insight. If the low flux excess significance is confirmed, then the third possibility will become likely, and all the more interesting, as it would clearly would represent a very important astrophysical information. Such a rapid decrease of the significance of a flux excess in a given direction would indeed be direct evidence for a strong departure from the generic expectations for UHECR models based on standard candle sources, or even suggest the existence of a UHECR component vanishing between $E_{\rm th}=$8~EeV and $E_{\rm th}=$32~EeV. It would certainly be helpful for this discussion to have access to the full Auger dataset, as it would increase the statistics by more than 50\% with respect to the dipole dataset, and allow to discuss the energy evolution between 8 and 32~EeV, thereby bringing additional constraints. This is further discussed in the next and final section.

\section{Summary and discussion}

In this paper, we extended the general study of the UHECRs anisotropies initiated in Paper~I by confronting the available observational data with a wide range of simulated datasets based on the generic assumption that the distribution of the UHECRs sources in the universe follows on average the distribution of galaxies, possibly with some bias in favor or against specific classes of sources. In this second part, we focused on small and intermediate angular scales, exploring anisotropies through different types of analyses already used by the Auger and TA collaborations, namely the search of a flux excess either in any direction (``blind search'', or BS) or in a predefined direction (e.g., Cen~A or, as a new type of study, in the direction of the maximum flux excess in a different energy range), as well as the search for similarities between the distribution of UHECRs and some catalogs of sources, through the so-called likelihood analysis.

In addition to the above-mentioned assumptions regarding the UHECR source distribution, we explored different scenarios varying several astrophysical parameters, namely the composition and the energy spectrum of the UHECRs at their source, the source density, the typical strength of the extragalactic magnetic field, the structure of the Galactic magnetic fields and the coherence length of its turbulent component. With simulated hundreds of datasets for each scenario, which allowed us to investigate the uncertainties associated with both the statistical fluctuations and the cosmic variance.

\subsection{BS and targeted flux excess}

{ Concerning the search of a flux excess in the UHECR skymap, we found that the values derived from the Auger data for both the maximum significance of a blind search and the maximum significance of a flux excess in the direction of Cen~A were individually relatively easy to reproduce with our baseline simulations (see Sect.~\ref{sec:BS}). Concerning the direction of the BS maximum, the dispersion found among our simulated datasets, which reproduce the anisotropy level observed by Auger, suggests that its ``true'' position cannot be currently estimated with a resolution better than $\sim15-20^\circ$. This implies that the very close proximity between the Auger BS maximum and the position of the radiogalaxy CenA is not astrophysically meaningful at this point and that this observation needs to be strengthened with significantly larger exposure measurements.
Moreover, the BS maximum direction observed by Auger lies in the periphery of the distributions obtained with our simulated datasets and thus appears marginally compatible with the expectation from our astrophysical models. In the case of the JF12+Planck GMF model and the baseline catalog, the flux excesses expected in the region of the sky near the Virgo cluster are however more in tension with the data and play an important role in the outcome of the likelihood analysis.}

%Concerning the search of a flux excess in the UHECR skymap, we found that the values derived from the Auger data for both the maximum significance of a blind search and the maximum significance of a flux excess in the direction of Cen~A were individually relatively easy to reproduce with our baseline simulations (see Sect.~\ref{sec:BS}), but that the pair of values $(\sigma_{\rm max}^{\rm BS},\,\sigma_{\rm max}^{\rm CenA})$ was not very common. This is mostly due to the fact that even though the region of the sky where the largest significances are expected to be found in our simulations are relatively close to { the actual BS maximum found by Auger (with a median angular distance between $\sim 15^\circ$ and $\sim25^\circ$ depending on the GMF model used) and then to CenA, the BS maximum is rarely found as close to the actual position of Cen~A as in the case of the Auger data (cf. Fig.~\ref{FigDistoCenBase} and \ref{FigDistoCenMC}). On the other hand, by performing the same BS analysis on arbitrarily large datasets, we found that for datasets with the statistics of Auger the location of the BS maximum flux excess has a dispersion of the order of $15$-$20^\circ$ around its position at infinite statistics (cf. Fig.~\ref{FigDistoInf}). These results suggest to use cautious when drawing any astrophysical conclusion from the current very close proximity between the BS maximum location and Cen~A in the Auger data.}

For a given source catalog, the fractional contributions of the various sources to the BS maximum flux excess depend on the GMF model assumed. For choices of the parameters that lead to a BS maximum significance comparable to that of Auger, the corresponding excess can be attributed to the cumulative contribution of a sizable number of sources, with the brightest source providing only a rather modest contribution to the relative flux excess, $r$ (cf. Fig.~\ref{FigContribVL1}). In the case of the JF12+Planck GMF model a strong contribution of the sources in Virgo cluster and the nearby structures (see also \citet{Ding2021}) is expected from our simulations, while it is not the case for the Sun+Planck GMF model as a result of a strong magnetic demagnification of the Virgo region. %It also follows that the position of the BS maximum is generally expected to be somewhat shifted towards the direction of Virgo when the JF12+Planck model is used.
When using our baseline catalog and the Sun+Planck GMF model, we find that Cen~A is in general the source contributing the most to the BS maximum flux excess, even though its central direction remains in general at an angular distance of the order of $\sim 20^\circ$ (median value) from the actual position of that Galaxy. This result provides an important reminder that the position of the maximum flux excess (even with very large statistics, such as obtained by combining our 300 Auger-like simulated datasets) is not necessarily located very close to the celestial position of the most contributing source.

\subsection{Likelihood analysis}

%One of the most striking features of the Auger dataset at the highest energy, is the absence of any notable flux excess in the region of Virgo cluster. This absence probably has a strong influence on the fact that the SBG catalog is preferred when the likelihood analysis is performed on Auger data as discussed in Sect.~\ref{Seclike}. This characteristics can be either at odd or in line with our simulations when the UHECR source distribution is assumed to follow that of the galaxies depending on which GMF model is used to produce our simulated dataset.
%In the case of the JF12+Planck GMF model, our baseline simulated datasets predict a strong contribution from the Virgo region which result in a BS maximum location systematically shifted northward in Galactic coordinates with respect to what is observed by Auger and to a better rejection of isotropy with the 2MRS catalog hypothesis when the likelihood analysis is performed. On the other, hand, when using the Sun+Planck GMF, the demagnification of the Virgo region results in a different outcome for the likelihood analysis, namely a better rejection of isotropy with the SBG source model and a BS maximum in most case shifted to lower galactic latitudes.  

One of the most striking features of the Auger dataset at the highest energy is the absence of any notable flux excess in the region of Virgo cluster. This absence probably has a strong influence on the fact that the SBG catalog is preferred when the likelihood analysis is performed on the Auger data, as discussed in Sect.~\ref{Seclike}. This feature can be either at odds or in line with our simulations when the UHECR source distribution is assumed to follow that of the galaxies, depending on which GMF model is assumed to produce the simulated datasets.

In the case of the JF12+Planck GMF model, our baseline simulated datasets predict a strong contribution from the Virgo region, which results in a direction of the BS maximum flux excess that is systematically shifted northwards in Galactic coordinates with respect to that obtained with the Auger data, and to a more significant rejection of isotropy based on the likelihood analysis with the 2MRS catalog hypothesis than with the SBG catalog. On the other, hand, when using the Sun+Planck GMF, the magnetic demagnification of the Virgo region results in the opposite outcome for the likelihood analysis, namely a more significant rejection of isotropy with the SBG source model and a BS maximum flux excess shifted in most case to lower Galactic latitudes.

To obtain the same outcome with the JF+Planck GMF model, one would have to make some additional hypotheses regarding the source distribution, such as assuming that large galaxy clusters, thus in particular the Virgo cluster, are much weaker sources of UHECRs than regions with less concentrated galaxies. In Sect.~\ref{Seclike}, we explored such a bias in a rather extreme form, assuming that such dense environments are simply devoid of any UHECR sources. More moderate versions of this bias could be justified on the basis of astrophysical scenarios where a high star formation rate plays an important role in the origin of UHECRs, since rich galaxy clusters tend to be significantly less active in this respect. Another justification could involve the trapping and destruction of UHECR nuclei in the environment of large galaxy clusters as discussed in detail in \citep{Kotera2009, Armengaud2005}. The latter could represent a viable possibility at the highest energies, say above 32~EeV, where the UHECRs responsible for the anisotropies are expected to be dominated by CNO and heavier nuclei in our models. It is however less likely to apply as much to protons and He nuclei, since they have much larger photo-nuclear interaction lengths and should thus be less affected. Their presence around 8~EeV, that is in the energy range where the dipolar modulation is most significant in the Auger data, should then still attract the dipole direction northwards, far from where it is actually observed by Auger (see the corresponding discussion in Paper~I).

It has also been argued in \citep{Kim2019} that the filamentary structure of the EGMF in the nearby universe could be responsible for the systematic deflection of UHECRs originating from the Virgo cluster toward completely different apparent arrival directions. Such a scenario was invoked to justify a potential major contribution of the UHECR events from the Virgo cluster to the TA hotspot \citep{TASpot2014, TAICRC2021, TANewSpot2021}. Such potential effects cannot be captured by our simulations, since we assume a homogeneous EGMF.

Undoubtedly, in the quest for the origin of UHECRs, it will be important to understand whether the lack of an excess of events in the region of the Virgo cluster is due to a bias in the distribution of UHECR sources with respect to that of the galaxies, irrespective of their membership of large clusters, or to the spatial distribution of the GMF and/or the EGMF. This will require better constraints on the intensity and structure of the GMF, which currently remain very uncertain (see in particular recent discussions in \citet{Unger2017,Unger2019,Boulanger2018}), and of the EGMF (see e.g., \citet{Heald2020,Beck2008} and references therein), as well as additional theoretical progress on the modeling of potential UHECR sources and their environment. In the current stage of knowledge, our results suggest that the greatest caution should be used when interpreting the differences in the p-values for isotropy rejection that are obtained with different catalogs to model the UHECR signal component.

Moreover, in Sect.~\ref{Sec:SFG-bias}, we used source catalogs obtained by forcing the presence of the SFGs used to model the signal component in \citep{AugerSFG2018} and found no significant improvement in the comparison between the data and the simulations in the framework of the likelihood analysis. This is true both from the point of view of the preferred model for the signal and the value of the signal fraction, $f_{\rm aniso}$, and angular scale, $\theta$. This result allowed us to emphasize a well-known shortcoming of the likelihood analysis, which relies on a simple Gaussian blurring to model the deflection of the UHECRs, discarding the systematic shifts of the image of the sources resulting from the crossing of the GMF (not mentioning possible magnification/demagnification effects). In the context of this simplifying hypothesis, even when the sources used to model the signal component are included in the source catalog used to produce the simulated datasets, the values of $f_{\rm aniso}$ and $\theta$ maximizing the likelihood should not be considered as an estimate of meaningful physical parameters, but rather as what they merely are, namely the set of parameters allowing to reject isotropy with the largest significance.

\subsection{UHECR source density}

Most of the models discussed in this paper assumed relatively large source densities, typically $\ga 10^{-3.5}$ $\rm Mpc^{-3}$. This is because we found that for typical values of the EGMF of the order of 1~nG and GMF coherence lengths varying from 50~pc to 500~pc, source densities as large as a few $10^{-3}$ $\rm Mpc^{-3}$ are indeed acceptable from the point of view of the resulting significance of the associated UHECR anisotropy, as estimated with the Auger statistic. In particular, similar values as with the Auger data can then easily be obtained for the significance of the BS maximum flux excess as well as for the likelihood p-values. This does not mean, however, that lower source densities should be considered as disfavored. Indeed, because the cosmic variance is larger in this case, some realizations of the lower density scenarios can result in simulated datasets with anisotropy levels similar to those found with our baseline catalog. This is mostly the case when the closest UHECR source happens to be rather distant compared to what is found for a typical realization. On the other hand, for realizations combining a low source density (say $10^{-4}$  $\rm Mpc^{-3}$ or lower) and a nearby closest source (e.g. as close as Cen~A), significantly larger EGMF (especially if those are confined to the local Universe) must be invoked to keep the level of anisotropy as low as observed.

In the absence of strong constraints on the structure and intensity of the EGMF in the local Universe, UHECR models assuming a dominant contribution of very few sources are viable in principle. Such scenarios could be favored in a context where the production of UHECRs is related to the presence of relatively rare objects, such as jetted AGNs. In that case, the source catalog compiled by \citep{Velzen2012} is interesting to use since its radio luminosity cut provides a volume-limited catalog up to $\sim$150~Mpc (with a source density of $\sim10^{-5}$ $\rm Mpc^{-3}$) from which to infer the distribution of potential sources as well as their associated UHECR luminosities. It suggests, in particular, that nearby sources such as CenA and/or FornaxA and/or M87 could make a considerable contribution to the overall UHECR flux at the highest energy. Although located at $\sim230$~Mpc away from Earth, Cygnus A has such a large intrinsic (radio) luminosity that it could also provide a strong contribution to the observed flux (but then only say below $\sim20$~EeV because of energy losses). Some scenarios assuming a dominant contribution of one or several of these nearby sources have recently been considered (see e.g., \citet{Linda2022, Matthews2018, Mollerach2019}). It will be interesting to investigate whether such models can provide a satisfactory account of the various aspects of the UHECR data, beyond one specific observable, for instance whether they can reproduce not only the position, but also the amplitude of the dipolar modulation (e.g. without overproducing the quadripolar modulation), as well as the overall level of anisotropy at high energy, without too much fine tuning of the dominant source(s) emission (spectral shape, composition, maximum energy, beaming, time evolution) and/or of the local universe magnetic field configuration. In parallel, further theoretical work on the acceleration of UHECRs in this type of sources will be welcome, as well as new progress on theoretical and observational constraints on the local universe magnetic field structure.

\subsection{Energy evolution}

Throughout this paper, we emphasized the importance of investigating the energy and angular evolution of the anisotropy signal, to reinforce and provide new constraints on the astrophysical models. Some of the existing, but currently non public data may already provide interesting information. To push this type of investigations described above up to, say, 80~EeV, a significant increase in statistics will be necessary. The levels of anisotropy (defined by the relative flux excess, $r$, introduced in Sects~\ref{SectContrib} and \ref{Sec:stat}) that this will allow to characterize are expected to be larger than those found at lower energy, because of the reduced UHECR horizon and the larger relative contribution of the nearby sources. Therefore, they may be expected to provide new constraints on the sources of UHECRs. Alternatively, as discussed in Paper~I, interesting constraints could be provided by the study of the light component (if any) above, say 20~EeV, where the horizon of He nuclei significantly drops and becomes restricted to the local universe, due to photo-disintegration interactions. This will be most valuable if the H/He abundance ratio is low at this energy, and if the separation from the heavier nuclei is efficient.  

In the previous section, we carried out a preliminary discussion of the energy evolution of the anisotropy below 32~EeV, using the Auger dataset published in the ``dipole paper'' \citep{AugerDip2017}. The examination of the Auger data above 8~EeV, shows that the flux excess in the direction of CenA has a maximum significance of $\sim2.4$ well below the one found scanning the data above 32~EeV. This energy evolution is in tension with the predictions of our simulations, in particular when our composition model A is used. 

Moreover, a BS maximum flux excess at ($l=266^\circ$,$b=-44^\circ$), somewhat above the direction of the Fornax cluster and south of where the dipole direction is found with this same dataset, is obtained. The existence of this higher-flux region clearly drags the direction of the observed dipolar component of the anisotropy toward the position where it is found in the 8--16~EeV energy bin, and can be seen as, at least partly, responsible for the discrepancy between the Auger data and our simulated datasets (see Paper~I)\footnote{In the case of the JF12+Planck GMF model, a flux excess is expected close to the direction of the Virgo cluster (see Fig.~20 of Paper I), further dragging the dipole toward the north and thus worsening the discrepancy.}. When one compares with the UHECR sky seen by Auger above 32~EeV, it is striking that the flux excess in this region has essentially disappeared.

It would be very interesting to know if the situation of this broad region is continuously evolving with energy, from which one may hope to see an astrophysically meaningful pattern emerge. Indeed, if one were to take at face value the results of Sect.~\ref{sec:energyEvol}, discarding for the sake of exploration the possibility of strong statistical fluctuations, one would be led to formulate some assumptions about possible amendments that would need to be made to the investigated scenarios. Among the possibilities, on may speculate that an extended Galactic component still contributes to the UHECR flux above the ankle, and then disappears. This would probably imply an extended magnetic halo able to confine particles at a higher energy that usually assumed, with additional consequences concerning the deflections of extragalactic UHECRs in the GMF. Another possibility could be the existence of a strong (possibly nearby) source with a particularly low energy cutoff. This would probably be more natural in the context of scenarios in which very few sources provide a large fraction of the flux observed above the ankle. %In this case, the spectral shape, composition and maximum energy of the UHECR output, as well as possible intermittency or time evolution of each of these dominant sources can have a strong impact on the observed anisotropies. {\color{red}[Cette dernière phrase est-elle vraiment utile ?]}
%, or if it essentially the result of statistical fluctuations associated with too small datasets, amplified by the search procedure, either in low, high or both energy ranges.

Although the potential implications of what would be a clear and consistent modification of the UHECR sky map as a function of energy are very interesting, it is important to stress that, at the moment, such a finding would clearly be {\it{a posteriori}}. It would thus be dangerous to draw any strong conclusion from it at this stage, even if a clear pattern, with a smooth and definite evolution with energy would appear to be present in the data. However, if such a pattern were to be observed in entirely independent datasets and could thus be considered a real, significant feature of the UHECR sky, then it could bear genuine astrophysical meaning, likely to shed long awaited light on the origin of (at least some of) the UHECRs. This is unfortunately not the case of the mere assessment of anisotropies, which by themselves do not tell us much about this origin and cannot be used in practice to constrain the proposed source models efficiently, especially in the absence of a reliable GMF model, as the results of the present paper show.

Given the {\it{a posteriori}} nature of such a discussion, it is likely that the current data will not allow to assess a behaviour such as the one sketched above (or other meaningful ones) with a high level of confidence. However, the public release of the currently collected data would allow the larger scientific community to push further such investigations. We understand the perceived risk of multiplying independent searches, compromising the statistical power of each given analysis. But since, to the best of our knowledge, the data have never been blinded within the Auger and TA collaborations, this is true anyway, whether the data are released or not. From this point of view, whether astrophysically meaningful patterns are present or not in the data, finding hints of them earlier can only be beneficial to the community in general, and allow meaningful future searches on independent datasets.

\subsection{Final comments}

In conclusion, the class of models that we have investigated in this paper and in Paper~I appeared to have difficulties to globally reproduce the Auger observations above { 8~EeV}. While, the observed anisotropy levels and significances are relatively easy to match by adjusting the free astrophysical parameters at our disposal (in particular the EGMF intensity of the GMF coherence length), some observed characteristics of the data, essentially related to the location of the predicted flux excesses and possibly to the energy evolution of the various anisotropy signals, proved to be much more challenging to reproduce even after taking into account the cosmic variance and statistical fluctuations of the simulated datasets. These discrepancies, which were also stated in Paper~I regarding the direction of the dipolar modulation, may call for a reconsideration of some of the key hypotheses underlying the astrophysical scenarios that we investigated, which rely on standard candle sources (at least in terms of spectral shape, composition and maximum energy) following the distribution of Galaxies and forming a single extragalactic component fully dominant above the ankle.

Unfortunately, as we have shown, the analysis of specific anisotropy signals can be very different depending on the assumed model of the GMF. In the absence of sufficiently strong constraints regarding this key ingredient of any modeling of the UHECR sky, it is thus difficult to judge to what extent the above-mentioned discrepancies are indicative of explicit failures of particular astrophysical assumptions, and how they could be used to make significant progress in the quest of the origin of UHECRs. As stressed all along these two companion papers, a better understanding and knowledge of cosmic magnetic fields appears crucial to push further the endeavour of deciphering the UHECR sky, which should go in parallel with improving the theoretical modeling of potential UHECR sources (in particular but not only from the point of view of the expected composition), which should help to narrow down the {\it{a priori}} range of the remaining astrophysical free parameters.

Finally concerning UHECR data themselves, given the relatively low level of anisotropies in the UHECR sky, as currently measured by the existing experiments, it appears unavoidable, if significant progress is to be made in this important field at the heart of high-energy astrophysics and astroparticle physics, that a new generation of detectors be constructed with much larger statistics and/or full sky coverage \citep{Coleman2023}.

\section{Note added in proof}

\begin{figure}[ht!]
   \centering
   \includegraphics[width=8.5cm]{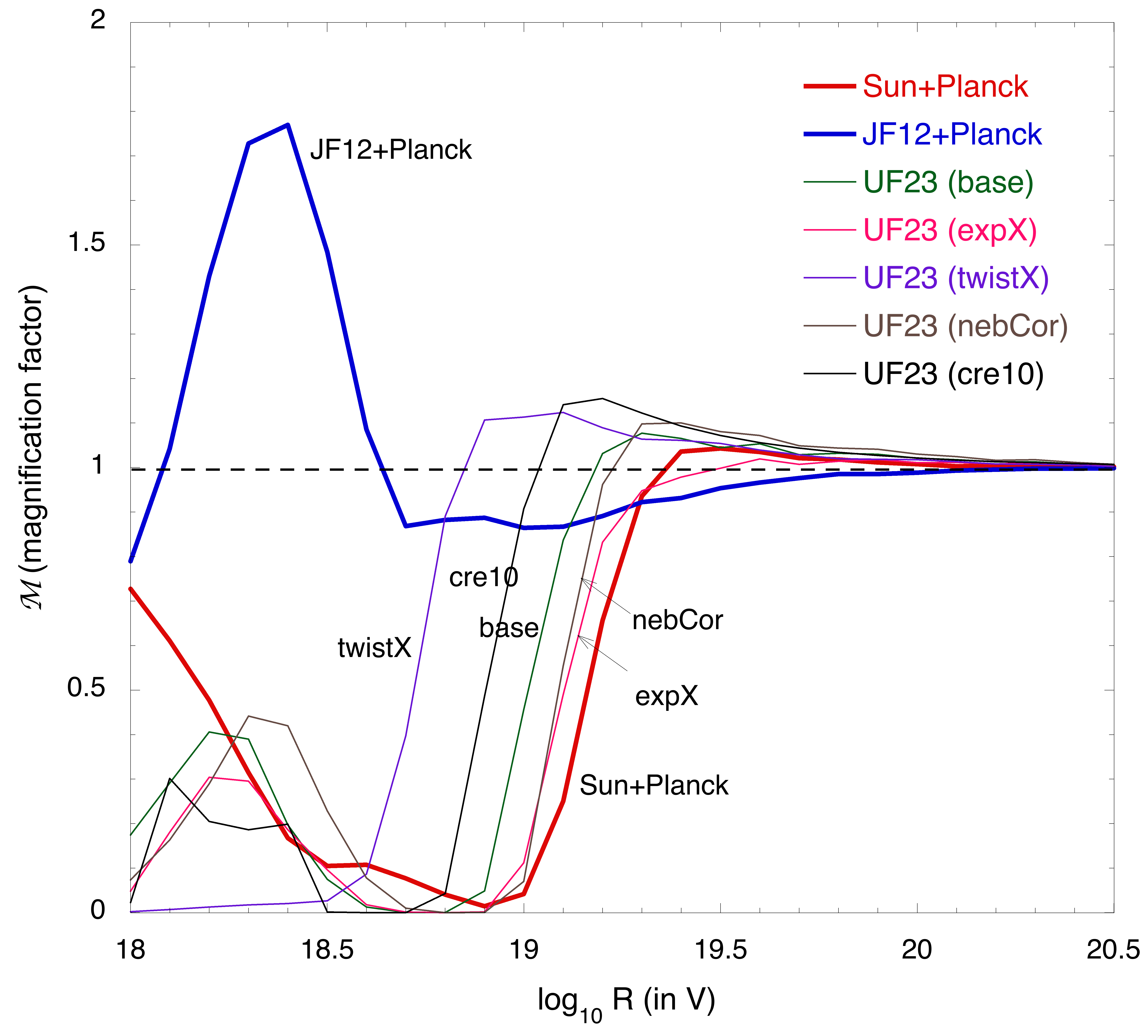}
      \caption{Magnification factors of the Virgo region ($\sim 15$ degrees around the center) as a function of particle rigidity, obtained with the different magnetic field models, as indicated. Five versions of the UF23 GMF model are shown (see \citet{Unger2023}).}
         \label{fig:magFactor}
\end{figure}

During the review process of the paper, new magnetic field models have been proposed (\citet{Unger2023}, referred to as the UF23 model), as an update of the JF12+Planck model that we used. We implemented these new models in our codes and found that the results obtained with the various versions proposed do not modify significantly the results presented here. In particular, it is confirmed that one key property of the GMF model is related to the possible magnification or demagnification of the region of the sky of the Virgo cluster, which directly impacts the position of the expected dipole, the amplitude of the quadrupole, as well as the interpretation of the outcome of the likelihood analysis and the position of the hottest spot at high energy. For reference, we show in Fig.~\ref{fig:magFactor} the magnification factor of the Virgo region as a function of rigidity, for the two GMF models used in this paper (JF12+Planck and Sun+Planck) as well as 5 of the 8 newly released models (the other 3 are omitted for clarity, but very similar to the so-called ``nebCor'' model). As can be seen, all the new models are strongly demagnifying Virgo up to a rigidity of $\sim 10^{19}$~V, except the so-called ``twistX'' version, which shows an earlier recovery. This is in sharp contrast with the original JF12+Planck model, and more similar to the Sun+Planck model. As a result, most of the new models, except twistX, favor an interpretation of the likelihood analysis similar to that obtained with the Sun+Planck model. In sum, the new models do not modify our main conclusions, in particular about the importance of examining the evolution of the anisotropy signal with energy, and further underline that the uncertainty on the GMF remains a major limitation for the astrophysical interpretation of the UHECR skymaps.

\appendix

\section{Anisotropies in the TA sky}

\subsection{TA ``hotspot'' and statistics}

We now examine the situation in the light of the anisotropy reported by the TA collaboration at the highest energies. Some of the latest results concerning the BS analysis were summarized in \citet{TAICRC2021}, where an update of the significance and position of the so-called ``hotspot'' reported in \cite{TASpot2014} is given, among other results. We note that the new anisotropy maximum significance of 5.1$\sigma$, located at equatorial coordinates (144$^\circ$, 40.5$^\circ$), was obtained without a scan on the energy threshold, inherited from the original study \citep{TASpot2014} to be 57~EeV in the TA energy scale, and with a limited and coarser scan of the angular scale, from 15$^\circ$ to 35$^\circ$ with steps of 5$^\circ$.  This results in a smaller penalty factor than for the Auger BS search, so that the claimed post-trial significance reported by TA is larger, namely 3.2$\sigma$, than the $\sim2\sigma$ significance reported by Auger \citep{AugerICRC2019}), even though the raw significance is actually larger in the latter case. To compare our simulations with the results of TA, we apply the same scanning procedure as them, i.e. fixing the energy threshold and examining only the above-mentioned five different angular scales.

To produce the simulated datasets, we applied the exposure of TA assuming maximum zenith angles for the detected UHECR events of 55$^\circ$, and accumulated events until we reached the same statistics as TA after a 10\% lowering of their energy scale to roughly correspond to that of Auger \citep{AugerTASpec2021}. This leads to datasets with 6000 events above 8 EeV \citep{TADip2020}, and brings the energy threshold for the BS analysis down to 51~EeV.

\begin{figure}
   \centering
   \includegraphics[width=7.5cm]{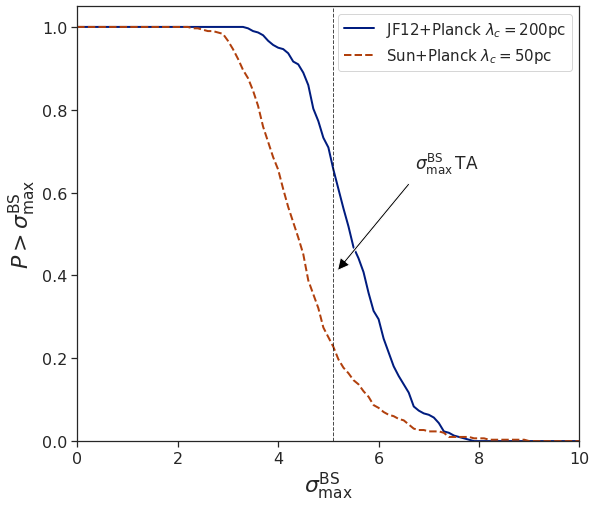}
\caption{Cumulative distributions of the largest flux excess significance from the BS analysis performed over $300+300$ TA-like simulated datasets, using sources catalogs in the mother catalog approach ($\rho=10^{-3}\,\rm Mpc^{-3}$) in which the presence of M82 is imposed, and assuming model A, for the JF12+Planck GMF model with $\lambda_{c}$=200~pc (in blue) and the Sun+Planck GMF model with $\lambda_{c}$=50~pc (in dark orange).}
         \label{FigTABS}
\end{figure}

\begin{figure}
   \centering
   \includegraphics[width=7.5cm]{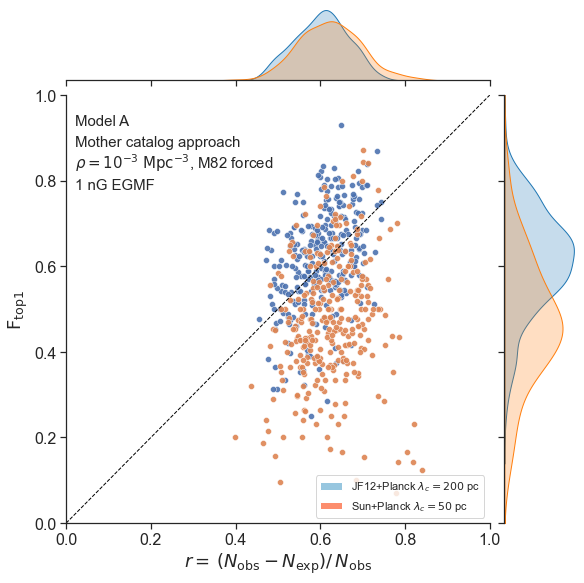}
%\caption{Results of the BS analysis performed on datasets simulated according to TA exposure for our baseline astrophysical model. The top panel shows the cumulative distribution of maximum significance obtained over 300 datasets for both the JF12+Planck ($\lambda_{c}$=200~pc) and the Sun+Planck GMF model($\lambda_{c}$=50~pc). The central and bottom panel show the distribution of the celestial position of the BS maxima for the JF12+Planck and the Sun+Planck models respectively, the position of the TA hotspot, the secondary maximum close to the direction of the Pisces-Perseus supercluster and the galaxy M82 are indicated by markers.}
\caption{Scatter plot of the fraction of events coming from the dominant source contributing to the BS maximum in TA sky vs. the relative flux excess, $r$, for the 300 realizations of the two models discussed in Fig.~\ref{FigTABS}.}
         \label{FigContribTA}
\end{figure}

\subsection{Significance of the BS maximum flux excess}

As for our comparison with the Auger data, we have produced a large amount of TA-like datasets from different underlying astrophysical model, i.e. different combinations of source parameters, source catalogs, source densities and GMF models. We only report here on the models providing datasets which are most similar to the actual TA data from the point of view of the BS maximum flux excess, i.e. with a maximum significance of the same order as that reported by TA. Those datasets were obtained from source distributions that included either M81 or M82, the latter being more naturally expected for UHECR models in which star-forming regions or starburst galaxies play an important role (see e.g., \citet{Anc1999}). These two galaxies are among the closest from the Milky-Way and located in a region of the sky where the GMF models that we considered result in magnification factors between $\sim$3 and 7, in the rigidity range of interest (see Figs.~3 and 4 of Paper~I).

In Fig.~\ref{FigTABS}, we show the cumulative distribution of the largest flux excess significance from the BS analysis performed over TA-like simulated datasets using an astrophysical based on  model A, with the mother catalog approach at a source density of $10^{-3}\,\rm Mpc^{-3}$, but with M82 forced to be present in each realization of the source catalog. All UHECR sources were also assumed to have the same intrinsic luminosity. The case of the JF12+Planck GMF model is shown in red, and that of the Sun+Planck model in blue, each with 300 independent realizations of the source distribution (except for the presence of M82). In both cases, BS maximum significances of the order or larger than that of TA are found to be rather common, although $\sim 3$ times more so with the JF12+Planck GMF for these particular parameters. This can be understood by noting that M82 appears to always make a strong contribution to the resulting anisotropy signal. Such a strong contribution of a single source is necessary to produce high values of the maximum significance for datasets of this size, which represent less than 20\% of the Auger statistics. In the present case, this can indeed be obtained from M82, even with such a high source density, because of the proximity of this source as well as its strong magnetic magnification with the GMF models under consideration. The reason why the Sun+Planck model provides lower values of the BS maximum significance with the quoted parameters is that the magnification is lower in this case, and at the same time the UHECR deflections from that region in the sky are somewhat larger, which results in a more diffuse image of the source in the range of rigidities of interest.

It is important to note that forcing the presence of M82 and/or M81 among the sources does not have a strong implication for the results of our previous study in the sky covered by Auger. 
%Indeed, it would be a mistake to assume that, because they are located outside of the Auger direct field of view, such sources would not contribute to the Auger flux. 
As a matter of fact, whenever they are present in a source catalog, these sources do make a sizeable contribution to the previously discussed simulated datasets (even though they are located outside of the Auger direct field of view), precisely because of their large magnification, which implies that the UHECRs from that region of the sky are distributed over a large fraction of the celestial sphere. This could be observed in our Auger-like simulated datasets, since M81 is indeed present in our baseline catalog, and M82 was imposed among the sources along with the other nearby star-forming galaxies, in the dedicated study presented in the previous section. However, the arrival directions of the UHECRs from M81 or M82 in the part of the sky observed by Auger happen to be nearly isotropic, so their contribution to the observed anisotropies is weak. 
%This allows to decouple both studies, 
So putting together the above results, we can conclude that a fair account of the  significances of the BS maximum flux excess reported both by Auger and TA can be obtained jointly in the general framework of our study, i.e. with the generic astrophysical scenarios investigated for a common choice of the parameters. However, the above-mentioned difficulty to account for the direction of the main anisotropies remains, and also applies in the TA sky, as we discuss in the next session.

Another interesting thing to note regarding the flux excess in the TA sky is that it can be almost entirely attribued to the contribution of one single source, in this case M82, contrary to what we saw in the case of the Auger sky, where the top 5 sources represented only $\sim 20$\% of the total number of events in the BS maximum flux excess window, and $\sim 50$\% of the flux excess (see Fig.~\ref{FigContribVL1}). By contrast, as shown in Fig.~\ref{FigContribTA}, the top source alone represents already between 40\% and 80\% of the events in the hotspot window in the case of the JF+Planck GMF model (and between 30\% and 70\% in the case of Sun+Planck), and can account for most, if not all, of the excess, even though the source density is the same as that used in the Auger sky and as large as $10^{-3}\,{\rm Mpc}^{-3}$ in this example. This difference between the weight of the top source in the Auger and TA skies could also be responsible for different spectra measured at the highest energies by the two experiments. However, an excess of the TA flux as large as currently reported when a simple and constant energy rescaling is performed between the two experiments still appears out of reach of these models (see \citet{GAPLP2017} for discussion). Note however that the current status and amplitude of the spectral differences between Auger and TA is still uncertain, since the necessity of an energy-dependent rescaling has been pointed out \citep{AugerTAspe2019, AugerTASpec2021, Tiny2021}.

\begin{figure}
   \centering
   \includegraphics[width=7.5cm]{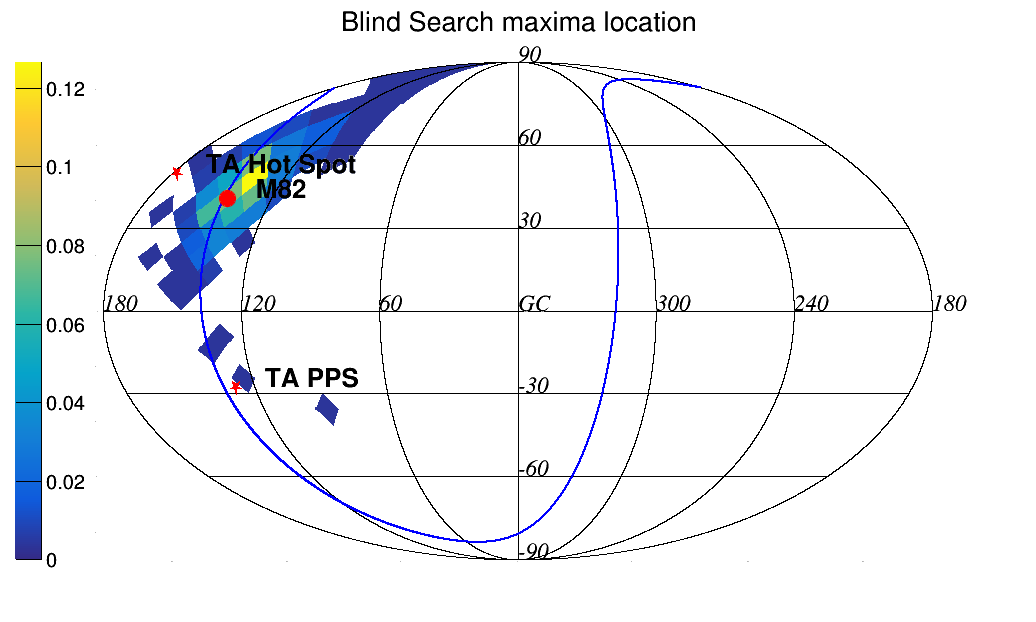}
   \includegraphics[width=7.5cm]{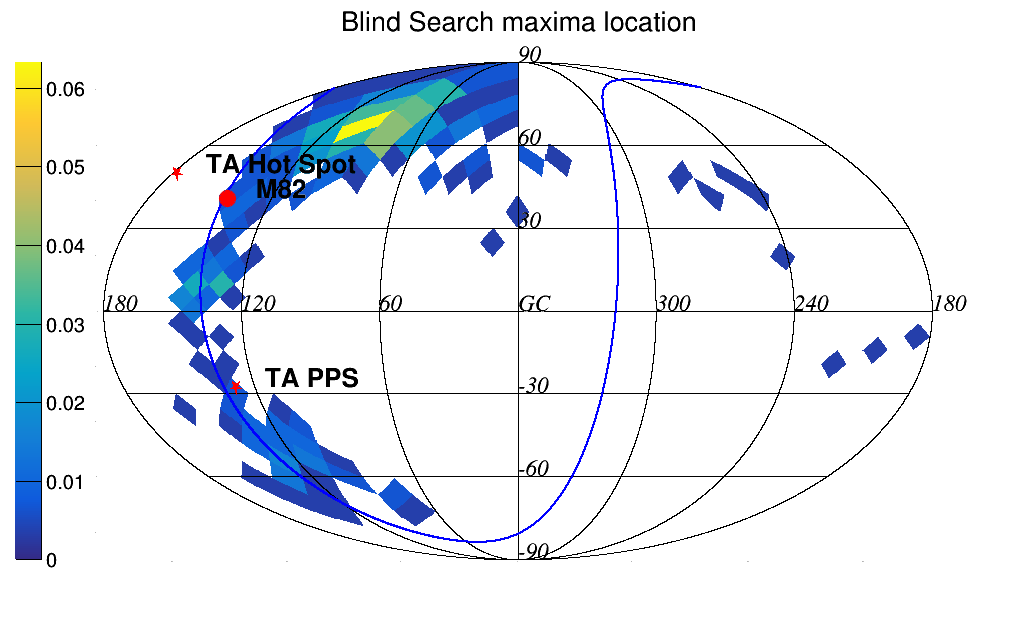}
%\caption{Results of the BS analysis performed on datasets simulated according to TA exposure for our baseline astrophysical model. The top panel shows the cumulative distribution of maximum significance obtained over 300 datasets for both the JF12+Planck ($\lambda_{c}$=200~pc) and the Sun+Planck GMF model($\lambda_{c}$=50~pc). The central and bottom panel show the distribution of the celestial position of the BS maxima for the JF12+Planck and the Sun+Planck models respectively, the position of the TA hotspot, the secondary maximum close to the direction of the Pisces-Perseus supercluster and the galaxy M82 are indicated by markers.}
\caption{Distribution of the direction of the BS maximum flux excesses for the same TA-like datasets as in Fig.~\ref{FigTABS} (see text) for the JF12+Planck (top) and the Sun+Planck (bottom) GMF models. The directions of the TA hotspot and of the secondary maximum close to the direction of the Pisces-Perseus supercluster are shown by red stars. The direction of the M82 galaxy is shown by a red circle. The colour scale indicates the fraction of the 300 datasets for which the BS maximum flux excess is found in the corresponding pixel.}
         \label{FigTABSPosition}
\end{figure}

\subsection{Direction of the BS maximum flux excess}

In Fig.~\ref{FigTABSPosition}, we show the distribution of the location of the BS maximum flux excess on the sky, for the same simulated datasets as discussed in the previous section. The top and bottom panels correspond to the JF12+Planck and Sun+Planck GMF models, respectively. The colour indicates the fraction of the 300 datasets that yield a BS maximum excess direction in the corresponding sky pixel. The position of the TA hotspot is shown (along the leftmost edge of the map). The position of M82 is also shown as a red circle.

As can be seen, for both GMF models, the position of the TA hotspot is not naturally accounted for by any of our simulated datasets. Although this result by itself does not rule out M82/M81 as a possible origin for the reported TA hotspot (see discussions in \citet{He16,Pfeffer16}), it does not support such a claim either, contrary to what could have seemed to be the case on the sole basis of the value of the maximum significance. As discussed above and in Paper~I, it appears once more that when proposing an astrophysical model to interpret the UHECR observations, it is not only important to take into account the large uncertainties regarding, in particular, the GMF model, but also to make sure that all observables are taken into account, which includes, in addition to the spectrum and the composition of the UHECRs, the amplitude of the various types of anisotropies as well as their distribution over the sky.

It is also worth remembering that the characterization of the anisotropy signal itself is still uncertain. Recently, the TA collaboration reported on the time evolution of the hotspot signal \citep{TAICRC2021}. Unfortunately, this evolution is established for the hotspot as it appears in its final position and at its final angular scale, rather than as it was identified in its original position and at its initial angular scale, as reported in \citet{TASpot2014}. But even so, the results suggest that the value of the maximum significance initially reported may have been affected by a rather strong positive fluctuation.

From this perspective, some caution may also be applied when considering the lower significance excess recently reported in  \citep{TANewSpot2021}, close to the direction of the Perseus-Pisces supercluster, with a different scan of the parameter space and thus different implied penalty factors. Therefore, we did not study further this secondary hotspot in the light of our models, although the corresponding position is indicated in the maps of Fig.~\ref{FigTABSPosition}, for direct visual comparison.

As rapid increase of the exposure in this part of sky is however to  be expected, resulting from the significant extension of the TA observatory that is currently ongoing \citep{TA*4ICRC2019}. This should clarify the level, direction and significance of the corresponding anisotropies at the highest energies.

\end{document}